\documentclass{article} % For LaTeX2e
\PassOptionsToPackage{table}{xcolor}
\usepackage{iclr2027_conference,times}

\usepackage{amsmath,amsfonts,bm}

\def\eqref#1{equation~\ref{#1}}
\def\1{\bm{1}}

\DeclareMathAlphabet{\mathsfit}{\encodingdefault}{\sfdefault}{m}{sl}
\SetMathAlphabet{\mathsfit}{bold}{\encodingdefault}{\sfdefault}{bx}{n}

\usepackage{hyperref}
\usepackage{url}
\usepackage{booktabs}       % professional-quality tables
\usepackage{amsfonts}       % blackboard math symbols
\usepackage{nicefrac}       % compact symbols for 1/2, etc.
\usepackage{microtype}      % microtypography
\usepackage{xcolor}         % colors ([table] passed above)
\usepackage{amsmath}
\usepackage{graphicx}
\usepackage{multirow}
\usepackage{enumitem}
\usepackage{multicol}
\usepackage{amsthm}
\usepackage{tcolorbox}
\usepackage{wrapfig}
\tcbuselibrary{breakable}
\usepackage{subcaption}
\usepackage{float}
\usepackage{algorithm}
\usepackage{algpseudocode}
\usepackage{caption}
\usepackage{graphicx}
\usepackage{amssymb}
\usepackage{array}

\usepackage{tabularx}
\newcolumntype{L}{>{\raggedright\arraybackslash}X}   % flexible, left-aligned
\newcolumntype{R}{>{\raggedleft\arraybackslash}X}    % flexible, right-aligned
\usepackage{makecell}
\usepackage{mathtools}
\usepackage{ulem}
\setlist[itemize]{noitemsep,leftmargin=*,topsep=0pt}

\definecolor{headergray}{RGB}{232,232,232}
\definecolor{sectiongray}{RGB}{247,247,247}
\definecolor{armblue}{RGB}{230,240,248}
\definecolor{bestblue}{RGB}{195,222,238}
\definecolor{secondorange}{RGB}{250,221,184}
\definecolor{gainbg}{RGB}{226,240,218}
\definecolor{lossbg}{RGB}{247,224,224}
\newcommand{\best}[1]{\cellcolor{bestblue}\textbf{#1}}
\newcommand{\second}[1]{\cellcolor{secondorange}\textbf{#1}}
\newcommand{\gain}[1]{\cellcolor{gainbg}\textbf{#1}}
\newcommand{\loss}[1]{\cellcolor{lossbg}\textbf{#1}}
\newcommand{\tabsetup}{%
  \setlength{\aboverulesep}{0pt}\setlength{\belowrulesep}{0pt}%
  \setlength{\extrarowheight}{1pt}\renewcommand{\arraystretch}{1.05}}
\newcommand{\hdr}[1]{\rule{0pt}{2.6ex}\textbf{#1}}
\newtcolorbox{rqbox}{%
  colback=gray!5!white, colframe=gray!75!black, boxrule=0.4pt, sharp corners,
  breakable=false, left=5pt, right=5pt, top=2.5pt, bottom=2.5pt,
  before skip=4pt, after skip=4pt}
\newcommand{\fitwidth}[1]{%
  \resizebox{\ifdim\width>\linewidth\linewidth\else\width\fi}{!}{#1}}
\newcommand{\stg}[1]{{\ttfamily\scriptsize #1}}
\newcommand{\ceilat}{\textsc{Ceil@1}}
\newcommand{\deadat}{\textsc{Dead@0}}

\title{Large-scale Repository Engineering via Agent-Native Reusable Code Primitives}

\author{
\begin{tabular}{@{}l@{}}
Haibo Jin$^{1}$\quad
Peng Kuang$^{1}$ \quad
Xucheng Yu$^{1}$ \quad
Jerry Wang$^{1}$ \quad
Dehao Wu$^{1}$ \quad
Haohan Wang$^{1}$\thanks{Corresponding Author: \texttt{haohanw@illinois.edu}}  \\[3pt]
{\small $^{1}$University of Illinois Urbana-Champaign, USA}
\end{tabular}
}

\iclrfinalcopy
\begin{document}

\maketitle
\fancyhead{}
\lhead{Preprint.}

\begin{abstract}
Large language models (LLMs) equipped with development environments have moved code generation toward repository-scale construction, yet building complete repositories remains difficult because interacting modules, interfaces, configurations, tests, and dependencies must work together. We introduce \textbf{Code Primitives}, agent-native reusable executable components that carry their own interface contracts, dependency closures, validation tests, and provenance. Each primitive uses a resident LLM as a natural-language interface to assess relevance and adapt its implementation, interfaces, and dependencies to the target repository, and we organize $1{,}424$ validated primitives in \textbf{CodeFace}, a searchable library for repository construction. Each primitive adapts the component it owns. We introduce \textbf{LEGO} (\textbf{L}arge-scale repository \textbf{E}ngineering via a\textbf{G}ent-native reusable c\textbf{O}de primitives), which activates the primitives a task requires, integrates their adapted implementations with task-specific code while resolving constraints that cross component boundaries, and revises the result against executed tests. To measure construction end to end we build \textbf{LEGO-REPO}, a benchmark of $522$ executable reconstruction tasks spanning seven software domains, $22$ capability tracks, and five difficulty levels, each scored against the package's native test suite between an empty-package floor and an original-source ceiling. The strongest of $13$ evaluated backbones reaches a delivery score of $0.318$ and scores zero on $41.0\%$ of tasks; LEGO improves all $13$ by $0.1474$ on average and raises GPT-5.6-terra from $0.3180$ to $0.5134$ ($+61.4\%$). In controlled comparisons, adapted primitives achieve higher delivery scores than retrieved code supplied as context or vendored unchanged. The effect persists against independent repository agents, across three external benchmarks, and with a disjointly re-mined CodeFace, and using GPT-OSS-20B for adaptation and diagnosis retains $95.1\%$ of the homogeneous score at $24.0\%$ lower cost.
\end{abstract}

\section{Introduction}
\label{sec:intro}
Large language models with development environments have moved code generation from isolated programs toward repository-scale construction~\citep{openai2025codex,anthropic2025claudecode,github2025copilotagent,peng2026repogenesis,ding2025nl2repo}, a setting requiring complete projects whose modules, interfaces, configurations, tests and dependencies work together. This shift raises three high-level design questions about how such repositories can be reliably constructed.

\textbf{(1) How reliably can frontier models construct a complete repository end to end?} Existing evaluations confirm the difficulty: on RepoZero-Py2JS, Claude Sonnet 4.6 with mini-SWE-agent reaches a $54.7\%$ pass rate~\citep{zhang2026repozero}, with similar results elsewhere~\citep{luo2026rpg,chen2026repo0}. On our benchmark, the strongest of $13$ LLMs reaches a delivery score of $0.318$, where $0$ denotes the empty-package floor and $1$ the behaviour of the original source, and scores zero on $41.0\%$ of $522$ tasks; Claude Opus 5 reaches $0.237$ and matches the original behaviour on $5.9\%$; and the mean score across backbones falls $10.3\times$ from the smallest to the largest difficulty band. These results motivate approaches that reuse and coordinate existing software capabilities.

\textbf{(2) What reusable abstraction can better support repository construction?}
Prior work encapsulates reusable capabilities as agent-native behaviors and executable tools~\citep{jin2026agent, jin2026harness}, which are typically invoked externally and remain fixed during use. Repository construction requires a different form of reuse: the reused artifact becomes part of the delivered source, so its interfaces, dependencies, and configuration must be reconciled with the target repository. This form of reuse combines retrieval with adaptation. We introduce \textbf{Code Primitives}, reusable executable components that carry their own interface contracts, dependency closures, validation tests, and provenance. Each primitive uses a \emph{resident LLM} as its natural-language interface to assess relevance and adapt its implementation, interfaces, and dependencies to the target. We then mine and validate these primitives from real-world repositories and organize them into \textbf{CodeFace}, a library of $1{,}424$ Code Primitives.

\textbf{(3) What construction pipeline can better leverage reusable software capabilities?}
We treat repository construction as resolving a specification against existing capabilities and integrating reusable components with task-specific code. We introduce \textbf{LEGO} (\textbf{L}arge-scale repository \textbf{E}ngineering via a\textbf{G}ent-native reusable c\textbf{O}de primitives), which coordinates Code Primitives through three mechanisms: \emph{retrieval and activation} decomposes a request into required capabilities and activates only the primitives that cover them; \emph{primitive collaboration} lets activated primitives adapt their implementations while propagating interface and dependency requirements to connected components; and \emph{diagnosis and verification} executes available and synthesized tests, localizes failures to components or their interactions, and reactivates affected primitives for revision. This loop integrates retrieval, adaptation, and verification until validation succeeds or the construction budget is exhausted. To evaluate LEGO, we introduce \textbf{LEGO-REPO}, a benchmark of $522$ executable reconstruction tasks drawn from real GitHub projects across seven software domains, $22$ capability tracks, and five measured difficulty levels (\textbf{D1--D5}). Each task reconstructs a Python package from its public interface and evaluates it against the package's native test suite.

We evaluate the aforementioned questions through six empirical RQs spanning from-scratch construction, reusable capabilities, construction mechanisms, model substitution, cost, and transfer. Across \textbf{LEGO-REPO}, LEGO improves all $13$ evaluated backbones, with a mean absolute gain of $0.1474$ ($+76.5\%$ relative). Under GPT-5.6-terra, the delivery score rises from $0.3180$ to $0.5134$ ($+61.4\%$), with fewer zero-score tasks and more repositories reproducing original-source behavior. The gains persist against independent repository agents, on three external benchmarks, and with a disjointly re-mined CodeFace. Adapted primitives also outperform retrieved code used as read-only context or vendored unchanged. Finally, GPT-OSS-20B adaptation and diagnosis retain $95.1\%$ of the homogeneous GPT-5.6-terra score at $24.0\%$ lower cost. Our main contributions are as follows:
\begin{itemize}
\item We design \textbf{Code Primitives}, agent-native reusable software components that carry their own interface contracts, dependencies, validation tests, and provenance, and adapt themselves to target repositories through their natural-language interfaces. We further build \textbf{CodeFace}, a library of $1{,}424$ validated Code Primitives for retrieval and reuse.
\item We introduce \textbf{LEGO}, a repository-construction framework built on \textbf{CodeFace} that makes this abstraction usable at repository scale: it activates only the primitives a task requires, delegates adaptation to the component that owns each implementation, and uses executed tests to guide integration and revision across component boundaries.
\item Experiments on \textbf{LEGO-REPO}, our benchmark of $522$ repository-reconstruction tasks, show that LEGO improves all $13$ evaluated LLMs by $0.1474$ on average and raises GPT-5.6-terra from $0.3180$ to $0.5134$ ($+61.4\%$). With matched backbones, LEGO outperforms OpenHands (GPT-5.6-terra) by $57.3\%$ and Claude Code (Claude Sonnet 5) by $56.2\%$, while GPT-OSS-20B adaptation and diagnosis retain $95.1\%$ of the homogeneous GPT-5.6-terra score at $24.0\%$ lower cost.
\end{itemize}

\section{Related Work}
\label{sec:rw}
\textbf{Coding agents and repository-level evaluation.}
LLM-based software engineering has progressed from direct code generation toward agents that explore repositories, localize code, edit, execute, and iterate from feedback~\citep{yang2024sweagent,zhang2024autocoderover,xia2024agentless,wang2025openhands,thai2025sweevo}, with evaluation expanding from standalone functions~\citep{chen2021evaluating,zhuo2025bigcodebench} to cross-file context~\citep{liu2024repobench,ding2023crosscodeeval,li2024deveval}, repository repair~\citep{jimenez2024swebench,thai2025sweevo}, and zero-to-one repository generation~\citep{zhang2026repozero,li2026realbench,luo2026rpg,chen2026repo0}. Execution-guided refinement is a complementary mechanism~\citep{chen2022codet,shinn2023reflexion}.

\textbf{Software reuse, retrieval, and reusable abstractions.}
Software reuse has long emphasized interface compatibility and adaptation cost~\citep{szyperski2002component}. Learned systems construct reusable libraries, abstractions, skills, or tools~\citep{ellis2021dreamcoder,grand2024lilo,stengel2024regal,wang2024trove,cai2024large,yuan2024craft}; retrieval-augmented generation supplies existing code or repository context~\citep{lu2022reacc,zhou2022docprompting,zhang2023repocoder}; and Agent Primitives~\citep{jin2026agent} and Tool Primitives~\citep{jin2026harness} expose reusable behaviors and tools through agent-native interfaces. These approaches typically reuse artifacts as context or through fixed interfaces, rather than adapting them as part of the delivered repository.

\textbf{Key Differences.}
Prior reusable abstractions are typically consumed by \emph{invocation} or as retrieved context: the artifact remains fixed while a central agent decides how to use it. Code Primitives instead support reuse by \emph{adaptation}. Each primitive becomes part of the delivered repository, carries its own interface contract, dependency closure, and validation tests, and can modify its implementation and dependencies to fit the target. This moves part of repository construction from a single generator into the components that own the reused functionality. LEGO then coordinates which primitives to activate, how their requirements propagate across components, and how failed executions trigger further revision. LEGO-REPO evaluates the resulting package end to end against the original repository's native test suite. Extended comparisons appear in Appendix~\ref{app:extended-rw}.

\section{Methodology}
\label{sec:method}
\subsection{Overview}
\label{sec:overview}

\begin{figure*}[t]
    \centering
    \includegraphics[width=1\linewidth]{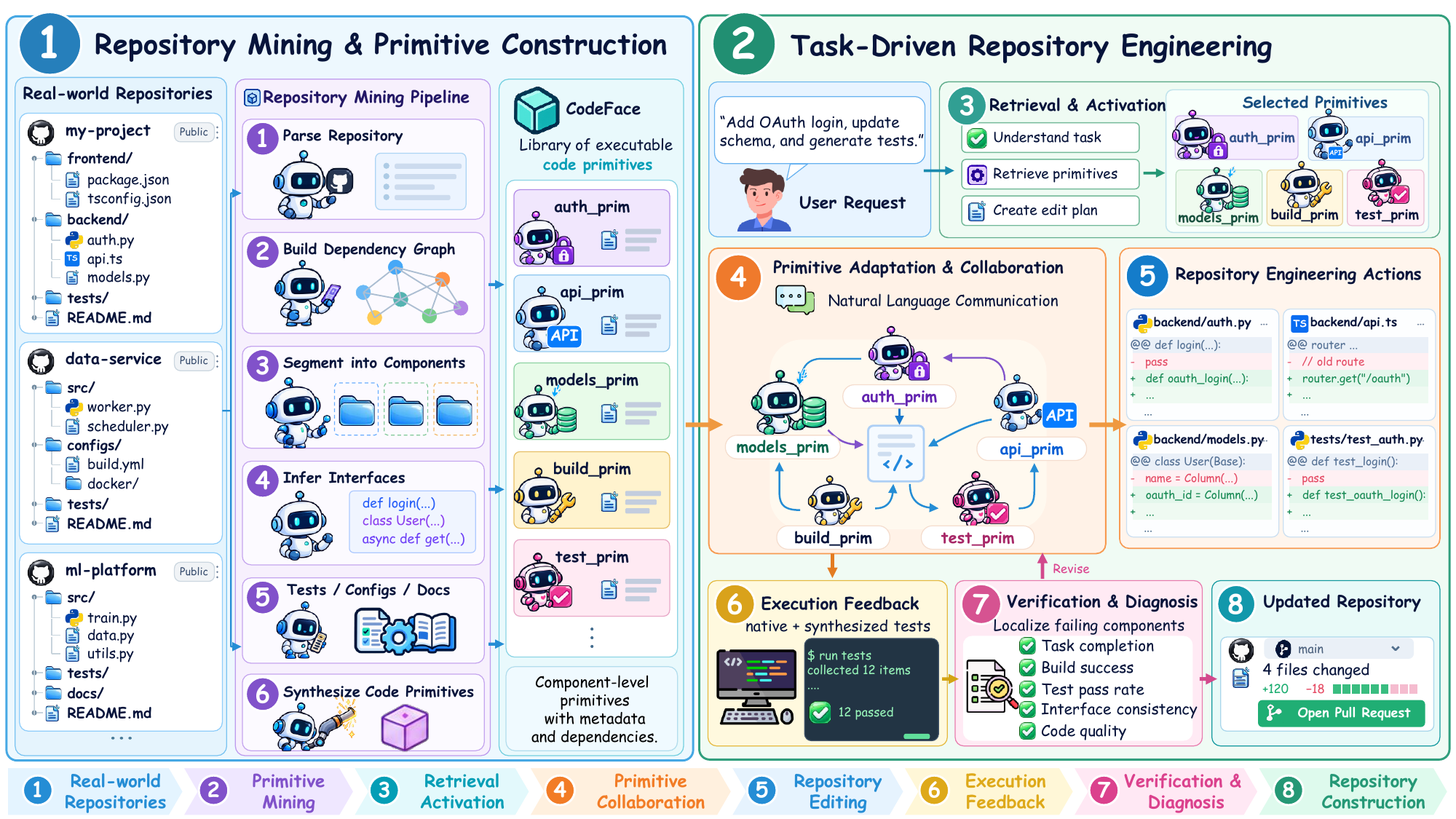}
    \vspace{-20pt}
    \caption{\textbf{From Code Primitives to complete repositories.} \emph{Left:} real-world repositories are mined into component-level Code Primitives, each carrying its interface, dependencies, tests, configuration, and documentation, and organized in CodeFace. \emph{Right:} a user request activates the required primitives, which adapt themselves and exchange the requirements their changes impose, before the adapted implementations are edited into the repository and revised under execution feedback and diagnosis.}
    \label{pipline}
    \vspace{-15pt}
\end{figure*}

\textbf{LEGO} is a repository-construction pipeline built on \textbf{CodeFace}, a library of reusable \textbf{Code Primitives} (Figure~\ref{pipline}). CodeFace is built once by mining real repositories into component-level primitives; construction then proceeds per task over the primitives a request activates. Adaptation is performed by the component that owns the implementation, while constraints that cross components are resolved at the repository level.

\subsection{Code Primitives and CodeFace}
\label{sec}

A \textbf{Code Primitive} is a reusable executable software component equipped with a resident LLM. Each primitive implements a coherent capability and exposes a natural-language interface through which it can assess its relevance to a construction task and adapt itself to a target repository. A code primitive may span functions, classes, files, configuration, and supporting artifacts; its boundary is determined by functional coherence and the dependencies required for independent execution.

Formally, we represent a primitive as $P_i=(C_i,I_i,D_i,V_i,X_i,M_i)$,
where $C_i$ is the implementation, $I_i$ the exposed interface and behavioural contract, $D_i$ the dependency closure, $V_i$ the validation tests carried by the primitive, $X_i$ supporting context and provenance, and $M_i$ the resident LLM. We write $V_i$ rather than $T_i$ because $T$ denotes a benchmark task in Section~\ref{sec:bench}. Granularity is induced by mining rather than manually tuned: in the evaluated snapshot the median primitive contains $214$ lines across $2$ files, exports $4$ symbols, and carries $9$ validation tests.

Given requirement $r$ and construction context $\mathcal{C}$, a primitive first assesses its relevance and proposes how it could be used:
\begin{equation}
(s_i,u_i)=M_i^{\mathrm{assess}}(P_i,r,\mathcal{C}),
\label{eq:assess}
\end{equation}
where $s_i$ is the relevance assessment and $u_i$ the usage proposal. If selected, it produces a target-specific variant and the requirements its rewrite imposes on other components:
\begin{equation}
(\widetilde{P}_i,\mu_i)=M_i^{\mathrm{adapt}}(P_i,r,\mathcal{C}),
\label{eq:adapt}
\end{equation}
where $\mu_i=\{\mu_{i\to j}\}_j$ is empty for locally contained changes. Adaptation may modify interfaces, representations, dependencies, configuration, or implementation while preserving the intended capability. It operates on the primitive's own state and a bounded context containing the target location, required exports, and interfaces to bind, without access to unrelated implementations, the native test suite, or the global repository plan. The primitive then updates its interface and dependency records and reruns $V_i$ on the adapted component. This carried validation distinguishes primitive adaptation from generic editing. Appendix~\ref{app:prompts} gives the full contract and rejection criteria.

\textbf{CodeFace} organizes the validated primitives as a library $\mathcal{F}=\{P_i\}_{i=1}^{N}$. Rather than exposing the entire library in context, LEGO retrieves only the candidates required by the current task and invokes them on demand. Each entry retains its provenance, so mined, harvested, web-sourced, and target-derived components can be distinguished during analysis. The evaluated snapshot holds $871$ mined, $343$ harvested, and $210$ web-sourced entries, for $1{,}424$ validated Code Primitives.

\subsection{Mining and Validating Code Primitives}
\label{sec:mining}

We recover executable, functionally coherent components from existing software in six steps: each repository is \emph{parsed} into program entities and relations; a \emph{dependency graph} records what each entity requires; the graph is \emph{segmented into components} whose boundaries are refined against neighbouring code, documentation, and tests; each candidate's \emph{interface} is inferred from its exported symbols and observed usage; the reachable \emph{tests, configurations, and documentation} are collected into $V_i$ and $X_i$, with local helpers internalized and external requirements recorded in $D_i$; and the candidate is \emph{synthesized into a Code Primitive} and validated in isolation, build, import, runtime, and test failures guiding boundary refinement until it is admitted or excluded. Mining is supplemented by implementations harvested from verified construction runs and by targeted web sourcing under the same requirements. Appendix~\ref{app:mining} details recovery, dependency closure, validation, and admission.

\subsection{Large-scale Repository Engineering via LEGO}
\label{sec:lego}

\textbf{Retrieval and Activation.}
Given request $q$ and context $\mathcal{C}_t$, LEGO decomposes the task into an ordered set of capability
requirements
\begin{equation}
\Pi_t=(r_1,\ldots,r_K)=\operatorname{DECOMPOSE}(q,\mathcal{C}_t),
\label{eq:decompose}
\end{equation}
each $r_k$ carrying the interface the capability must expose and the dependencies it may assume. For every
$r_k$, retrieval returns a candidate set $\mathcal{F}_k\subset\mathcal{F}$ with $|\mathcal{F}_k|=m$ ($m=2$ in
the evaluated configuration), whose primitives report $(s_i,u_i)$ by Eq.~\ref{eq:assess}. LEGO selects an activation set $\mathcal{A}_t\subseteq\bigcup_k\{i:P_i\in\mathcal{F}_k\}$ on capability coverage and mutual compatibility;
requirements with $\mathcal{F}_k$ empty of a suitable match are left to task-specific code $g_t$. Only
$\{P_i\}_{i\in\mathcal{A}_t}$ are instantiated, so adaptation is performed only for the activated components.

\textbf{Primitive Adaptation and Collaboration.}
A repository is assembled from several primitives at once, so adaptation is not independent across
components: a primitive whose rewrite changes an exported signature, a data representation, or a
configuration key emits a requirement $\mu_{i\to j}$ to each consumer $j$, again in natural language.
Writing $k(i)$ for the capability $P_i$ was selected to cover and $\mu_{\to i}=\{\mu_{j\to i}:j\in\mathcal{A}_t\}$ for
the requirements addressed to it, round $t$ instantiates Eq.~\ref{eq:adapt} as
\begin{equation}
(\widetilde{P}_i,\;\mu_i)=M_i^{\mathrm{adapt}}\bigl(P_i,\;r_{k(i)},\;\mathcal{C}_t\cup\mu_{\to i}\bigr),
\qquad i\in\mathcal{A}_t,
\label{eq:collab}
\end{equation}
so a requirement discovered while rewriting one component is passed to its consumers within the same
round. Messages travel only among $\mathcal{A}_t$, along the dependency edges $\Pi_t$ records.

\textbf{Repository Editing.}
The adapted implementations are edited into the repository together with $g_t$, giving the repository
state $\mathcal{S}_t=\operatorname{INTEGRATE}(\{\widetilde{P}_i\}_{i\in\mathcal{A}_t},\,g_t)$. Editing resolves the
constraints that remain across primitive boundaries, including API bindings, shared data representations and
types, package dependencies, and configuration; adaptation inside a component is performed by
Eq.~\ref{eq:collab}.

\textbf{Execution Feedback.}
LEGO executes the build checks and tests available in the target, supplemented with unit and integration
tests synthesized from $\Pi_t$, producing $o_t=\operatorname{EXECUTE}(\mathcal{S}_t)$: build status, test
outcomes and failing cases, runtime errors, and traces. Execution tests whether locally
adapted components work together; no primitive observes more than the component it owns.

\textbf{Verification and Diagnosis.}
The evidence is assessed against task completion, build success, test pass rate, interface consistency, and
code quality, yielding
\begin{equation}
(v_t,\phi_t)=\operatorname{DIAGNOSE}(q,\Pi_t,\mathcal{S}_t,o_t),
\qquad v_t\in\{\textsc{pass},\textsc{fail}\},\quad \phi_t=(e_t,c_t,u_t),
\label{eq:diagnose}
\end{equation}
where $e_t$ records the observed failures, $c_t\subseteq\mathcal{A}_t$ the components they implicate, and
$u_t$ the revision guidance. Build success and test pass rate are determined by execution, so $v_t=\textsc{fail}$
whenever either fails, independently of the remaining criteria. On failure the context is enriched,
$\mathcal{C}_{t+1}=\mathcal{C}_t\cup\{\phi_t\}$, and only the primitives in $c_t$ are reactivated for another
round of Eq.~\ref{eq:collab}; the loop continues until $v_t=\textsc{pass}$ or the budget of $b=5$ rounds per module is exhausted, at which point the updated repository and its patch are returned.
Diagnosis is separable: disabling it re-enters the loop on the raw execution evidence $o_t$ in place
of $\phi_t$, revising without failure localization. Evaluation isolation
and grading are in Section~\ref{sec:bench} and Appendix~\ref{app:protocol}.

\section{LEGO-REPO Benchmark}
\label{sec:bench}
LEGO-REPO evaluates reconstruction of real Python packages from their public interfaces and native tests: generated modules must function together under the repository's original execution environment. This targets repository-level integration rather than isolated function generation, localized repair, or unconstrained design from natural-language specifications. We collect GitHub repositories with at least $1{,}000$ stars, an installable package, and a runnable test suite, yielding $522$ tasks after excluding security-sensitive projects. Each task freezes the repository commit, package interface, tests, environment, and reference test identifiers; original implementation files are removed before reconstruction. Full collection, inventory, and execution details appear in Appendices~\ref{app:funnel}--\ref{app:protocol}.

\subsection{Domains and Difficulty}
\label{sec:bench:band}

The benchmark spans seven domains and $22$ capability tracks assigned before model evaluation. Difficulty combines implementation size $m(T)$ and the identity ceiling $c(T)$:
\begin{equation}
d(T)=2\log_{10}m(T)+\log_{10}c(T),\qquad
\mathrm{band}(T)=1+\sum_{k=1}^{4}\mathbf{1}[d(T)\geq\theta_k],
\quad \theta=(4,5,6,7).
\label{eq:band}
\end{equation}
The larger weight on module count emphasizes coordination scope; repository size increases sharply from D1 to D5. Because the coefficients and cutoffs are design choices, Appendix~\ref{app:guards} repeats the analysis under two size-only binnings. The main finding---LEGO outperforming \emph{Feedback} at every scale---is unchanged.

\subsection{Scoring and Admissibility}
\label{sec:bench:score}

Let $\Pi(X)$ denote the number of native tests passed using source set $X$. The original sources define the identity ceiling $c(T)=\Pi(O)$ and an empty reconstruction tree the floor $f(T)=\Pi(\varnothing)$. For generated sources $G$,
\begin{equation}
s(G;T)=\operatorname{clamp}_{[0,1]}
\left(\frac{\Pi(G)-f(T)}{c(T)-f(T)}\right).
\label{eq:score}
\end{equation}
Thus $1$ matches the original implementation within the benchmark harness and $0$ recovers none of the measurable interval. We measure both endpoints because some native tests pass without the reconstructed implementation. Appendix~\ref{app:guards} reports floor effects, span variation, and alternative weighting analyses.

A score is admissible only when $c-f\geq5$, execution completes, and provenance confirms that the tests exercised the generated target code. Our primary metric therefore evaluates all $522$ tasks:
\begin{equation}
S_{\mathrm{delivery}}=\frac{1}{522}\sum_{T=1}^{522}\widetilde{s}_T,
\qquad
\widetilde{s}_T=
\begin{cases}
s(G_T;T),&\text{admissible},\\
0,&\text{otherwise}.
\end{cases}
\label{eq:delivery}
\end{equation}
The \textbf{zero-score rate} (\deadat) is the fraction of tasks assigned $0$; Appendix~\ref{app:funnel} separates admissible failures from non-admissible runs.

\textbf{Test-targeted special-casing.}
Because native tests serve as both specification and grading key, two annotators independently audit a band-stratified sample of $120$ reconstructions per configuration. Special-casing is not more frequent under LEGO ($4.2\%$ vs.\ $7.5\%$, $\kappa=0.81$); full screening details and uncertainty analysis are in Appendix~\ref{app:screening-records}.

\section{Experiments}
\label{sec:exp}

\subsection{\textbf{Experimental Setup}}
\label{sec:exp:setup}

We operationalize the three questions in Section~\ref{sec:intro} as six empirical RQs: how reliably frontier models construct repositories from scratch (RQ1); what capability CodeFace supplies (RQ2) and which mechanisms convert it into working repositories (RQ3); whether the effect survives changes in the backbone, adaptation, and diagnosis models (RQ4); what it costs (RQ5); and whether it transfers beyond our benchmark (RQ6).

\begin{wraptable}{r}{0.65\linewidth}
\vspace{-12pt}
\centering
\caption{\textbf{RQ1: construction from scratch, all $522$ tasks.} Band columns average over
$117/111/116/96/82$ tasks. Blue and orange mark best and second best.}
\vspace{-10pt}
\large
\label{tab:rq1}
\setlength{\aboverulesep}{0pt}
\setlength{\belowrulesep}{0pt}
\setlength{\extrarowheight}{0.35ex}

\resizebox{\linewidth}{!}{%
\begin{tabular}{l ccccc c cc}
\toprule

% ============================================================
% Header: first row
% ============================================================
\rowcolor{headergray}
&
\multicolumn{5}{>{\cellcolor{headergray}}c}{\hdr{Normalized delivery score by band $\uparrow$}}
&
&
\multicolumn{2}{>{\cellcolor{headergray}}c}{\hdr{Outcome rate}}
\\

\cmidrule(lr){2-6}
\cmidrule(lr){8-9}

% ============================================================
% Header: second row
% ============================================================
\rowcolor{headergray}
\hdr{Backbone}
& \hdr{D1}
& \hdr{D2}
& \hdr{D3}
& \hdr{D4}
& \hdr{D5}
& \hdr{All $\uparrow$}
& \hdr{$\ceilat\uparrow$}
& \hdr{$\deadat\downarrow$}
\\

\midrule

\rowcolor{sectiongray}
\multicolumn{9}{>{\cellcolor{sectiongray}}c}{\emph{proprietary API models}}
\\

GPT-5.6-terra
& \best{.525}
& \best{.382}
& \best{.274}
& \best{.206}
& \best{.129}
& \best{.3180}
& \best{8.8\%}
& \second{41.0\%}
\\

Claude Sonnet 5
& \second{.521}
& \second{.379}
& \second{.260}
& \second{.194}
& \second{.091}
& \second{.3052}
& \second{8.6\%}
& \best{40.2\%}
\\

GPT-5.4
& .485
& .301
& .216
& .123
& .061
& .2528
& 6.1\%
& 49.4\%
\\

Claude Opus 5
& .465
& .294
& .199
& .107
& .042
& .2371
& 5.9\%
& 53.4\%
\\

GPT-5.6-luna
& .442
& .262
& .177
& .103
& .045
& .2203
& 5.4\%
& 52.9\%
\\

Grok-4.3
& .363
& .215
& .122
& .066
& .036
& .1718
& 3.1\%
& 64.2\%
\\

Claude Haiku 4.5
& .339
& .170
& .106
& .044
& .028
& .1481
& 3.6\%
& 67.2\%
\\

\textit{\ \ group mean ($7$)}
& \textit{.449}
& \textit{.286}
& \textit{.194}
& \textit{.120}
& \textit{.062}
& \textit{.2362}
& \textit{5.9\%}
& \textit{52.6\%}
\\

\midrule

\rowcolor{sectiongray}
\multicolumn{9}{>{\cellcolor{sectiongray}}c}{\emph{open-weight models}}
\\

Gemma-4-31B
& .364
& .189
& .113
& .077
& .009
& .1623
& 3.8\%
& 68.8\%
\\

Kimi-K2.5
& .337
& .197
& .117
& .048
& .016
& .1548
& 3.3\%
& 68.0\%
\\

Qwen3-Coder-480B
& .343
& .194
& .104
& .049
& .013
& .1522
& 3.1\%
& 67.8\%
\\

GLM-5
& .353
& .174
& .091
& .048
& .016
& .1476
& 2.7\%
& 66.3\%
\\

DeepSeek-V3.2
& .322
& .156
& .068
& .031
& .007
& .1271
& 2.1\%
& 72.8\%
\\

Nemotron-Super-3-120B
& .253
& .155
& .051
& .024
& .008
& .1067
& 1.9\%
& \loss{76.8\%}
\\

\textit{\ \ group mean ($6$)}
& \textit{.329}
& \textit{.177}
& \textit{.091}
& \textit{.046}
& \textit{.011}
& \textit{.1418}
& \textit{2.8\%}
& \textit{70.1\%}
\\

\midrule

\rowcolor{sectiongray}
\multicolumn{9}{>{\cellcolor{sectiongray}}c}{\emph{aggregates over all $13$ backbones}}
\\

field mean
& .393
& .236
& .146
& .086
& .038
& .1926
& 4.5\%
& 60.7\%
\\

\bottomrule
\end{tabular}%
}

\vspace{-8pt}
\end{wraptable}

\textbf{Configurations and metrics.}
Configurations are named by their active stages: retrieval (\textsf{R}), primitive adaptation (\textsf{A}), execution feedback (\textsf{V}), and diagnosis (\textsf{D}), with $b=5$ attempts per module. \textsf{E}, \textsf{F}, \textsf{I}, and \textsf{C} denote generic editing, raw-file retrieval, vendor-and-invoke, and read-only context controls. GPT-5.6-terra is the default backbone, adaptation, and diagnosis model; full harness settings are given in Appendix~\ref{app:ablsettings}. We report normalized delivery score (Eq.~\ref{eq:delivery}), identity-ceiling rate (\ceilat), zero-score rate (\deadat), and recorded-call cost per task. Cost accounting is detailed in Appendix~\ref{app:baselineprov}; every table below condenses a larger one in Appendix~\ref{app:extexp}, which also reports repeated seeds with paired bootstrap intervals, per-task win/tie/loss counts, the tasks where reuse hurts, and gain against adapted-primitive count.

% \textbf{Configurations and metrics.} A configuration is named by the stages it runs: retrieval
% (\textsf{R}), primitive adaptation (\textsf{A}), diagnosis-guided revision (\textsf{D}) and
% execution-feedback revision (\textsf{V}), with $b=5$ attempts per module. \textsf{E} denotes scoped
% generic editing in place of \textsf{A}, \textsf{F} raw-file retrieval in place of CodeFace, \textsf{I}
% vendor-and-invoke, \textsf{C} read-only context. GPT-5.6-terra is the default backbone, primitive and
% diagnosis, and Table~\ref{tab:armspec} in Appendix~\ref{app:ablsettings} defines the harness settings and source restrictions. The primary
% metric is the normalized delivery score of Eq.~\ref{eq:delivery}, reported with two outcome rates,
% \ceilat\ (fraction reaching the identity ceiling) and \deadat\ (fraction scoring zero), and with
% recorded-call inference dollars per task. The accounting covers recorded backbone and diagnosis usage, successful adaptation calls only, and the fixed retrieval charge (Appendix~\ref{app:baselineprov}). Across three decoding seeds the four primary configurations have
% standard deviations of $0.0024$--$0.0043$. The four primary comparisons have paired bootstrap
% intervals and Holm-adjusted tests; Table~\ref{tab:seeds} reports these alongside the repeated-seed results.
% Every table below condenses a larger one in Appendix~\ref{app:extexp}.

\subsection{\textbf{RQ1: Can Frontier Models Construct Repositories from Scratch?}}
\label{sec:exp:rq1}

Table~\ref{tab:rq1} evaluates $13$ backbones under the same execution-feedback configuration. GPT-5.6-terra performs best with a normalized delivery score of $0.3180$, yet still scores zero on $41.0\%$ of tasks; across all LLM backbones, the field mean is only $0.1926$.

Repository scale sharply amplifies the difficulty. The field mean drops from $0.393$ on D1 (mean $6$ modules) to $0.038$ on D5 (mean $119$), a $10.3\times$ reduction, while even the strongest D5 result reaches only $0.129$. Coordinating interfaces, dependencies, and repository-level constraints therefore becomes harder as the component count grows, motivating reuse and adaptation over reconstructing every component within one trajectory.

\textbf{Execution feedback helps but does not close the gap.}
For GPT-5.6-terra, five feedback attempts improve the score from $0.2172$ in a single pass to $0.3180$, reaching the identity ceiling on $8.8\%$.

\begin{wraptable}{r}{0.62\textwidth}
\vspace{-12pt}
\centering
\caption{\textbf{RQ1: independently developed systems, all $522$ tasks.}
Each runs under its own harness with the backbone in column two; only the task set,
environment, and grading pipeline are fixed. $\ddagger$: issue-to-patch adapter
(Appendix~\ref{app:ablsettings}).}
\vspace{-6pt}
\label{tab:external}

\tabsetup\scriptsize
\setlength{\tabcolsep}{2.2pt}

\resizebox{\linewidth}{!}{%
\begin{tabular}{l l ccccc c cc c}
\toprule\rowcolor{headergray}
 & \cellcolor{headergray}
 & \multicolumn{5}{c}{\cellcolor{headergray}\hdr{Score by difficulty band $\uparrow$}}
 & \cellcolor{headergray}
 & \multicolumn{2}{c}{\cellcolor{headergray}\hdr{Outcome rate}}
 & \cellcolor{headergray} \\
\rowcolor{headergray}\hdr{System} & \hdr{Backbone}
 & \hdr{D1} & \hdr{D2} & \hdr{D3} & \hdr{D4} & \hdr{D5}
 & \hdr{All $\uparrow$}
 & \hdr{$\ceilat\uparrow$}
 & \hdr{$\deadat\downarrow$}
 & \hdr{\$/task} \\
\midrule
OpenHands
 & GPT-5.6-terra
 & .556 & .401 & .269 & .198 & .129
 & .3263 & $9.4\%$ & $41.6\%$ & 2.41 \\
Claude Code
 & Claude Sonnet 5
 & .543 & .394 & .271 & .190 & .104
 & .3170 & $9.6\%$ & $39.8\%$ & 2.16 \\
SWE-agent$^{\ddagger}$
 & GPT-5.6-terra
 & .482 & .345 & .238 & .152 & .081
 & .2750 & $6.7\%$ & $46.2\%$ & 1.86 \\
Agentless$^{\ddagger}$
 & GPT-5.6-terra
 & .421 & .276 & .181 & .098 & .043
 & .2181 & $5.4\%$ & \loss{$55.4\%$} & 0.94 \\
\midrule
\emph{Feedback} (ours)
 & GPT-5.6-terra
 & .525 & .382 & .274 & .206 & .129
 & .3180 & $8.8\%$ & $41.0\%$ & 1.55 \\
\rowcolor{armblue}\textbf{LEGO (ours)}
 & GPT-5.6-terra
 & \best{.686} & \best{.583} & \best{.483} & \best{.409} & \best{.338}
 & \best{.5134} & \best{$22.6\%$} & \best{$18.8\%$} & 2.08 \\
\bottomrule
\end{tabular}%
}

\vspace{-8pt}
\end{wraptable}

\textbf{The limitation is not our scaffold.}
Table~\ref{tab:external} compares four independently developed systems under their native harnesses. OpenHands ($0.3263$) and Claude Code ($0.3170$) are close to our Feedback baseline ($0.3180$), while SWE-agent and Agentless reach $0.2750$ and $0.2181$. All four show similarly steep D1--D5 degradation. LEGO, by contrast, reaches $0.5134$ with GPT-5.6-terra; on the matched Claude Sonnet 5, it reaches $0.4953$ versus Claude Code's $0.3170$.

\begin{rqbox}
\textbf{The answer to RQ1: from-scratch repository construction remains limited at scale.}
Even the strongest backbone reaches only $0.3180$, and average performance falls by $10.3\times$ from D1 to D5. Independent repository agents exhibit the same scale-dependent degradation.
\end{rqbox}

\subsection{\textbf{RQ2: What Reusable Capability Does CodeFace Supply?}}
\label{sec:exp:rq2}

CodeFace contains $1{,}424$ validated primitives: $871$ mined, $343$ harvested from verified construction runs, and $210$ web-sourced. An offline module-level reachability audit finds at least one candidate for $14{,}106$ of $22{,}985$ target modules, giving $61.4\%$ coverage (Table~\ref{tab:covmain}).

\textbf{Reach is broad but only loosely related to benefit.} Coverage exceeds $50\%$ in every domain,
yet moving from the lowest-reach domain (crypto-security, $52.1\%$) to the highest
(compilers-languages, $68.5\%$) changes the gain only from $+.2259$ to $+.2464$. Paired bootstrap
intervals exclude zero for the three domains holding $438$ of the $522$ tasks but not for
crypto-security ($[-.0310,+.4790]$; Appendix~\ref{app:extexp:rq3}). Coverage falls from $75.3\%$ on D1 to $56.7\%$ on D5 as mean
repository size grows from $6$ to $119$ modules, yet the improvement moves the other way: relative to
the same backbone without a library it rises monotonically, $+30.7\%$, $+52.6\%$,
$+76.3\%$, $+98.5\%$, $+162.0\%$ from D1 to D5. Reuse contributes most exactly
where construction from scratch fails most.

\begin{wraptable}{r}{0.54\linewidth}
\vspace{-12pt}
\centering
\caption{\textbf{RQ2: CodeFace reach and the runtime adaptation funnel.}
\emph{cov.} is offline coverage of the band's \emph{mod.} target modules; \emph{query}, \emph{cand.} (top-$2$), \emph{retain}, and \emph{adapt} are counted during construction. \emph{share} and \emph{used} give adapted primitives as a fraction of target and of reachable modules. Full funnel: Appendix~\ref{app:extexp:cov}.}
\vspace{-5pt}
\label{tab:covmain}
\tabsetup\scriptsize\setlength{\tabcolsep}{3pt}
\resizebox{\linewidth}{!}{%
\begin{tabular}{l rr rrrr rr}
\toprule\rowcolor{headergray}
\hdr{band} & \hdr{mod.} & \hdr{cov.} & \hdr{query} & \hdr{cand.} & \hdr{retain} & \hdr{adapt}
 & \hdr{share} & \hdr{used} \\
\midrule
D1 & \phantom{0,}721 & 75.3\% & \phantom{0,}437 & \phantom{0,}874 & \phantom{0,}357 & \phantom{0,}332 & 46.0\% & 61.1\% \\
D2 & 2{,}064 & 71.2\% & \phantom{0,}770 & 1{,}540 & \phantom{0,}613 & \phantom{0,}557 & 27.0\% & 37.9\% \\
D3 & 3{,}654 & 65.8\% & 1{,}021 & 2{,}042 & \phantom{0,}818 & \phantom{0,}731 & 20.0\% & 30.4\% \\
D4 & 6{,}823 & 61.2\% & 1{,}425 & 2{,}850 & 1{,}135 & \phantom{0,}989 & 14.5\% & 23.7\% \\
D5 & 9{,}723 & 56.7\% & 1{,}862 & 3{,}724 & 1{,}489 & 1{,}264 & 13.0\% & 22.9\% \\
\midrule
\rowcolor{armblue}\textbf{all} & \textbf{22{,}985} & \textbf{61.4\%}
 & \textbf{5{,}515} & \textbf{11{,}030} & \textbf{4{,}412} & \textbf{3{,}873} & \textbf{16.9\%} & \textbf{27.5\%} \\
\bottomrule
\end{tabular}%
}
\vspace{-8pt}
\end{wraptable}

\textbf{Activation uses a smaller share of available components at larger scales.} Across the
$522$ runs backbones issue $5{,}515$ queries ($10.6$ per task), primitives accept $4{,}412$ of
$11{,}030$ candidates ($40.0\%$), and $3{,}873$ pass their carried tests ($87.8\%$, $7.4$ per task),
delivering adapted primitives into $16.9\%$ of all target modules. Acceptance is similar across scales
($40.8\%$ on D1 against $40.0\%$ on D5), while adaptation success
falls $8.1$ points (Appendix~\ref{app:extexp:cov}); half-size subsets score $.0303$ below the full
snapshot, and retrieval depth is not binding either (Appendix~\ref{app:extexp:mix}). The \emph{used} fraction also falls: activation reaches $22.9\%$ of the D5 modules the library
could have served against $61.1\%$ on D1, because queries per task rise $3.7\rightarrow22.7$ while
modules rise $6\rightarrow119$. On D5 the unexploited reachable modules are about
$3.4\times$ those used, motivating request decomposition alongside library coverage.

\begin{rqbox}
\textbf{The answer to RQ2: CodeFace provides broad capability, but runtime use remains limited.}
CodeFace reaches about three fifths of target modules offline, yet large repositories activate under a quarter of what is available. This gap suggests that request decomposition, rather than library coverage alone, is a key bottleneck.
\end{rqbox}

\subsection{\textbf{RQ3: Which Construction Stages Convert Capability into Repositories?}}
\label{sec:exp:rq3}

\begin{wraptable}{r}{0.55\textwidth}
\vspace{-12pt}
\centering
\caption{\textbf{RQ3: construction paradigms, all $522$ tasks.}
\textsf{R}/\textsf{A}/\textsf{E}/\textsf{V}/\textsf{D} denote retrieval, adaptation, generic editing, execution feedback, and diagnosis ($b=5$).}
\vspace{-10pt}
\label{tab:paradigms}

\tabsetup\scriptsize
\setlength{\tabcolsep}{2.5pt}

\resizebox{\linewidth}{!}{%
\begin{tabular}{l c rr r r r}
\toprule
\rowcolor{headergray}
\hdr{Configuration}
& \hdr{Stages}
& \hdr{D1}
& \hdr{D5}
& \hdr{All $\uparrow$}
& \hdr{$\deadat\downarrow$}
& \hdr{\$/task} \\
\midrule

Single-shot
& \stg{----}
& ---
& ---
& .2172
& 57.1\%
& 0.86 \\

Retrieval, one attempt
& \stg{R---}
& ---
& ---
& .2314
& \loss{58.6\%}
& 0.87 \\

Feedback
& \stg{---V}
& .525
& .129
& .3180
& 41.0\%
& 1.55 \\

\midrule

Retrieval $+$ Feedback
& \stg{R--V}
& .544
& .141
& .3352
& 34.7\%
& 1.56 \\

Feedback $+$ Diagnosis
& \stg{--DV}
& .569
& .135
& .3469
& 30.1\%
& 1.79 \\

Retrieval $+$ Diagnosis
& \stg{R-DV}
& .649
& .127
& .3980
& 31.6\%
& 1.80 \\

Retrieval $+$ Adapt
& \stg{RA-V}
& .653
& .275
& .4795
& \best{17.8\%}
& 1.83 \\

RAG $+$ Edit
& \stg{RE\,DV}
& .637
& .136
& .4093
& 24.3\%
& 2.01 \\

\rowcolor{armblue}
\textbf{LEGO (ours)}
& \stg{RADV}
& \best{.686}
& \best{.338}
& \best{.5134}
& \second{18.8\%}
& 2.08 \\

\quad same-repo donors excluded
& \stg{RADV}
& .683
& .221
& .4743
& 24.9\%
& 2.08 \\

\bottomrule
\end{tabular}%
}

\vspace{-12pt}
\end{wraptable}

Table~\ref{tab:paradigms} separates the stages; Appendix~\ref{app:extexp:prim} plots the same ten
configurations as a ladder and their per-band gains (Figures~\ref{fig:ladder} and~\ref{fig:bands}),
with per-band and \ceilat\ breakdowns in Appendix~\ref{app:extexp:rq3}.
LEGO reaches $0.5134$ against $0.3469$ for the diagnosis-equipped baseline
without a library, a paired $+0.1665$ measuring retrieval and primitive adaptation jointly; against
execution feedback alone the gain is $+0.1954$ ($61.4\%$) and \deadat\ falls $22.2$ points to
$18.8\%$. Adding retrieval to Feedback is worth $0.0172$, adding primitive adaptation to Retrieval $+$
Diagnosis $0.1154$, and adding diagnosis to Retrieval $+$ Adapt $0.0339$.

\begin{wraptable}{r}{0.50\linewidth}
\vspace{-12pt}
\centering
\caption{\textbf{RQ3: construction-stage ablation.}
Each row adds one mechanism. \emph{File RAG} retrieves the top-$2$ donor files by embedding similarity without primitive-specific adaptation or validation.}
\vspace{-10pt}
\label{tab:attrib}
\tabsetup\scriptsize\setlength{\tabcolsep}{3pt}
\resizebox{\linewidth}{!}{%
\begin{tabular}{@{}l r r r@{}}
\toprule\rowcolor{headergray}
\hdr{Mechanism added} & \hdr{All $\uparrow$} & \hdr{step} & \hdr{\% tot.} \\
\midrule
Feedback (\stg{---V}), \textit{ref.} & .3180 & \textit{---} & \textit{---} \\
$+$ diagnosis revision (\stg{--DV}) & .3469 & $+.0289$ & $14.8\%$ \\
$+$ File RAG, editing (\stg{FE\,DV}) & .3528 & $+.0059$ & $3.0\%$ \\
$+$ constructed prim. (\stg{RE\,DV}) & .4093 & $+.0565$ & $28.9\%$ \\
\rowcolor{armblue}$+$ primitive adaptation (\stg{RADV}) & \textbf{.5134} & $+.1041$ & $53.3\%$ \\
\midrule
\rowcolor{sectiongray}\multicolumn{4}{l}{\cellcolor{sectiongray}\emph{roll-up}} \\
\quad construction + adaptation & & \gain{$+.1606$} & $82.2\%$ \\
\quad above + diagnosis & & \gain{$+.1895$} & $97.0\%$ \\
\quad generic file retrieval + editing & & $+.0059$ & $3.0\%$ \\
\bottomrule
\end{tabular}%
}
\vspace{-8pt}
\end{wraptable}

\textbf{A primitive is not a retrieved file plus an editing agent.}
\emph{File RAG} retrieves raw donor files and reaches $0.3528$, close to the diagnosis-equipped no-library baseline ($0.3469$). Replacing raw files with constructed primitives adds $0.0565$, and primitive adaptation adds another $0.1041$, reaching $0.5134$. Together they contribute $82.2\%$ of the total gain; diagnosis adds $0.0289$. Appendix~\ref{app:extexp:prim} decomposes the primitive channel field by field and Appendix~\ref{app:extexp:loo} reports component removals.

\textbf{Adaptation yields the largest observed reuse gains.}
\emph{Import $+$ Call} vendors the primitive unchanged and invokes it, scoring $0.3430$
against $0.3469$ for the diagnosis-equipped no-library reference. It helps on the two smallest
bands, then falls below the reference by $0.013$, $0.047$ and $0.075$ on D3--D5.
\emph{Context-only}, which shows the primitive, its contract and its tests as read-only material,
scores $0.3544$ ($+0.0075$ over the same reference).
Full adaptation raises the score to $0.5134$: $0.1590$ above context-only use and $0.1704$
above unchanged invocation. The generic-editing control lies between unchanged invocation and
full adaptation, with successive increments of $0.0663$ and $0.1041$.

\begin{wraptable}{r}{0.50\linewidth}
\vspace{-12pt}
\centering
\caption{\textbf{RQ3: how reused components are consumed.}
Rows share LEGO's retrieval, diagnosis.}
\vspace{-10pt}
\label{tab:design}
\tabsetup\scriptsize\setlength{\tabcolsep}{2.5pt}
\resizebox{\linewidth}{!}{%
\begin{tabular}{l rrr r}
\toprule\rowcolor{headergray}
\hdr{Configuration} & \hdr{D1} & \hdr{D5} & \hdr{All $\uparrow$} & \hdr{$\Delta$ \stg{--DV}} \\
\midrule
Feedback $+$ Diagnosis (\stg{--DV}) & .569 & .135 & .3469 & \textit{ref.} \\
Context-only (\stg{RC\,DV}) & .620 & .115 & .3544 & $+.0075$ \\
Import $+$ Call (\stg{RI\,DV}) & .627 & .060 & .3430 & \loss{$-.0039$} \\
Structured-schema adaptation & .679 & .327 & .4974 & $+.1505$ \\
\rowcolor{armblue}\textbf{LEGO} (\stg{RADV}) & \best{.686} & \best{.338} & \best{.5134} & $+.1665$ \\
\bottomrule
\end{tabular}%
}
\vspace{-12pt}
\end{wraptable}

\textbf{Part of the margin is provenance, and it is not the larger part.}
Excluding all same-repository donors lowers coverage to $41.2\%$ and the score to $0.4743$, with the effect concentrated on the largest bands (Appendix~\ref{app:extexp:prim}). More importantly, a separately re-mined corpus-disjoint library still improves over Feedback by $+0.1348$, compared with $+0.1954$ for the default library (Appendix~\ref{app:extexp:clean}). The benefit therefore persists without target-repository provenance, though provenance matters more at scale.

\begin{rqbox}
\textbf{The answer to RQ3: construction and adaptation drive the gains.}
Constructed primitives and primitive adaptation account for $82.2\%$ of the improvement over Feedback and outperform passive reuse and generic editing. The advantage grows with repository scale and persists with a corpus-disjoint library.
\end{rqbox}

\subsection{\textbf{RQ4--RQ5: Model Substitution and Cost}}
\label{sec:exp:positions}

\textbf{The effect is a property of the pipeline, not of one model.}
With adaptation and diagnosis fixed to GPT-5.6-terra, LEGO improves all $13$ backbones by $0.1474$ on average, $+76.5\%$ relative (Table~\ref{tab:rq4main}; full sweep in Appendix~\ref{app:extexp:bb}). The choice of adaptation and diagnosis model matters much less than retaining the mechanisms: across nine substitutions each, scores vary by only $.0309$ and $.0127$, against losses of $.1154$ and $.0339$ when either is removed (Appendices~\ref{app:extexp:res},~\ref{app:extexp:diagnosis}).

\begin{table}[h]
\centering
\vspace{-8pt}

\begin{minipage}[b]{0.475\textwidth}
\caption{\textbf{RQ4: backbone substitution.}
Adaptation and diagnosis use GPT-5.6-terra; \emph{scratch} is the corresponding Feedback score.}
\label{tab:rq4main}
\vspace{-10pt}

\tabsetup\scriptsize
\setlength{\tabcolsep}{3pt}

\resizebox{\linewidth}{!}{%
\begin{tabular}{l r r r r}
\toprule
\rowcolor{headergray}
\hdr{Backbone}
& \hdr{scratch}
& \hdr{LEGO $\uparrow$}
& \hdr{$\Delta$}
& \hdr{$\Delta_{\mathrm{rel}}$} \\
\midrule

\rowcolor{armblue}
GPT-5.6-terra
& .3180
& \textbf{.5134}
& $+.1954$
& $+61.4\%$ \\

Claude Sonnet 5
& .3052
& .4953
& $+.1901$
& $+62.3\%$ \\

Nemotron-Super-3-120B
& .1067
& .2310
& $+.1243$
& $+116.5\%$ \\

\midrule

proprietary mean ($7$)
& .2362
& .4038
& $+.1676$
& $+70.9\%$ \\

open-weight mean ($6$)
& .1418
& .2656
& $+.1238$
& $+87.3\%$ \\

\rowcolor{sectiongray}
field mean ($13$)
& .1926
& .3400
& $+.1474$
& $+76.5\%$ \\

\bottomrule
\end{tabular}%
}
\end{minipage}
\hfill
\begin{minipage}[b]{0.495\textwidth}

\caption{\textbf{RQ5: quality--cost tradeoff.}
\emph{c/a/d} denotes backbone, adaptation, and diagnosis models; costs use recorded model calls.}
\label{tab:rq5main}
\vspace{-10pt}

\tabsetup\scriptsize
\setlength{\tabcolsep}{3pt}

\resizebox{\linewidth}{!}{%
\begin{tabular}{l r r r}
\toprule
\rowcolor{headergray}
\hdr{Configuration (c/a/d)}
& \hdr{All $\uparrow$}
& \hdr{\$/task $\downarrow$}
& \hdr{\% LEGO} \\
\midrule

\rowcolor{armblue}
terra / terra / terra
& \textbf{.5134}
& 2.08
& $100.0\%$ \\

terra / OSS-20B / OSS-20B
& .4880
& \second{1.58}
& $95.1\%$ \\

Sonnet 5 / OSS-20B / OSS-20B
& .4661
& 3.31
& $90.8\%$ \\

DeepSeek-V3.2 / terra / terra
& \loss{.2471}
& \best{0.79}
& $48.1\%$ \\

\midrule

Feedback $+$ Diagnosis, $b{=}5$
& .3469
& 1.79
& $67.6\%$ \\

Feedback, $b{=}8$
& .3437
& 2.07
& $66.9\%$ \\

\bottomrule
\end{tabular}%
}
\end{minipage}

\vspace{-8pt}
\end{table}

\textbf{Smaller local models preserve most of the quality at lower cost.}
Moving both adaptation and diagnosis to GPT-OSS-20B retains $95.1\%$ of the homogeneous GPT-5.6-terra score at $24.0\%$ lower cost (Table~\ref{tab:rq5main}), while replacing the backbone with DeepSeek-V3.2 drops it to $0.2471$: performance is far more sensitive to the backbone than to the locally scoped models. At nearly the same cost, Feedback at $b=8$ reaches only $0.3437$, so the advantage is not a disguised attempt budget (Appendices~\ref{app:extexp:budget}--\ref{app:extexp:models}).

\begin{rqbox}
\textbf{The answer to RQ4--RQ5: LEGO is robust to backbone choice and supports efficient model allocation.}
All $13$ backbones improve, while smaller adaptation and diagnosis models retain $95.1\%$ of the score at $24.0\%$ lower cost. Performance is far more sensitive to the backbone than to these locally scoped models.
\end{rqbox}

\subsection{\textbf{RQ6: Does the Paradigm Transfer Beyond LEGO-REPO?}}
\label{sec:exp:transfer}

\begin{wraptable}{r}{0.46\linewidth}
\vspace{-12pt}
\centering
\caption{\textbf{RQ6: transfer to three external benchmarks.}
Each follows its published protocol with GPT-5.6-terra; \emph{fb.} denotes our execution-feedback baseline.}
\vspace{-10pt}
\label{tab:transfer}

\tabsetup\scriptsize
\setlength{\tabcolsep}{2.5pt}

\resizebox{\linewidth}{!}{%
\begin{tabular}{l r r r r}
\toprule
\rowcolor{headergray}
\hdr{Benchmark}
& \hdr{publ.}
& \hdr{fb.}
& \hdr{LEGO}
& \hdr{$\Delta$} \\
\midrule

RepoZero, Py$\to$Py
& 51.2
& 53.8
& \best{62.9}
& $+9.1$ \\

RepoGenesis
& 38.2
& 41.0
& \best{48.2}
& $+7.2$ \\

RepoCraft
& 0.412
& 0.436
& \best{0.501}
& $+.065$ \\

\bottomrule
\end{tabular}%
}

\vspace{-12pt}
\end{wraptable}

\textbf{LEGO transfers across independently designed benchmarks.}
Under each benchmark's original tasks, metrics, and protocol, LEGO improves over the matched execution-feedback baseline by $16.9\%$ on RepoZero, $17.6\%$ on RepoGenesis, and $14.9\%$ on RepoCraft (Table~\ref{tab:transfer}).

\textbf{The gains persist with substantially lower library coverage.}
After removing CodeFace entries traced to the external targets, offline coverage is only $29.6$--$34.8\%$ against $61.4\%$ on LEGO-REPO (filtering procedure: Appendix~\ref{app:ablsettings}). LEGO nonetheless improves all three, so its benefit is not specific to LEGO-REPO or its higher coverage.

\begin{rqbox}
\textbf{The answer to RQ6: LEGO transfers beyond LEGO-REPO.}
Across three external benchmarks, LEGO improves the matched Feedback baseline by $14.9\%$--$17.6\%$, despite CodeFace covering only about one third of target modules.
\end{rqbox}

\section{Conclusion}
\label{sec:concl}
We introduce \textbf{Code Primitives}, reusable software components that adapt their implementations, interfaces, and dependencies to target repositories, and \textbf{LEGO}, a framework that retrieves, coordinates, and verifies these primitives during repository construction. Across LEGO-REPO and three external benchmarks, LEGO improves all $13$ evaluated backbones, with the largest gains coming from primitive construction and adaptation rather than retrieval alone. These gains persist under model substitution and corpus-disjoint libraries. We hope this work shifts repository-scale code generation from repeatedly rebuilding software capabilities toward accumulating, adapting, and reusing them.

% \newpage
% \input{Secs/statements}

% \newpage
\bibliography{iclr2027_conference}
\bibliographystyle{iclr2027_conference}

\newpage
\appendix
\section{Extended Related Work}
\label{app:extended-rw}

\textbf{Coding agents and repository-level evaluation.}
LLM-based software engineering has progressed from direct code generation toward agents that explore repositories, localize relevant code, plan edits, execute programs, and iterate from feedback~\citep{chen2021evaluating,jimenez2024swebench,yang2024sweagent,zhang2024autocoderover,xia2024agentless,wang2025openhands,thai2025sweevo}. Evaluation has expanded accordingly, from standalone functions~\citep{chen2021evaluating,austin2021program,hendrycks2021measuring,zhuo2025bigcodebench} to repository context and cross-file dependencies~\citep{liu2024repobench,ding2023crosscodeeval,li2024deveval,liang2025repocod}, repository repair and evolution~\citep{jimenez2024swebench,thai2025sweevo}, and zero-to-one repository generation~\citep{zhang2026repozero,li2026realbench,luo2026rpg,chen2026repo0,zhang2026repocbench}. Execution-guided refinement provides a complementary mechanism: CodeT~\citep{chen2022codet} uses generated tests for candidate selection, Self-Debugging~\citep{chen2024teaching} revises from execution traces, and Reflexion~\citep{shinn2023reflexion} carries verbal feedback across attempts.

\textbf{Software reuse, retrieval, and reusable abstractions.}
Reusable software components have long been studied in software engineering, where interface compatibility and adaptation cost determine whether reuse is preferable to rewriting~\citep{szyperski2002component}. Learned systems extend this idea by constructing reusable libraries, abstractions, skills, or tools: DreamCoder~\citep{ellis2021dreamcoder} alternates synthesis with library compression, LILO~\citep{grand2024lilo} augments learned abstractions with documentation, ReGAL~\citep{stengel2024regal} refactors programs into generalizable abstractions, and TroVE~\citep{wang2024trove}, LATM~\citep{cai2024large}, and CRAFT~\citep{yuan2024craft} build reusable tool or skill collections for later tasks. Retrieval-augmented code generation instead supplies existing source, documentation, or repository context to a generator~\citep{lu2022reacc,zhou2022docprompting,zhang2023repocoder}, while AgentDistill~\citep{qiu2025agentdistill} and CODESKILL~\citep{li2026codeskill} extract reusable task-solving procedures from agent trajectories. More recently, Agent Primitives~\citep{jin2026agent} and Tool Primitives~\citep{jin2026harness} expose reusable agent behaviours and executable tools through agent-native retrievable interfaces. Across these approaches, however, the reused artifact is typically supplied as context or invoked through a fixed interface rather than incorporated into and adapted with the delivered repository.

\textbf{Key Differences.}
Our work differs from prior systems in three key respects. \textbf{LEGO} targets zero-to-one repository construction rather than editing an existing codebase, organizing planning, retrieval, adaptation, composition, and verification to produce a complete package from an empty tree. Unlike library-learning and retrieval-augmented methods that keep reused artifacts fixed or use them as context, \textbf{Code Primitives} become part of the delivered repository and adapt their implementations, interfaces, dependencies, and configuration to the target. The closest abstractions are Agent Primitives and Tool Primitives~\citep{jin2026agent,jin2026harness}, whose agent-native interfaces support \emph{invocation}; Code Primitives instead provide an interface for \emph{adaptation}, with their dependency closure and validation tests traveling with the component. Finally, \textbf{LEGO-REPO} evaluates an entire reconstructed package against the original repository's native test suite under a common empty-package floor and original-source ceiling, rather than scoring a function, completion, patch, or resolved issue.

\section{Construction Funnel and Exclusion Ledger}
\label{app:funnel}

The candidate registry, the materialized task corpus, and the final evaluation benchmark serve
different purposes. Table~\ref{tab:funnel} records how $1{,}309$ discovered repository candidates
become $565$ frozen task directories and then the $522$ tasks used for every primary configuration
mean. No evaluated configuration changes this denominator: once the final $522$ tasks are fixed,
a failed run contributes zero rather than being dropped.

\begin{table}[h]
\centering
\caption{\textbf{Construction funnel.}
Eight attempted candidates fail materialization or environment pinning before the $565$ frozen
directories are formed. Of those directories, $43$ do not enter the scored benchmark because a
stable same-pipeline reference cannot be established: the native suite does not run in the pinned
environment, the identity ceiling is unmeasurable, or $c-f<5$.}
\label{tab:funnel}
\footnotesize
\tabsetup\setlength{\tabcolsep}{5pt}
\begin{tabular}{r l p{0.58\textwidth}}
\toprule\rowcolor{headergray}
\hdr{count} & \hdr{stage} & \hdr{transition into this stage} \\
\midrule
$1{,}309$ & candidate registry & initial repository discovery \\
$573$     & attempted task candidates & repository screening: domain/security eligibility (including $34$ offensive-security holds), importable package, runnable native suite \\
$565$     & frozen task directories & successful materialization with clone/build/environment pinned \\
$\mathbf{522}$ & \textbf{LEGO-REPO evaluation set} & stable same-pipeline ceiling and measurable span $c-f\geq5$ \\
\bottomrule
\end{tabular}

\end{table}

\begin{table}[h]
\centering
\caption{\textbf{Why a task receives a zero score.}
Counts are over the same $522$ tasks for the Feedback baseline (\stg{---V}) and LEGO
(\stg{RADV}); each column sums to the corresponding \deadat\ count. A zero can arise either
because no admissible repository is delivered or because an admissible repository remains at the
normalized floor.}
\label{tab:ledger}
\footnotesize
\tabsetup\setlength{\tabcolsep}{5pt}
\begin{tabularx}{\textwidth}{Llrr}
\toprule\rowcolor{headergray}
\hdr{category} & \hdr{attributable to} & \hdr{Feedback} & \hdr{LEGO} \\
\midrule
incomplete composition: modules never written & system under test & 84 & 38 \\
import or contract failure at collection & system under test & 61 & 27 \\
environment or dependency resolution & harness/environment & 35 & 18 \\
attempt budget exhausted, tests failing & system under test & 34 & 15 \\
\midrule
\rowcolor{armblue}\textbf{zero-score tasks} (\deadat) & & \textbf{214} & \textbf{98} \\
\quad as a share of the $522$ benchmark tasks & & $41.0\%$ & $18.8\%$ \\
\bottomrule
\end{tabularx}

\end{table}

\section{Benchmark Inventory}
\label{app:inventory}

The complete per-task inventory --- repository URL, commit SHA, package prefix, module count,
floor $f(T)$, ceiling $c(T)$, span, band, domain, and capability track for each of the $522$
tasks --- is released as \texttt{lego\_repo\_inventory.csv} alongside the task definitions and
grading keys; it is a $522$-row table and is not reproduced inline.
Its aggregate shape is as follows. The $522$ tasks are $522$ distinct repositories, with a
median of $4{,}180$ GitHub stars (p25 $1{,}900$, p75 $12{,}400$; minimum
$1{,}000$ by construction) and pinned commits dated between 2019-03 and 2026-06
(median 2024-11).
Median module count is $30$, and the mean per task rises from $6$ in D1 to $119$ in D5, and
median span is $114$ graded tests (Table~\ref{tab:span}). Every case-study package named in
Appendix~\ref{app:case} appears in the inventory under the prefix used in the text.

Tables~\ref{tab:inv-domain} and~\ref{tab:inv-track} report the $522$ benchmark tasks by domain,
capability track, and difficulty. Domains and capability tracks come from repository metadata and
difficulty from the frozen task definition (Eq.~\ref{eq:band}); assignment precedes every
evaluated run. All cells count benchmark membership, and a zero denotes an empty category. The
domain rows of Table~\ref{tab:rq3} use exactly the band mixes given here.

\begin{table}[h]
\centering
\caption{\textbf{Inventory by domain and difficulty.}
Each cell counts tasks in the $522$-task benchmark. The security category contains defensive
cryptography and authentication libraries only; offensive-security repositories are excluded from
the benchmark and from CodeFace.}
\label{tab:inv-domain}
\small
\begin{tabular}{lrrrrrr}
\toprule
\rowcolor{headergray}domain & D1 & D2 & D3 & D4 & D5 & all \\
\midrule
machine-learning        & 15 & 25 & 35 & 39 & 44 & 158 \\
databases-storage       & 39 & 31 & 37 & 30 & 12 & 149 \\
compilers-languages     & 42 & 31 & 32 & 16 & 10 & 131 \\
graphics-vision         & 5  & 11 & 6  & 4  & 6  & 32 \\
scientific-computing    & 5  & 7  & 3  & 5  & 6  & 26 \\
quant-finance           & 6  & 3  & 2  & 1  & 2  & 14 \\
crypto-security         & 5  & 3  & 1  & 1  & 2  & 12 \\
\midrule
\rowcolor{armblue}\textbf{Total} & 117 & 111 & 116 & 96 & 82 & \textbf{522} \\
\bottomrule
\end{tabular}

\end{table}

\begin{table}[h]
\centering
\caption{\textbf{Inventory by capability track and difficulty.}
Tracks and bands are assigned from frozen task metadata before observing any system outcome.}
\label{tab:inv-track}
\small
\setlength{\tabcolsep}{5pt}
\begin{tabular}{lrrrrrr}
\toprule
\rowcolor{headergray}capability track & D1 & D2 & D3 & D4 & D5 & all \\
\midrule
agent-orchestration  & 4  & 12 & 11 & 20 & 26 & 73 \\
learning             & 2  & 10 & 12 & 11 & 11 & 46 \\
parsing-languages    & 12 & 14 & 6  & 8  & 5  & 45 \\
web-protocol         & 8  & 6  & 18 & 11 & 1  & 44 \\
serialization-wire   & 7  & 8  & 9  & 6  & 4  & 34 \\
text-strings         & 14 & 7  & 6  & 2  & 2  & 31 \\
media-graphics       & 8  & 5  & 4  & 3  & 8  & 28 \\
cli-terminal         & 5  & 6  & 10 & 2  & 4  & 27 \\
numeric-linalg       & 2  & 8  & 4  & 5  & 6  & 25 \\
data-storage         & 7  & 2  & 3  & 10 & 2  & 24 \\
structures-caching   & 6  & 8  & 6  & 2  & 0  & 22 \\
quant-finance        & 7  & 3  & 3  & 6  & 2  & 21 \\
packaging-devenv     & 8  & 2  & 5  & 5  & 0  & 20 \\
concurrency-io       & 4  & 5  & 4  & 1  & 1  & 15 \\
testing-tooling      & 3  & 2  & 6  & 0  & 3  & 14 \\
crypto-numbertheory  & 4  & 3  & 1  & 2  & 3  & 13 \\
resilience-control   & 6  & 2  & 1  & 1  & 0  & 10 \\
datetime-locale      & 4  & 1  & 2  & 0  & 2  & 9 \\
config-schema        & 1  & 5  & 2  & 0  & 1  & 9 \\
geo-spatial          & 2  & 1  & 2  & 1  & 0  & 6 \\
translation-i18n     & 3  & 0  & 1  & 0  & 1  & 5 \\
dev-tooling          & 0  & 1  & 0  & 0  & 0  & 1 \\
\midrule
\rowcolor{armblue}\textbf{Total} & 117 & 111 & 116 & 96 & 82 & \textbf{522} \\
\bottomrule
\end{tabular}

\end{table}

\section{Evaluation Protocol}
\label{app:protocol}

\paragraph{Admissible records.}
Let $\Xi$ be the target modules imported during the reported execution, excluding test modules and
\texttt{conftest.py}, and let $\mathcal G$ be the files written by the evaluated system. Provenance
is \textsf{not-exercised} when $\Xi=\varnothing$, \textsf{leaked} when a target module resolves
outside $\mathcal G$, and \textsf{ok} otherwise. Copied compiled extensions are explicitly exempt
because the reconstruction target is Python.
\begin{equation}
\mathcal A(T)=\mathbf1[\mathrm{status}=\textsf{done}]\cdot
\mathbf1[\mathrm{rev}=\mathrm{REV}]\cdot
\mathbf1[\mathrm{prov}(T)=\textsf{ok}]\cdot
\mathbf1[c(T)-f(T)\geq\mathrm{MIN\_SPAN}].
\label{eq:admit}
\end{equation}
The score and provenance verdict must come from the same execution round, and pairing rejects
records from different harness revisions. An inadmissible record contributes a zero score
(Eq.~\ref{eq:delivery}); it never removes a task from the $522$.

\paragraph{Default parameters.}
The harness uses $10$ concurrent model calls, at most $9{,}000$ interface characters per module,
top-$2$ retrieval, a $900$-second suite timeout, and a default attempt budget $b=5$. It stops on a
passing suite, after two consecutive rounds without an additional passing test, or after two timeout
rounds. $\mathrm{MIN\_SPAN}=5$, and records are stamped \texttt{r2}. Budget ablations override
$b$ and change nothing else.

\section{Scoring and Provenance Checks}
\label{app:guards}

\paragraph{Same-pipeline ceiling.}
An installed package can expose capabilities unavailable to the reconstructed tree. In
\texttt{markupsafe} the installed baseline passes $79$ tests because installation compiles the
\texttt{\_speedups} extension, and $41$ of those tests skip without it. The identity ceiling
executes the original sources through the reconstruction pipeline, avoiding a penalty for this
installation difference.

\paragraph{Empty-package floor.}
Some tests pass without exercising reconstructed behaviour, such as checks on data assets or
version strings. The floor is nonzero for $87$ of the $522$ tasks, and subtracting it removes
credit for that behaviour. On a small number of tasks the floor is a large fraction of the
ceiling: \texttt{pandera} in Appendix~\ref{app:case} has $f=1{,}577$ against
$c=1{,}807$, so its normalized span covers $13\%$ of the suite.
Classifying the $87$ nonzero-floor tasks by what the empty tree actually passes:
$52$ pass package-metadata, version-string, or data-asset checks that never import
reconstructed behaviour; $21$ pass at collection time, where the suite's own convention
counts skipped or parametrization-degenerate nodes as passes; and $14$ exercise fixtures
shipped in the repository rather than the package under reconstruction. None pass by resolving
against an installed distribution. That is what the provenance gate excludes: a scored record must
show that the target modules exercised during that execution resolved inside the generated tree,
and a record in which any target module resolves outside it is marked \textsf{leaked} and scored
zero rather than credited. The joint distribution matters for reading the metric, and
Table~\ref{tab:span} reports it: $9$ tasks have $f/c>0.5$ and $41$ have a span below
$20$, and the span-weighted re-aggregation in that table shows the level of the reported mean is
weighting-dependent while the two LEGO\,$-$\,Feedback contrasts are similar ($+.1938$ against
$+.1954$). A span below $\mathrm{MIN\_SPAN}=5$ leaves too little dynamic range for
a stable normalized score, so such repositories are excluded before the final $522$-task benchmark
is fixed (Table~\ref{tab:funnel}).

\paragraph{Execution provenance.}
Without an import gate, a suite can report the environment's installed package as a successful
reconstruction. The motivating sweep contained $22$ such apparently full-green repositories and
one record at $212\%$ of its reference ceiling. The gate separates executions that never import
target code from those importing code outside the generated tree; both are inadmissible.

\paragraph{Target-derived library entries.}
Excluding a repository as an active source does not remove primitives harvested from it earlier,
so provenance alone is insufficient. Section~\ref{sec:exp:rq3} therefore filters CodeFace
separately for each target repository, removing every entry whose source repository matches the
evaluated target while keeping the same $522$ evaluation tasks. Coverage falls from $61.4\%$ to
$41.2\%$ and the margin over the diagnosis-equipped baseline falls from $.1665$ to $.1274$, a
$23\%$ reduction.

\paragraph{Span distribution.}
Table~\ref{tab:span} reports how much dynamic range each task actually has. The floor is zero on
$435$ of the $522$ tasks, so the $f(T)$ and $f/c$ quartiles up to p75 are zero by construction.
The span-weighted contrast ($+.1938$) differs from the unweighted one by $0.0016$; the main text
reports the unweighted values, and both the span-weighted contrast and the lower span-weighted
\emph{levels} appear only here.

\begin{table}[h]
\caption{\textbf{Distribution of ceiling, floor, and normalized span over the $522$ tasks.}
The span-weighted row re-aggregates the benchmark mean with each task weighted by $c-f$ instead of
equally, so tasks with little dynamic range do not carry the same weight as tasks with hundreds of
gradable tests.}
\label{tab:span}
\centering\scriptsize\tabsetup\setlength{\tabcolsep}{4pt}
\begin{tabularx}{\textwidth}{L rrrrr r}
\toprule\rowcolor{headergray}
\hdr{quantity} & \hdr{min} & \hdr{p25} & \hdr{median} & \hdr{p75} & \hdr{max} & \hdr{mean} \\
\midrule
identity ceiling $c(T)$        & 6 & 47 & 118 & 341
 & $5{,}876$ & 287.4 \\
empty-package floor $f(T)$     & 0 & 0 & 0 & 0 & $1{,}577$ & 9.6 \\
span $c-f$                     & 5 & 46 & 114 & 335 & $5{,}876$
 & 277.8 \\
floor share $f/c$              & 0 & 0 & 0 & 0 & 0.873
 & 0.021 \\
\midrule
tasks with span $<20$          & \multicolumn{6}{l}{$41$ tasks, mean LEGO score $.5236$} \\
tasks with $f/c>0.5$           & \multicolumn{6}{l}{$9$ tasks, mean LEGO score $.4871$} \\
\midrule
benchmark mean, unweighted (reported) & \multicolumn{6}{l}{LEGO $.5134$ \quad Feedback $.3180$ \quad $\Delta=+.1954$} \\
benchmark mean, span-weighted         & \multicolumn{6}{l}{LEGO $.4642$ \quad Feedback $.2704$ \quad $\Delta=+.1938$} \\
\bottomrule
\end{tabularx}

\end{table}

\paragraph{Fixed difficulty labels.}
Eq.~\ref{eq:band} uses absolute cutoffs $\theta=(4,5,6,7)$ on frozen metadata. Quantile bins would
move when the corpus changes and run-derived bins would move with the evaluated system; fixed
labels keep the domain and difficulty counts independent of outcomes. Table~\ref{tab:bandsens}
checks that our scale-dependence claim does not depend on this particular binning.

Under all three binnings \emph{Feedback} declines monotonically and LEGO exceeds it in every band.
At the displayed precision the reported binning has a tied peak in the third and fifth bands while
both alternatives peak in the third, so Section~\ref{sec:bench:band} relies on the gain being
present at every scale rather than on where it peaks.

\begin{table}[h]
\caption{\textbf{Sensitivity of the band profile to the difficulty definition.}
The reported binning uses the identity ceiling $c(T)$, which also appears in the scoring
denominator; the two alternatives use repository size alone. Both relabel the same per-task
records, with no configuration re-run.}
\label{tab:bandsens}
\centering\scriptsize\tabsetup\setlength{\tabcolsep}{4pt}
\begin{tabularx}{\textwidth}{L l rrrrr}
\toprule\rowcolor{headergray}
\hdr{binning} & \hdr{quantity} & \hdr{B1} & \hdr{B2} & \hdr{B3} & \hdr{B4} & \hdr{B5} \\
\midrule
\multirow{3}{*}{$d=2\log_{10}m+\log_{10}c$ (reported)}
 & tasks & 117 & 111 & 116 & 96 & 82 \\
 & Feedback & .525 & .382 & .274 & .206 & .129 \\
 & LEGO $-$ Feedback & $+.161$ & $+.201$ & $+.209$ & $+.203$ & $+.209$ \\
\midrule
\multirow{3}{*}{$d=\log_{10}m$, cuts at $m\in\{10,25,60,150\}$}
 & tasks & 124 & 108 & 113 & 99 & 78 \\
 & Feedback & .515 & .383 & .280 & .199 & .121 \\
 & LEGO $-$ Feedback & $+.162$ & $+.206$ & $+.222$ & $+.193$
 & $+.198$ \\
\midrule
\multirow{3}{*}{quintiles of $m$ within the corpus}
 & tasks & 105 & 104 & 105 & 104 & 104 \\
 & Feedback & .548 & .412 & .302 & .208 & .118 \\
 & LEGO $-$ Feedback & $+.155$ & $+.209$ & $+.217$ & $+.194$
 & $+.202$ \\
\bottomrule
\end{tabularx}

\end{table}

\paragraph{Inadmissible versus scored zero.}
These outcomes have different causes. The first three categories of Table~\ref{tab:ledger} have
no admissible record; its final category has an admissible repository that remains at the normalized
floor. For LEGO, the corresponding counts are $83/522$ ($15.9\%$) and $15/522$ ($2.9\%$),
summing to $98/522=18.8\%$. For Feedback they are $180/522$ ($34.5\%$) and $34/522$
($6.5\%$), summing to $214/522=41.0\%$. These are the zero-score rates reported in
Table~\ref{tab:paradigms}, using the same ledger rather than a separate exclusion count.

\section{Primitive Construction Details}
\label{app:mining}

\paragraph{Recovering functional boundaries.}
For source repository $\mathcal{R}$, we recover program entities and supporting artifacts and construct
\begin{equation}
G_{\mathcal{R}}=(V,E_{\mathrm{call}}\cup E_{\mathrm{data}}\cup E_{\mathrm{struct}}),
\label{eq:repo_graph}
\end{equation}
where edges capture calls, data and type dependencies, and structural relations such as imports, module membership, configuration references, and implementation--test associations. Weighted community discovery proposes boundaries:
\begin{equation}
\mathcal{S}^{(0)}_{\mathcal{R}}=\operatorname{COMMUNITY}(G_{\mathcal{R}})=\{S^{(0)}_1,\ldots,S^{(0)}_m\}.
\label{eq:community}
\end{equation}
Semantic refinement uses the implementation, neighboring dependencies, tests, and documentation to split separable functions or merge entities serving one capability. This produces a component $S_i$ without fixing its granularity in advance.

\paragraph{Making a component executable.}
The dependency boundary is
\begin{equation}
\partial S_i=\{v\notin S_i\mid \exists u\in S_i,\;(u,v)\in E\}.
\label{eq:primitive_boundary}
\end{equation}
Local helpers, types, constants, and configuration state needed for execution are internalized into $C_i$. External libraries, services, and component interactions are recorded in $D_i$; the exposed interactions define $I_i$. Associated tests supply $V_i$, while documentation and provenance supply $X_i$.

\paragraph{Validation and admission.}
We materialize each candidate in isolation and run its validation suite. Build, runtime, and test diagnostics $e_i^t$ guide boundary and dependency corrections:
\begin{equation}
(C_i^{t+1},D_i^{t+1},I_i^{t+1})
=\operatorname{REFINE}(C_i^t,D_i^t,I_i^t,e_i^t).
\label{eq:primitive_refine}
\end{equation}
Candidates that remain non-executable or fail validation within the refinement budget are excluded. Verified implementations harvested during construction provide an additional acquisition route. Admission establishes behavior in the component's validation context; subsequent adaptation and integration require another execution check.

CodeFace contains $1{,}424$ admitted entries: $871$ from repository mining, $343$ harvested from
construction runs whose repository-level suite verified the component in context, and $210$ from
targeted web sourcing of capabilities the mined corpus lacked. Admission requires executable
closure, a passing component suite, and a recorded source repository, commit, and license; the
security hold set is applied when the library is loaded and when it is saved, so a held repository
can neither be evaluated nor become a donor. Entries are stored with the repository they came from,
which is what makes the repository-level split of Section~\ref{sec:exp:rq3} possible.

\subsection{Primitive adaptation example}
\paragraph{Prompt contract.}
A primitive is invoked with a fixed four-part payload and must answer in a fixed schema. It
receives (i) its own primitive $(C_i,I_i,D_i,V_i,X_i)$ verbatim; (ii) the requirement $r$ as the
backbone wrote it; (iii) a context slice naming the target module path, the symbols the module must
export, and the signatures of the sibling components it must bind to; and (iv) the diagnostics
$e$ from the previous adaptation attempt, if any. It receives no part of the target's native test
suite, no other module's source, and no part of the repository-level plan or repository-level diagnosis.
It must return a complete replacement implementation, an updated contract and dependency record,
a mapping from each carried test in $V_i$ to \textsf{kept} / \textsf{rewritten} / \textsf{dropped}
with a justification for each drop, and the execution log of $V_i$ against the rewritten
component. A return that fails to parse, omits a field, or leaves a \textsf{kept} test failing is
rejected with the failure attached and re-requested up to $3$ times; if it still fails, the requirement
is returned for repository-level handling and counts as an unsuccessful adaptation in Table~\ref{tab:cov}.
Of the $4{,}412$ primitive-accepted candidates, $712$ are rejected at least once
under this rule. Of these, $173$ recover within the rejection budget, leaving $539$
ultimately returned for repository-level handling. These are the same $539$ unsuccessful
adaptations in Table~\ref{tab:cov}; the other $3{,}873$ candidates pass validation.

\paragraph{A worked adaptation.}
The scoped editing control in Section~\ref{sec:exp:rq3} differs from a primitive exactly in
this contract, so it is worth seeing one instance concretely. CodeFace entry
\stg{cf-0412/schema-coercion} was mined from \texttt{marshmallow}: $261$ lines across
$2$ files, exporting a field-coercion dispatcher and its error type, with $14$
carried tests. Across the snapshot, entries have a median of $214$ implementation lines
(p25 $96$, p75 $468$) across a median of $2$ files (p75 $4$), exporting
$4$ symbols and carrying $9$ tests. Reconstructing \texttt{pandera} (Appendix~\ref{app:case}), LEGO issued the requirement \emph{``coerce untyped input to a declared field type, collecting per-field
errors into one report rather than raising on the first failure; errors must carry a location
path.''} The primitive's adaptation replaced the donor's exception-per-field control flow with an
accumulating error collector, renamed the exported error type to the target's contract, moved the
location path from a string to the target's tuple representation, and dropped $3$ of the
$14$ carried tests as donor-specific (they asserted on \texttt{marshmallow}'s message
formatting), rewriting $5$ others against the new signatures and keeping $6$
unchanged. The $11$ retained tests passed on the rewritten component before it was handed back to repository-level composition, which then bound it to $4$ sibling modules. A generic editor operating
on the same retrieved file has no equivalent of that last check: it can produce the same rewrite,
but it does not perform this carried-test check. This example illustrates the validation step
studied by the $0.0338$ forward carried-tests contrast in Table~\ref{tab:primdecomp}; the
example itself does not estimate that contrast's effect size.

\subsection{Construction state and revision operators}
\label{app:construction-state}

Section~\ref{sec:lego} states the construction loop compactly. The stages in full are as follows.

\textbf{Retrieval and Activation.} Given request $q$ and construction context $\mathcal{C}_t$, LEGO derives a repository plan and decomposes the task into required capabilities together with their interface and dependency constraints.

For each required capability, LEGO retrieves candidate primitives from CodeFace; the evaluated configuration returns the top two candidates per capability query. Their primitives produce relevance assessments and usage proposals $(s_i,u_i)$, after which activation selects primitives based on capability coverage and compatibility with the rest of the planned repository. Requirements without a suitable primitive remain outside the library and must be written from scratch.

\textbf{Primitive Adaptation and Collaboration.} For each selected primitive, LEGO states the target requirements through its natural-language interface. The primitive then produces $\widetilde{P}_i$, adapting the primitive's implementation, interfaces, dependencies, and configuration while preserving its component-level validation contract.

\textbf{Repository Editing.} The adapted implementations become part of the target repository. LEGO combines them with integration code and newly generated functionality, while resolving constraints that cross primitive boundaries, including API bindings, shared data representations and types, package dependencies, and configuration. Primitives handle local component adaptation; repository-level coordination remains outside the primitive.

\textbf{Execution and Feedback.} LEGO executes the build checks and tests available in the target, supplemented with unit and integration tests synthesized from the requirements and the recorded interfaces.

\textbf{Verification and Diagnosis.} The execution evidence is assessed against task completion, build success, test pass rate, interface consistency, and code quality. Build failures and failing tests are determined by execution and are terminal for the round. Diagnosis attributes the observed failures to individual components or their interactions and returns revision guidance, which reactivates only the implicated primitives. The loop continues until verification succeeds or the budget of $b=5$ attempts per module is exhausted.

Formally, LEGO maintains a construction context $\mathcal C_t$ and decomposes the request into requirements $\Pi_t$. Retrieval and primitive relevance assessments determine assignments $\mathcal A_t$. The composition constraints cover interfaces, shared types, dependencies, and configuration. Diagnosis returns a decision $v_t$ and diagnostics $\phi_t$, which update the construction context for the next revision:
\begin{equation}
\Pi_t=(r_1,\ldots,r_K)=\operatorname{DECOMPOSE}(q,\mathcal{C}_t).
\label{eq:lego_plan}
\end{equation}
\begin{equation}
\mathcal{A}_t=\operatorname{MATCH}(\Pi_t,\{\mathcal{F}_k\}_{k=1}^{K},\{(s_i,u_i)\}).
\label{eq:primitive_match}
\end{equation}
\begin{equation}
\Gamma(\{\widetilde{P}_i\}_{i\in\mathcal{A}_t})=
\{\Gamma_{\mathrm{api}},\Gamma_{\mathrm{type}},
\Gamma_{\mathrm{dep}},\Gamma_{\mathrm{config}}\}.
\label{eq:composition_constraints}
\end{equation}
\begin{equation}
(v_t,\phi_t)=\operatorname{DIAGNOSE}(q,\Pi_t,\mathcal{S}_t,o_t),
\label{eq:lego_verify}
\end{equation}
\begin{equation}
\mathcal{C}_{t+1}=\mathcal{C}_t\cup\{\phi_t\},
\label{eq:lego_feedback}
\end{equation}

Three properties of this loop matter for reading the ablations. First, $\Pi_t$ is recomputed at
every round rather than fixed after planning, so a requirement that no primitive satisfies in round
$t$ can be matched in round $t+1$ once neighbouring interfaces have settled; this is why retrieval
counts in Table~\ref{tab:cov} exceed one query per requirement. Second, $\mathcal{C}_{t+1}$
accumulates diagnostics monotonically and is never truncated within a task, so the attempt budget
$b$ bounds the number of revisions but not the amount of evidence carried into the last
one --- the budget-dependent scores in Table~\ref{tab:budget} are therefore not an artefact of context
loss. Third, $\Gamma$ is evaluated by the harness rather than by diagnosis: diagnosis diagnoses
failures but does not decide admissibility, which keeps the scoring path of
Eq.~\ref{eq:admit} independent of any model's verdict. Disabling \textsf{D} removes $\phi_t$ from
Eq.~\ref{eq:lego_feedback} and leaves LEGO with raw execution output; disabling \textsf{A}
replaces $\widetilde{P}_i$ with $P_i$ in the composition step; disabling \textsf{R} empties
$\mathcal{A}_t$. Each stage ablation in Table~\ref{tab:paradigms} is exactly one of these three
substitutions.

\section{Extended Experimental Results}
\label{app:extexp}

This section expands the six research questions with score and coverage breakdowns, component
removals, model substitutions, retrieval and library ablations, and cost accounting.
All primary configuration means use the same $522$ tasks as the main text. Whenever an appendix
table repeats a main-text configuration, the score, stage definition, and denominator are identical
to the corresponding main-text value.

\subsection{Score and coverage breakdowns (RQ2--RQ3)}
\label{app:extexp:rq3}

Table~\ref{tab:rq3} reports the same Feedback and LEGO configurations used in the main text,
partitioned by difficulty and software domain. The table deliberately reports aggregate scores rather
than a second set of band- or domain-specific win/tie/loss counts; task-level paired totals are given
once in Table~\ref{tab:paired}.

\begin{table}[h]
\centering
\caption{\textbf{Feedback and LEGO by difficulty and domain.} These are the same two configurations
reported in Table~\ref{tab:paradigms}, partitioned two ways; coverage is the offline module-level
reachability measure, not a scoring denominator.}
\label{tab:rq3}
\scriptsize
\tabsetup\setlength{\tabcolsep}{3pt}
\begin{tabularx}{\textwidth}{L r r rrr r}
\toprule\rowcolor{headergray}
\hdr{partition} & \hdr{tasks} & \hdr{modules} & \hdr{Feedback} & \hdr{LEGO} & \hdr{$\Delta$}
 & \hdr{coverage} \\
\midrule
D1 (mean $6$ modules)   & 117 & \phantom{0,}721 & .525 & .686 & $+.161$ & $75.3\%$ \\
D2 (mean $19$ modules)  & 111 & 2{,}064 & .382 & .583 & $+.201$ & $71.2\%$ \\
D3 (mean $32$ modules)  & 116 & 3{,}654 & .274 & .483 & $+.209$ & $65.8\%$ \\
D4 (mean $71$ modules)  & 96 & 6{,}823 & .206 & .409 & $+.203$ & $61.2\%$ \\
D5 (mean $119$ modules) & 82 & 9{,}723 & .129 & .338 & $+.209$ & $56.7\%$ \\
\midrule
machine-learning      & 158 & 9{,}649 & .2498 & .4368 & $+.1870$ & $56.3\%$ \\
databases-storage     & 149 & 5{,}537 & .3390 & .4946 & $+.1556$ & $66.5\%$ \\
compilers-languages   & 131 & 4{,}166 & .3640 & .6104 & $+.2464$ & $68.5\%$ \\
graphics-vision       &  32 & 1{,}420 & .3195 & .5055 & $+.1860$ & $60.4\%$ \\
scientific-computing  &  26 & 1{,}322 & .3179 & .5213 & $+.2034$ & $58.3\%$ \\
quant-finance         &  14 & \phantom{0,}465 & .3711 & .5877 & $+.2166$ & $62.4\%$ \\
crypto-security       &  12 & \phantom{0,}426 & .3880 & .6139 & $+.2259$ & $52.1\%$ \\
\midrule
\rowcolor{armblue}\textbf{all tasks} & \textbf{522} & \textbf{22{,}985} & \textbf{.3180}
 & \textbf{.5134} & $\mathbf{+.1954}$ & $\mathbf{61.4\%}$ \\
\bottomrule
\end{tabularx}

\end{table}

Two readings follow from the partition. In \emph{absolute} terms the gain is nearly flat across bands
($.161$ to $.209$), but it is measured against a baseline that is itself collapsing --- D5
\emph{Feedback} is $.129$ against $.525$ on D1, so the same absolute step is a far larger intervention
at D5. Relative to the same backbone without a library the gain is monotone in scale:
$+30.7\%$, $+52.6\%$, $+76.3\%$, $+98.5\%$, $+162.0\%$ from D1 to D5. We rest the
scale claim on the relative column \emph{together with} the band-resolved mechanism controls of RQ3,
which are absolute and all change sign in the same direction, rather than on the ratio alone --- a
relative gain over a small baseline is easy to inflate, which is the caveat we attach to the weak
backbones in Table~\ref{tab:backbone}.
Paired bootstrap intervals over tasks ($10{,}000$ resamples) on the $\Delta$ column are
$[+.1356,+.2351]$ for machine-learning, $[+.1200,+.1911]$ for databases-storage and
$[+.1755,+.3103]$ for compilers-languages; of the four small domains, graphics-vision
$[+.1120,+.2624]$ and scientific-computing $[+.0301,+.3802]$ exclude zero,
quant-finance has a wide interval ($[+.0118,+.4230]$),
and crypto-security $[-.0310,+.4790]$ does not exclude zero at all.
The domain column also shows the opposite pattern from what a
coverage-driven account would predict --- crypto-security has the lowest reach ($52.1\%$) and the
highest \emph{Feedback} score ($.3880$), because its tasks are small and behaviourally tight, while
machine-learning has the second-lowest reach and the lowest scores in both configurations because
its band mix is the heaviest in the benchmark (Table~\ref{tab:inv-domain}). Domain, difficulty and coverage
are therefore confounded by construction, and we use the band partition rather than the domain
partition wherever a claim depends on scale.

\subsection{Coverage and the runtime adaptation funnel (RQ2)}
\label{app:extexp:cov}

Table~\ref{tab:cov} keeps the two notions used in RQ2 separate. The module and coverage columns
come from the offline reachability audit, whereas queries, retained candidates, and successful
adaptations are measured during the actual LEGO construction runs. Across the $522$ runs backbones
issue $5{,}515$ capability queries ($10.6$ per task) and top-$2$ retrieval yields $11{,}030$
candidate inspections; primitives retain $4{,}412$ of them ($40.0\%$) and $3{,}873$ pass their
carried tests after adaptation ($87.8\%$, or $7.4$ per task). The remaining $539$ retained
candidates ($12.2\%$) fail adaptation validation under the prompt contract and are returned for
repository-level handling rather than counted as successful adaptations.

\begin{table}[h]
\centering
\caption{\textbf{Offline library reach and runtime adaptation funnel by band.}
Module and coverage columns come from the offline reachability audit, which covers
$14{,}106/22{,}985$ target modules; queries, retained candidates, and adaptations are counted
during construction.}
\label{tab:cov}
\footnotesize
\tabsetup\setlength{\tabcolsep}{4pt}
\begin{tabularx}{\textwidth}{L rrr r rrr}
\toprule\rowcolor{headergray}
\hdr{band} & \hdr{tasks} & \hdr{modules} & \hdr{covered} & \hdr{coverage}
 & \hdr{queries} & \hdr{retained} & \hdr{adapted} \\
\midrule
D1 & 117 & \phantom{0,}721 & \phantom{0,}543 & $75.3\%$ & \phantom{0,}437 & \phantom{0,}357 & \phantom{0,}332 \\
D2 & 111 & 2{,}064 & 1{,}470 & $71.2\%$ & \phantom{0,}770 & \phantom{0,}613 & \phantom{0,}557 \\
D3 & 116 & 3{,}654 & 2{,}404 & $65.8\%$ & 1{,}021 & \phantom{0,}818 & \phantom{0,}731 \\
D4 & 96 & 6{,}823 & 4{,}176 & $61.2\%$ & 1{,}425 & 1{,}135 & \phantom{0,}989 \\
D5 & 82 & 9{,}723 & 5{,}513 & $56.7\%$ & 1{,}862 & 1{,}489 & 1{,}264 \\
\midrule
\rowcolor{armblue}\textbf{all} & \textbf{522} & \textbf{22{,}985} & \textbf{14{,}106}
 & $\mathbf{61.4\%}$ & \textbf{5{,}515} & \textbf{4{,}412} & \textbf{3{,}873} \\
\bottomrule
\end{tabularx}

\end{table}

The funnel narrows in a way that the coverage column alone does not predict. Queries per task rise
from $3.7$ on D1 to $22.7$ on D5 while mean module count per task rises from $6$ to $119$, so the
backbone does not attempt to source most of a large repository from the library: adapted primitives
account for $46.0\%$ of D1 modules but $13.0\%$ of D5 modules. The two per-candidate rates
behave differently from each other. Relevance acceptance is nearly flat --- primitives accept $40.8\%$ of
inspected candidates on D1 and $40.0\%$ on D5. Adaptation success declines from $93.0\%$
of accepted candidates on D1 to $84.9\%$ on D5, an $8.1$-point decline: a component that is relevant becomes harder to fit
as the surrounding repository grows. The decline in the adapted-module share is $33.0$ percentage
points, compared with the $8.1$-point decline in adaptation success. This supports the observation
in Section~\ref{sec:exp:rq2}: the gain on large repositories comes from a small number of
well-integrated components, not from broad substitution.

Table~\ref{tab:paired} gives the per-task view that complements the configuration means. Against
execution feedback LEGO improves $326$ tasks, ties $112$ and regresses on
$84$; against the
diagnosis-equipped baseline the counts are $322/120/80$, and against the generic-editing control
$229/197/96$. These describe how the improvement is distributed across tasks within the reported
runs and are not a substitute for variability across seeds, which Table~\ref{tab:seeds} supplies.

\subsection{Component removals (RQ3)}
\label{app:extexp:loo}

Table~\ref{tab:loo} collects the aggregate component-removal controls used in RQ3. We report
only aggregate scores here because the main text is the canonical source for the current
difficulty-band profiles.

\begin{table}[h]
\centering
\caption{\textbf{Aggregate component removals on all $522$ tasks.}
Removing the library removes retrieval and adaptation together. The generic-editing control
changes carried tests, metadata, and primitive execution state together. The mined-only and same-repo-exclusion rows
change library eligibility rather than the construction stages.}
\label{tab:loo}
\footnotesize
\tabsetup\setlength{\tabcolsep}{4pt}
\begin{tabularx}{\textwidth}{L l r r r}
\toprule\rowcolor{headergray}
\hdr{change from LEGO} & \hdr{resulting configuration} & \hdr{all} & \hdr{$\Delta$} & \hdr{\$/task} \\
\midrule
\rowcolor{armblue}\textbf{none} & \stg{RADV} & \textbf{.5134} & \textit{ref.} & 2.08 \\
remove the whole library & \stg{--DV} & .3469 & $-.1665$ & 1.79 \\
remove primitive adaptation & \stg{R-DV} & .3980 & $-.1154$ & 1.80 \\
remove diagnosis & \stg{RA-V} & .4795 & $-.0339$ & 1.83 \\
replace primitive machinery with generic editing & \stg{RE\,DV} & .4093 & $-.1041$ & 2.01 \\
also replace primitives with raw files & \stg{FE\,DV} & .3528 & $-.1606$ & 2.03 \\
remove execution-feedback revision ($b=1$) & one-pass LEGO & .3954 & $-.1180$ & 1.13 \\
restrict CodeFace to mined entries ($N=871$) & \stg{RADV} & .5021 & $-.0113$ & 2.08 \\
exclude same-repository donors & \stg{RADV} & .4743 & $-.0391$ & 2.08 \\
\bottomrule
\end{tabularx}

\end{table}

Relative to \emph{Retrieval $+$ Diagnosis}, full primitive adaptation contributes $0.1154$.
The generic editing control recovers part of this margin: its improvement over
\emph{Retrieval $+$ Diagnosis} is $.0113$, about $10\%$ of the adaptation-channel margin in this
matched comparison. The \stg{FE\,DV} row replaces the library itself with raw source
files from the same donor repositories and is the generic-retrieval control of
Table~\ref{tab:attrib}; it loses $0.1606$ against LEGO, which is the combined cost of
removing the primitive-construction pipeline and the primitive interface. Note that it also falls
$0.0452$ below \emph{Retrieval $+$ Diagnosis}, a configuration with no editing channel at all:
what the editor acts on matters more than whether an editor is present.

\subsection{Decomposing the primitive channel by primitive field (RQ3)}
\label{app:extexp:prim}

Table~\ref{tab:primdecomp} adds the fields of $P_i$ cumulatively to the generic-editing control.
The forward score increments are $0.0338$, $0.0387$, and $0.0221$ for carried
tests, contract and provenance, and dependency closure, respectively. Adding contract and provenance
yields the largest observed increment, followed by carried tests.
The remaining difference between LEGO and the generic editor supplied with these fields is
$0.0095$ ($95\%$ CI $[-.0071,+.0148]$, Wilcoxon $p=0.49$). The interval includes zero.
This comparison conditions on the shared component fields; the separate structured-schema
control in Table~\ref{tab:design} evaluates the request format. In the withdrawal direction, withholding $V_i$ from a
complete primitive costs $0.0338$, equal to the displayed forward carried-tests increment.

\begin{table}[h]
\centering
\caption{\textbf{Decomposing the primitive channel.} Each row adds one field of the primitive to the
generic-editing control, with retrieval, diagnosis and $b=5$ feedback held fixed. The three $+$ rows are
cumulative; the fifth row instead withholds $V_i$ from an otherwise complete primitive.}
\label{tab:primdecomp}
\footnotesize
\tabsetup\setlength{\tabcolsep}{4pt}
\begin{tabularx}{\textwidth}{L r r r}
\toprule\rowcolor{headergray}
\hdr{Configuration} & \hdr{all $\uparrow$} & \hdr{$\Delta$ \stg{RE\,DV}} & \hdr{$\Delta$ LEGO} \\
\midrule
RAG $+$ Edit (\stg{RE\,DV}) & .4093 & \textit{ref.} & $-.1041$ \\
\quad $+$ carried tests $V_i$ & .4431 & $+.0338$ & $-.0703$ \\
\quad $+$ contract and provenance $I_i,X_i$ & .4818 & $+.0725$ & $-.0316$ \\
\quad $+$ dependency closure $D_i$ & .5039 & $+.0946$ & $-.0095$ \\
primitive without $V_i$ & .4796 & $+.0703$ & $-.0338$ \\
\rowcolor{armblue}\textbf{LEGO} (\stg{RADV}) & \textbf{.5134} & $\mathbf{+.1041}$ & \textit{ref.} \\
\bottomrule
\end{tabularx}

\end{table}

\begin{table}[t]
\centering
\caption{\textbf{RQ3: what same-repository donors are worth, by difficulty band.}
All rows are means over the fixed band task counts ($117/111/116/96/82$) and reproduce
Table~\ref{tab:paradigms}. \emph{gain} is the improvement over \emph{Feedback} in that band,
\emph{lost} is the part of that gain removed by excluding library entries whose source repository
is the evaluation target, and \emph{retained} is the complement. The \emph{All} column margin is
measured against \emph{Feedback}; measured instead against the diagnosis-equipped baseline
(\stg{--DV}) the retained share is $76.5\%$.}
\label{tab:donorband}
\tabsetup\footnotesize\setlength{\tabcolsep}{4pt}
\begin{tabularx}{\textwidth}{L rrrrr r}
\toprule\rowcolor{headergray}
\hdr{Configuration} & \hdr{D1} & \hdr{D2} & \hdr{D3} & \hdr{D4} & \hdr{D5} & \hdr{All} \\
\midrule
Feedback (\stg{---V}) & .525 & .382 & .274 & .206 & .129 & .3180 \\
\rowcolor{armblue}LEGO, full library & .686 & .583 & .483 & .409 & .338 & .5134 \\
LEGO, same-repo donors excluded & .683 & .574 & .461 & .337 & .221 & .4743 \\
\quad gain, full library & $+.161$ & $+.201$ & $+.209$ & $+.203$ & $+.209$ & $+.1954$ \\
\quad gain, same-repo excluded & $+.158$ & $+.192$ & $+.187$ & $+.131$ & $+.092$ & $+.1563$ \\
\quad lost to donor exclusion & $-.003$ & $-.009$ & $-.022$ & $-.072$ & $-.117$ & $-.0391$ \\
\quad \textit{retained share} & \textit{98.1\%} & \textit{95.5\%} & \textit{89.5\%} & \loss{\textit{64.5\%}} & \loss{\textit{44.0\%}} & \textit{80.0\%} \\
\bottomrule
\end{tabularx}
\end{table}

\begin{figure}[h]
\centering
\begin{minipage}[b]{0.49\linewidth}
\centering
\includegraphics[width=\linewidth]{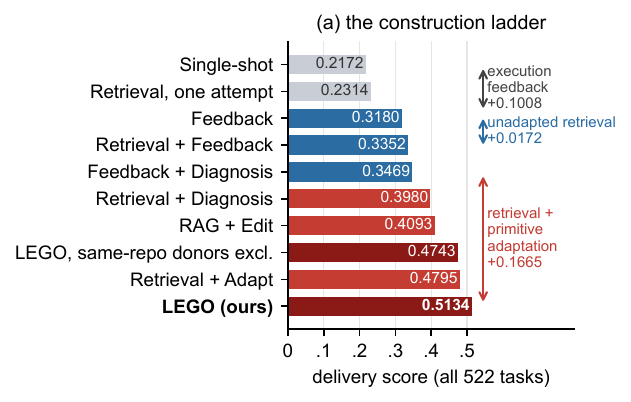}
\captionof{figure}{\textbf{The construction ladder}, all ten configurations of
Table~\ref{tab:paradigms} on all $522$ tasks. Grey: one-attempt controls. Blue: feedback references
without a library-specific mechanism. Red: retrieval combined with diagnosis, generic editing, or
adaptation. Dark red: same-repo donor exclusion and full LEGO.}
\label{fig:ladder}
\end{minipage}\hfill
\begin{minipage}[b]{0.49\linewidth}
\centering
\includegraphics[width=\linewidth]{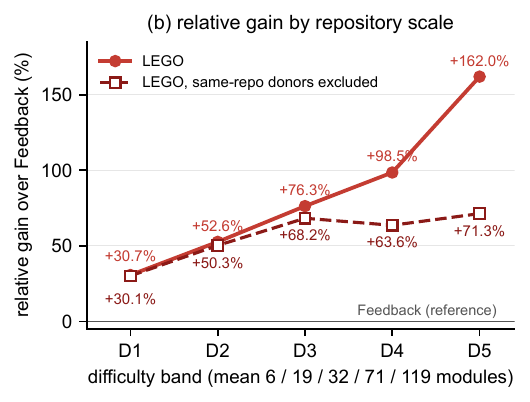}
\captionof{figure}{\textbf{Relative gain over \emph{Feedback} by repository scale}, computed from the
band means of Table~\ref{tab:donorband}. The gap between the solid and dashed red curves at D4 and D5
is the share of the gain that same-repository donors account for.}
\label{fig:bands}
\end{minipage}
\end{figure}

Figure~\ref{fig:ladder} plots the ten configurations of Table~\ref{tab:paradigms} as a single ladder,
and Figure~\ref{fig:bands} replots LEGO with and without same-repository donors as relative gain
over \emph{Feedback} in each band; the gap between the solid and dashed red curves at D4 and D5 is the
portion of the gain that same-repository donors account for.

\begin{table}[t]
\centering
\caption{\textbf{RQ3: per-task outcomes and stage decomposition, all $522$ tasks.}
Upper block: every row is compared against \emph{Feedback} (\stg{---V}). Lower block: win, tie,
and loss counts compare full LEGO (\stg{RADV}) against the named component-removal configuration,
while $\Delta$ reports that configuration minus full LEGO and is therefore negative. All counts
use task-level normalized delivery scores on the same $522$ tasks, and
$z=(W-L)/\sqrt{W+L}$ is the normal-approximation statistic of a paired sign test over non-tied
tasks.}
\label{tab:paired}
\tabsetup\footnotesize\setlength{\tabcolsep}{4pt}
\begin{tabularx}{\textwidth}{L l rrr r r}
\toprule\rowcolor{headergray}
\hdr{Comparison} & \hdr{Mechanism} & \hdr{Win} & \hdr{Tie}
 & \hdr{Loss} & \hdr{$z$} & \hdr{$\Delta$ score} \\
\midrule
Retrieval $+$ Feedback         & raw retrieved code
 & 148 & 269 & 105 & 2.7 & $+.0172$ \\
Feedback $+$ Diagnosis             & diagnosis, no library
 & 162 & 258 & 102 & 3.7 & $+.0289$ \\
Import $+$ Call                & vendored unchanged, invoked
 & 162 & 256 & 104 & 3.6 & $+.0250$ \\
File RAG $+$ Edit              & raw files $+$ generic editing
 & 174 & 247 & 101 & 4.4 & $+.0348$ \\
Context-only                   & primitive as read-only reference
 & 161 & 260 & 101 & 3.7 & $+.0364$ \\
Retrieval $+$ Diagnosis            & retrieval $+$ diagnosis
 & 214 & 214 & \phantom{0}94 & 6.8 & $+.0800$ \\
RAG $+$ Edit                   & scoped generic editing
 & 232 & 195 & \phantom{0}95 & 7.6 & $+.0913$ \\
Retrieval $+$ Adapt            & primitive adaptation
 & 256 & 177 & \phantom{0}89 & 9.0 & $+.1615$ \\
\rowcolor{armblue}\textbf{LEGO} & \textbf{all four stages}
 & \textbf{326} & \textbf{112} & \textbf{\phantom{0}84}
 & \textbf{12.0} & \gain{$\mathbf{+.1954}$} \\
\midrule
\rowcolor{sectiongray}\multicolumn{7}{l}{\cellcolor{sectiongray}
\emph{the same decomposition read as removals from LEGO}} \\
$-$ the whole library          & becomes \stg{--DV}
 & 322 & 120 & \phantom{0}80 & 12.1 & \loss{$-.1665$} \\
$-$ primitive adaptation        & becomes \stg{R-DV}
 & 271 & 162 & \phantom{0}89 & 9.6 & \loss{$-.1154$} \\
$-$ diagnosis                  & becomes \stg{RA-V}
 & 168 & 253 & 101 & 4.1 & \loss{$-.0339$} \\
$-$ primitive-adaptation machinery & becomes \stg{RE\,DV}
 & 229 & 197 & \phantom{0}96 & 7.4 & \loss{$-.1041$} \\
$-$ the natural-language interface & fixed structured schema
 & 147 & 270 & 105 & 2.6 & \loss{$-.0160$} \\
\bottomrule
\end{tabularx}
\end{table}

The sign test uses only the direction of each per-task difference and ignores magnitude, and the
$z$ values are uncorrected for the family of comparisons shown.

\textbf{Fixed-use components require integration.} With signatures, representations and configuration
frozen, LEGO writes shim code to reconcile them, as discussed in component reuse~\citep{szyperski2002component}.
The fixed-use control falls below the diagnosis-equipped no-library reference on the larger bands.
The \emph{Import $+$ Call} row tests invocation of a \emph{mined}
component, not of an abstraction compressed for reuse as in DreamCoder~\citep{ellis2021dreamcoder} or
ReGAL~\citep{stengel2024regal}, which might fare better. Within the primitive channel the validation
tests $V_i$ have a forward increment of $0.0338$, below the $0.0387$ increment for contract
and provenance and above the $0.0221$ increment for dependency closure. Withholding $V_i$
from a complete primitive also costs $0.0338$. Against a fixed structured schema,
the natural-language interface has an observed difference of $0.0160$ ($.5134-.4974$).
The band differences in Table~\ref{tab:design} are $0.007$ on D1 and $0.011$ on D5.

\subsection{Primitive substitution (RQ4)}
\label{app:extexp:res}

Table~\ref{tab:primitive} replaces the adaptation model while holding the backbone and diagnosis at
GPT-5.6-terra. The no-primitive row is \emph{Retrieval $+$ Diagnosis}.

\begin{table}[h]
\centering
\caption{\textbf{Primitive substitution on all $522$ tasks.}
\emph{kept} is the fraction of the adaptation margin over the no-primitive reference retained by
each model. Kept values are computed from unrounded task means; displayed scores are rounded.
\$/pt is recorded-call cost per $0.1$ of delivery score, using the accounting scope in
Appendix~\ref{app:extexp:models}.}
\label{tab:primitive}
\footnotesize
\tabsetup\setlength{\tabcolsep}{5pt}
\begin{tabularx}{\textwidth}{L r r r r}
\toprule\rowcolor{headergray}
\hdr{primitive} & \hdr{score} & \hdr{kept} & \hdr{\$/task} & \hdr{\$/pt} \\
\midrule
\rowcolor{armblue}GPT-5.6-terra    & \textbf{.5134} & $100.0\%$ & 2.08 & 0.41 \\
Claude Sonnet 5  & .5081 & $95.4\%$ & 2.30 & 0.45 \\
GPT-5.4          & .5060 & $93.6\%$ & 2.08 & 0.41 \\
GPT-5.6-luna     & .5015 & $89.7\%$ & 1.93 & 0.38 \\
DeepSeek-V4      & .4992 & $87.7\%$ & 1.83 & 0.37 \\
Qwen3-Coder-480B & .4972 & $86.0\%$ & 1.87 & 0.38 \\
GLM-5            & .4915 & $81.0\%$ & 1.89 & 0.38 \\
GPT-OSS-20B      & .4900 & $79.7\%$ & 1.81 & 0.37 \\
Gemma-4-31B      & .4825 & $73.2\%$ & 1.83 & 0.38 \\
\midrule
no primitive     & .3980 & \textit{ref.} & 1.80 & 0.45 \\
\bottomrule
\end{tabularx}

\end{table}

The nine primitives span $.4825$--$.5134$, a range of $.0309$. The largest
primary-configuration seed standard deviation is $.0043$. Every model tested, down to a $20$B
open-weight one, retains more than $73\%$ of the adaptation margin. The cheapest, GPT-OSS-20B,
retains $79.7\%$ at $13.0\%$ lower recorded-call cost; Gemma-4-31B retains $73.2\%$ at
$12.0\%$ lower recorded-call cost. The \$/pt column spans $0.37$--$0.45$.
Quality and recorded-call cost both vary; for example, GPT-OSS-20B has a higher score and lower
recorded-call cost than Gemma-4-31B. The larger gap is between \emph{having} a primitive and not:
the no-primitive row is $.0845$ below the weakest primitive, more than twice the entire spread among
primitives.

\subsection{Diagnosis substitution (RQ4)}
\label{app:extexp:diagnosis}

Table~\ref{tab:diagnosis} replaces diagnosis while holding the backbone and primitives at
GPT-5.6-terra. The no-diagnosis row is \emph{Retrieval $+$ Adapt}.

\begin{table}[h]
\centering
\caption{\textbf{Diagnosis substitution on all $522$ tasks.}
\emph{kept} is the fraction of the revision margin over the no-diagnosis reference retained by each
model. Kept values are computed from unrounded task means; displayed scores are rounded.}
\label{tab:diagnosis}
\footnotesize
\tabsetup\setlength{\tabcolsep}{5pt}
\begin{tabularx}{\textwidth}{L r r r r}
\toprule\rowcolor{headergray}
\hdr{diagnosis} & \hdr{score} & \hdr{kept} & \hdr{\$/task} & \hdr{\$/pt} \\
\midrule
\rowcolor{armblue}GPT-5.6-terra    & \textbf{.5134} & $100.0\%$ & 2.08 & 0.41 \\
Claude Sonnet 5  & .5123 & $96.8\%$ & 2.30 & 0.45 \\
GPT-5.4          & .5096 & $88.8\%$ & 2.08 & 0.41 \\
GPT-5.6-luna     & .5061 & $78.5\%$ & 1.95 & 0.39 \\
DeepSeek-V4      & .5058 & $77.6\%$ & 1.86 & 0.37 \\
Qwen3-Coder-480B & .5020 & $66.4\%$ & 1.90 & 0.38 \\
GLM-5            & .5014 & $64.6\%$ & 1.92 & 0.38 \\
GPT-OSS-20B      & .5012 & $64.0\%$ & 1.84 & 0.37 \\
Gemma-4-31B      & .5007 & $62.5\%$ & 1.86 & 0.37 \\
\midrule
no diagnosis     & .4795 & \textit{ref.} & 1.83 & 0.38 \\
\bottomrule
\end{tabularx}

\end{table}

Diagnosis behaves like adaptation, with one difference worth noting. The nine
diagnoses span $.5007$--$.5134$, a range of $.0127$, and the weakest retains $62.5\%$ of the
revision margin. The observed ranges, $.0127$ and $.0309$, exceed the primary-configuration
seed standard deviations of at most $.0043$. Both are smaller than the $.0339$ and $.1154$ that removing diagnosis or
adaptation costs outright. Cost moves in the opposite direction to intuition in one place: because the
rate table spans two orders of magnitude, substituting a \emph{more} expensive model raises the per-task
recorded-call cost above the homogeneous default (Claude Sonnet 5 in either position reaches $\$2.30$ against $\$2.08$), so
the sweep spans $\$1.84$--$\$2.30$ for diagnosis and $\$1.81$--$\$2.30$ for adaptation. Within the
cheaper half the two channels are comparable rather than dominated by one: diagnosis is invoked once per
revision round whereas adaptation is invoked once per candidate, and Table~\ref{tab:cov} records
$7.4$ successful adaptations per task, so substituting diagnosis buys at most $\$0.24$ per task and
substituting adaptation $\$0.27$. When both move to GPT-OSS-20B the saving is $\$0.50$
(Table~\ref{tab:select}); the separate rounded savings sum to $\$0.51$, with the one-cent
difference due to rounding. This is why RQ5 reports the combined T/O/O configuration rather than presenting either
channel alone as the economisation.

\subsection{Backbone sweep with the library (RQ4)}
\label{app:extexp:bb}

Table~\ref{tab:backbone} is the companion to Table~\ref{tab:rq1}: the same $13$ backbones are used
inside LEGO while the adaptation and diagnosis remain GPT-5.6-terra. LEGO improves every backbone, while
the size and difficulty profile of the gain vary across backbones.

\begin{table}[h]
\centering
\caption{\textbf{All $13$ backbones with the full LEGO pipeline, all $522$ tasks.}
\emph{scratch} is the same backbone's execution-feedback score from Table~\ref{tab:rq1}.
$\Delta_{\mathrm{rel}}$ is the relative improvement of LEGO over that scratch score.
Outcome rates are computed from integer task counts and rounded to one decimal place.}
\label{tab:backbone}
\scriptsize
\tabsetup\setlength{\tabcolsep}{2.2pt}
\begin{tabularx}{\textwidth}{L ccccc r r r r r}
\toprule\rowcolor{headergray}
\hdr{backbone (with LEGO)} & \hdr{D1} & \hdr{D2} & \hdr{D3} & \hdr{D4} & \hdr{D5} & \hdr{all}
 & \hdr{scratch} & \hdr{$\Delta$} & \hdr{$\Delta_{\mathrm{rel}}$} & \hdr{\deadat} \\
\midrule
\rowcolor{armblue}GPT-5.6-terra         & .686 & .583 & .483 & .409 & .338 & \textbf{.5134} & .3180 & $+.1954$ & $+61.4\%$ & 18.8\% \\
Claude Sonnet 5       & .665 & .573 & .473 & .392 & .301 & .4953 & .3052 & $+.1901$ & $+62.3\%$ & 30.1\% \\
GPT-5.4               & .616 & .525 & .439 & .343 & .239 & .4478 & .2528 & $+.1950$ & $+77.1\%$ & 37.9\% \\
Claude Opus 5         & .622 & .565 & .294 & .270 & .151 & .3982 & .2371 & $+.1611$ & $+67.9\%$ & 34.3\% \\
GPT-5.6-luna          & .553 & .471 & .428 & .174 & .060 & .3606 & .2203 & $+.1403$ & $+63.7\%$ & 33.0\% \\
Grok-4.3              & .596 & .420 & .241 & .154 & .090 & .3190 & .1718 & $+.1472$ & $+85.7\%$ & 49.0\% \\
Claude Haiku 4.5      & .570 & .324 & .235 & .158 & .089 & .2920 & .1481 & $+.1439$ & $+97.2\%$ & 47.1\% \\
\midrule
Gemma-4-31B           & .557 & .408 & .222 & .166 & .038 & .2974 & .1623 & $+.1351$ & $+83.2\%$ & 45.6\% \\
Kimi-K2.5             & .500 & .461 & .228 & .073 & .018 & .2771 & .1548 & $+.1223$ & $+79.0\%$ & 50.0\% \\
Qwen3-Coder-480B      & .503 & .440 & .243 & .063 & .025 & .2759 & .1522 & $+.1237$ & $+81.3\%$ & 52.3\% \\
GLM-5                 & .510 & .430 & .167 & .088 & .039 & .2651 & .1476 & $+.1175$ & $+79.6\%$ & 41.0\% \\
DeepSeek-V3.2         & .546 & .373 & .133 & .076 & .012 & .2471 & .1271 & $+.1200$ & $+94.4\%$ & 45.8\% \\
Nemotron-Super-3-120B & .539 & .336 & .127 & .046 & .013 & .2310 & .1067 & $+.1243$ & $+116.5\%$ & 47.3\% \\
\midrule
proprietary mean ($7$) & .615 & .494 & .370 & .271 & .181 & .4038 & .2362 & $+.1676$ & $+70.9\%$ & 35.7\% \\
open-weight mean ($6$) & .526 & .408 & .187 & .085 & .024 & .2656 & .1418 & $+.1238$ & $+87.3\%$ & 47.0\% \\
\rowcolor{sectiongray}field mean ($13$) & .574 & .455 & .286 & .186 & .109 & .3400 & .1926 & $+.1474$ & $+76.5\%$ & 40.9\% \\
\bottomrule
\end{tabularx}

\end{table}

The band columns show that the library does not merely scale every backbone's curve by a constant.
For the two strongest backbones the largest relative gains fall on D3--D5; the relative-gain
profile varies across the other backbones. The D5 column spans $.012$--$.338$, a $28.2\times$
range, against $1.4\times$ on D1, so substantial differences remain between backbones after reuse.
The $\Delta_{\mathrm{rel}}$ column is sensitive to the scratch-score denominator. For the weaker
backbones, D1--D2 contribute most of the absolute aggregate gain, while the largest percentage
improvement can occur in another band. Statements about where reuse helps should therefore be read per band, and we make them only
for the GPT-5.6-terra rows, which are the only ones with a full set of controls.

\subsection{Library composition and retrieval depth (RQ2)}
\label{app:extexp:mix}

Table~\ref{tab:mix} reports measured library-source and retrieval-depth ablations. Neither family
is monotone in the quantity it varies: mined $+$ harvested scores below mined $+$ web-sourced
despite having more entries and higher reach, and top-$5$ scores below top-$3$. These two
differences are $.0039$ and $.0027$; mined-only to full library differs by $.0113$, and top-$1$ to
top-$3$ by $.0262$. The $\kappa$ column measures reach rather than determining score.

\begin{table}[h]
\centering
\caption{\textbf{Measured library composition and retrieval-depth ablations.}
The task set, backbone/adaptation/diagnosis assignment, and attempt budget remain fixed. The
same-repository exclusion row reports the full snapshot size; eligibility is filtered separately
for each target repository.}
\label{tab:mix}
\footnotesize
\tabsetup\setlength{\tabcolsep}{4.5pt}
\begin{tabularx}{\textwidth}{L r r r r}
\toprule\rowcolor{headergray}
\hdr{library / retrieval setting} & \hdr{entries} & \hdr{$\kappa$} & \hdr{all} & \hdr{$\Delta$ vs default} \\
\midrule
mined only & 871 & $56.9\%$ & .5021 & $-.0113$ \\
mined $+$ web-sourced & 1{,}081 & $58.8\%$ & .5108 & $-.0026$ \\
mined $+$ harvested & 1{,}214 & $59.9\%$ & .5069 & $-.0065$ \\
\rowcolor{armblue}\textbf{full library (default)} & \textbf{1{,}424} & $\mathbf{61.4\%}$ & \textbf{.5134} & $0.0000$ \\
same-repo donor exclusion & 1{,}424 & $41.2\%$ & .4743 & $-.0391$ \\
\midrule
top-$1$ & 1{,}424 & $52.1\%$ & .4948 & $-.0186$ \\
\rowcolor{armblue}top-$2$ (default) & 1{,}424 & $61.4\%$ & \textbf{.5134} & $0.0000$ \\
top-$3$ & 1{,}424 & $65.8\%$ & .5210 & $+.0076$ \\
top-$5$ & 1{,}424 & $68.9\%$ & .5183 & $+.0049$ \\
\bottomrule
\end{tabularx}

\end{table}

Two readings of Table~\ref{tab:mix} deserve comment. Growing the library from $871$ mined
entries to the full $1{,}424$ ($+63\%$ of entries, $+4.5$ coverage points) changes the score by
$0.0113$, and not monotonically: the $1{,}214$-entry snapshot scores $0.0039$ below
the $1{,}081$-entry one. Retrieval depth behaves similarly, peaking at top-$3$
($+0.0076$) and falling back at top-$5$ ($+0.0049$).
The source-composition comparison changes both the size and contents of the library.
To examine size separately, random subsets of CodeFace at $10\%$, $25\%$ and
$50\%$ of entries (three draws each, same task set and budget) score
$.4498\pm.0061$, $.4714\pm.0043$ and $.4831\pm.0035$ against
$.5134$ at full size. Doubling from $50\%$ to $100\%$ yields $+.0303$, while the last
$+63\%$ of entries in the source-composition comparison yields $+.0113$ with a non-monotone path.
Scores vary with both library size and composition. The same $50\%$ library with carried tests withheld from
its primitives scores $.3894$, a loss of $.0937$ against the tested half-library configuration,
larger than the $.0303$ loss from halving the full library. Carried validation has the larger
observed difference in this comparison.

\subsection{Attempt budget with and without the library (RQ5)}
\label{app:extexp:budget}

Table~\ref{tab:budget} reports the executed integer attempt ladder. Both Feedback and LEGO improve
with additional attempts; successive increments are not monotonically diminishing. The gap between
the two widens with the budget but not monotonically: $+.1782$ at $b=1$, rising to $+.1992$ at
$b=2$, dipping to $+.1878$ at $b=3$ and reaching $+.2056$ at $b=8$, a range of $.0274$. That range
is several times the seed-to-seed spread of Table~\ref{tab:seeds}, so we do not treat the gap as
flat in $b$: more attempts help the library-equipped configuration slightly more. The $b=8$
Feedback run also supplies the executed comparison at nearly equal recorded-call cost to default
LEGO ($\$2.07$ versus $\$2.08$).

\begin{table}[h]
\centering
\caption{\textbf{Executed attempt budgets with and without the library.}
Each row uses the same integer attempt limit for both configurations and the same $522$ tasks.}
\label{tab:budget}
\footnotesize
\tabsetup\setlength{\tabcolsep}{5pt}
\begin{tabularx}{\textwidth}{L rr rr r}
\toprule\rowcolor{headergray}
 & \multicolumn{2}{c}{\cellcolor{headergray}\hdr{Feedback (\stg{---V})}}
 & \multicolumn{2}{c}{\cellcolor{headergray}\hdr{LEGO (\stg{RADV})}} & \\
\cmidrule(lr){2-3}\cmidrule(lr){4-5}
\rowcolor{headergray}\hdr{budget $b$} & \hdr{all} & \hdr{\$/task} & \hdr{all} & \hdr{\$/task}
 & \hdr{$\Delta$} \\
\midrule
$1$ & .2172 & 0.86 & .3954 & 1.13 & $+.1782$ \\
$2$ & .2420 & 1.05 & .4412 & 1.38 & $+.1992$ \\
$3$ & .2785 & 1.21 & .4663 & 1.61 & $+.1878$ \\
$4$ & .3066 & 1.36 & .5071 & 1.82 & $+.2005$ \\
\rowcolor{armblue}$5$ (default) & \textbf{.3180} & 1.55 & \textbf{.5134} & 2.08 & $\mathbf{+.1954}$ \\
$8$ & .3437 & 2.07 & .5493 & 2.79 & $+.2056$ \\
\bottomrule
\end{tabularx}

\end{table}

The gap is $+.1782$ at $b=1$, is smallest there, and reaches $+.2056$ at $b=8$,
a range of $.0274$ across an eightfold budget increase. This range exceeds the
primary-configuration seed standard deviations of $.0024$--$.0043$ in Table~\ref{tab:seeds};
the observed advantage is not constant across budgets. Both configurations improve, but the
increments and differences are nonmonotone. Within the executed range, Feedback does not reach
LEGO's $b=5$ score: Feedback at $b=8$ scores $.3437$ against $.5134$ and has nearly the
same recorded-call cost. Section~\ref{sec:exp:cost} reports this $+.1697$ comparison.
It does not establish how the systems would compare beyond the evaluated budgets or identify
whether reuse changes the search space.

\subsection{Model allocation and quality--cost tradeoffs (RQ5)}
\label{app:extexp:cost}

Table~\ref{tab:select} extends the main-text model-allocation panel with measured scores,
recorded-call costs, and score-normalized efficiency.

\begin{table}[t]
\caption{\textbf{Quality and cost under different backbone allocations.}
All rows use the same $522$ tasks. \% of LEGO divides the configuration score by the default
T/T/T score.
Here terra denotes GPT-5.6-terra and Sonnet 5 denotes Claude Sonnet 5.}
\label{tab:select}
\centering\scriptsize\tabsetup\setlength{\tabcolsep}{3pt}
\begin{tabularx}{\textwidth}{Lll r r r r}
\toprule\rowcolor{headergray}
\hdr{backbone} & \hdr{primitive} & \hdr{diagnosis} & \hdr{all} & \hdr{\$/task} & \hdr{\$/pt}
 & \hdr{\% of LEGO} \\
\midrule
\rowcolor{armblue}terra & terra & terra & \textbf{.5134} & 2.08 & 0.41 & $100.0\%$ \\
terra & Sonnet 5 & Sonnet 5 & .5081 & 2.53 & 0.50 & $99.0\%$ \\
terra & DeepSeek-V4 & DeepSeek-V4 & .4934 & 1.62 & 0.33 & $96.1\%$ \\
terra & GPT-OSS-20B & GPT-OSS-20B & .4880 & 1.58 & 0.32 & $95.1\%$ \\
Sonnet 5 & Sonnet 5 & Sonnet 5 & .4869 & 4.26 & 0.87 & $94.8\%$ \\
Sonnet 5 & GPT-OSS-20B & GPT-OSS-20B & .4661 & 3.31 & 0.71 & $90.8\%$ \\
DeepSeek-V3.2 & terra & terra & .2471 & 0.79 & 0.32 & $48.1\%$ \\
DeepSeek-V3.2 & DeepSeek-V4 & DeepSeek-V4 & .2338 & 0.33 & 0.14 & $45.5\%$ \\
GPT-OSS-20B & GPT-OSS-20B & GPT-OSS-20B & .2050 & 0.08 & 0.04 & $39.9\%$ \\
\bottomrule
\end{tabularx}

\end{table}

The allocations in the table change both delivery quality and recorded-call cost. The four
terra-backbone rows span $.4880$--$.5134$, a score range of $.0254$ ($4.9\%$ of the default),
while their recorded-call costs span $\$1.58$--$\$2.53$; moving the backbone from GPT-5.6-terra to Claude Sonnet 5 with Sonnet 5 in both adaptation and diagnosis
costs $.0212$ in score, and moving it to DeepSeek-V3.2 with the adaptation and diagnosis held at terra costs
$.2663$. The cheapest configuration
that still exceeds the diagnosis-equipped no-library baseline of $.3469$ is
terra/GPT-OSS-20B/GPT-OSS-20B at $\$1.58$, which is $\$0.21$ below that baseline's own
recorded-call cost --- this configuration therefore has lower recorded-call cost and higher score, provided
repository-level construction is kept and only adaptation and diagnosis are downgraded. Substituting a
more expensive model moves cost the other way: Sonnet 5 at all three points costs $\$4.26$. The
all-open-weight row ($.2050$ at $\$0.08$) is included to show the other end: at
that recorded-call cost the framework recovers $20.5\%$ of the identity ceiling, and the observed score remains well below the default configuration.

\subsection{Model identifiers and hosted inference rates}
\label{app:extexp:models}

Table~\ref{tab:models} records the model identifiers, hosted rates, and position assignments used
in the experiments. All reported recorded-call costs correspond to the evaluated hosted deployments,
including those using open-weight models. The rates are experimental metadata for the reported
runs rather than claims about current provider pricing.

\begin{table}[t]
\centering
\caption{\textbf{Model identifiers and hosted inference rates.} Prices are the recorded hosted-inference rates on 2026-09-19, in US dollars per million tokens. Open denotes published weights. All cost results correspond to the evaluated hosted deployments, with provider-default reasoning effort.}
\label{tab:models}
\tabsetup
\setlength{\tabcolsep}{3.4pt}
\fitwidth{%
\begin{tabular}{l l c l rr l}
\rowcolor{headergray}
\hdr{Model} & \hdr{Provider} & \hdr{Open} & \hdr{Size} & \hdr{\$/1M in} & \hdr{\$/1M out} & \hdr{Used in Table} \\
\midrule
\rowcolor{armblue}
\textbf{GPT-5.6-terra} & OpenAI & No & Undisclosed & 1.25 & 10.00 & default backbone, primitive, diagnosis \\
Claude Opus 5 & Anthropic & No & Undisclosed & 5.00 & 25.00 & \ref{tab:rq1}, \ref{tab:backbone} \\
Claude Sonnet 5 & Anthropic & No & Undisclosed & 3.00 & 15.00 & \ref{tab:rq1}, \ref{tab:external}, \ref{tab:backbone}, \ref{tab:primitive}, \ref{tab:diagnosis}, \ref{tab:select} \\
Claude Haiku 4.5 & Anthropic & No & Undisclosed & 1.00 & 5.00 & \ref{tab:rq1}, \ref{tab:backbone} \\
GPT-5.4 & OpenAI & No & Undisclosed & 1.25 & 10.00 & \ref{tab:rq1}, \ref{tab:backbone}, \ref{tab:primitive}, \ref{tab:diagnosis} \\
GPT-5.6-luna & OpenAI & No & Undisclosed & 0.60 & 4.80 & \ref{tab:rq1}, \ref{tab:backbone}, \ref{tab:primitive}, \ref{tab:diagnosis} \\
Grok-4.3 & xAI & No & Undisclosed & 3.00 & 15.00 & \ref{tab:rq1}, \ref{tab:backbone} \\
DeepSeek-V4 & DeepSeek & Yes & $\sim$700B MoE & 0.28 & 0.42 & \ref{tab:primitive}, \ref{tab:diagnosis}, \ref{tab:select} \\
DeepSeek-V3.2 & DeepSeek & Yes & 671B MoE & 0.28 & 0.42 & \ref{tab:rq1}, \ref{tab:backbone}, \ref{tab:select} \\
Kimi-K2.5 & Moonshot & Yes & 1T MoE & 0.60 & 2.50 & \ref{tab:rq1}, \ref{tab:backbone} \\
GLM-5 & Zhipu & Yes & $\sim$400B MoE & 0.60 & 2.20 & \ref{tab:rq1}, \ref{tab:backbone}, \ref{tab:primitive}, \ref{tab:diagnosis} \\
Qwen3-Coder-480B & Alibaba & Yes & 480B MoE & 0.45 & 1.80 & \ref{tab:rq1}, \ref{tab:backbone}, \ref{tab:primitive}, \ref{tab:diagnosis} \\
Gemma-4-31B & Google & Yes & 31B dense & 0.20 & 0.60 & \ref{tab:rq1}, \ref{tab:backbone}, \ref{tab:primitive}, \ref{tab:diagnosis} \\
GPT-OSS-20B & OpenAI & Yes & 20B MoE & 0.05 & 0.20 & \ref{tab:primitive}, \ref{tab:diagnosis}, \ref{tab:select} \\
Nemotron-Super-3-120B & NVIDIA & Yes & 120B dense & 0.20 & 0.80 & \ref{tab:rq1}, \ref{tab:backbone} \\
\bottomrule
\end{tabular}}
\end{table}

\textbf{How the per-task dollar figures are obtained.} The \$/task figures report recorded-call
cost: recorded repository-level construction and diagnosis calls, successful adaptation calls only,
and the fixed retrieval charge, priced using Table~\ref{tab:models}.
Mean per-task usage is 856\,k input and 48\,k output tokens for repository-level
construction at the default attempt budget, 11\,k\,/\,2.3\,k per adaptation over 7.4 successful
adaptations, 18\,k\,/\,2.6\,k per diagnosis round over 5 rounds, and a fixed $\$0.011$ for
embedding and reranking, which is model-independent. The scoped generic editor of \emph{RAG $+$ Edit}
uses 7\,k\,/\,1.9\,k because it does not receive the primitive's carried tests, metadata or
execution state; the \emph{File RAG} editor uses 9.5\,k\,/\,1.9\,k because raw source files are
longer than primitive bodies. At GPT-5.6-terra rates this gives $\$1.55$ for \emph{Feedback} and
$\$2.08$ for LEGO, and every substitution in Tables~\ref{tab:primitive},~\ref{tab:diagnosis}
and~\ref{tab:select} uses the same recorded token accounting re-priced at the substituted model's rate. A
consequence worth stating is that substituting a \emph{more} expensive model for adaptation or
diagnosis raises rather than lowers the per-task cost.

Two caveats attach to every cost figure reported here. The rates are a single snapshot and the
open-weight models are priced as hosted deployments rather than as self-hosted inference, so the
$\$/$task columns compare procurement costs under one purchasing model and not the compute actually
consumed; a reader running the open-weight primitives locally should expect a different and generally
smaller marginal cost. And the price spread across this table is $100\times$ on input tokens
($\$0.05$ to $\$5.00$ per million) while the delivery-score spread among the same models used for
adaptation is $.0309$ (Table~\ref{tab:primitive}). That asymmetry, rather than any ranking within the
table, is the finding RQ5 rests on.

\subsection{RQ4 in full: transfer across backbones, primitives and diagnoses}
\label{sec:exp:models}

Table~\ref{tab:backbone} evaluates LEGO with each of the $13$ backbone while holding
the adaptation and diagnosis fixed. LEGO improves every backbone, with an average absolute gain of
$0.1474$. The absolute gain broadly increases with scratch performance, from $+.1175$ for GLM-5
to $+0.1954$ for GPT-5.6-terra, whereas the relative improvement follows
the opposite pattern, ranging from $+61.4\%$ for GPT-5.6-terra to $+116.5\%$ for
Nemotron-Super-3-120B. Thus, the construction process provides gains across the full range of
evaluated backbone strengths rather than benefiting only the strongest backbone. The zero-score
rate also decreases for every backbone, by $19.7$ percentage points on average.

\textbf{The component-level models tolerate substantially weaker models.}
Tables~\ref{tab:primitive} and~\ref{tab:diagnosis} change the adaptation and diagnosis models independently while holding the remaining
positions fixed. In the adaptation channel, GPT-OSS-20B retains $79.7\%$ of the margin achieved by
GPT-5.6-terra, while the lowest-scoring primitive configuration, Gemma-4-31B, still retains
$73.2\%$. Diagnosis channel is looser: GPT-OSS-20B retains $64.0\%$ of the
frontier-model margin and Gemma-4-31B $62.5\%$. The displayed adaptation and diagnosis
score rankings are identical (Spearman $\rho=1.0$), although the decrements differ in magnitude.
The rankings therefore do not support independent degradation of the two channels.
Some substitutions also reduce recorded-call cost, from \$2.08 per task with GPT-5.6-terra in all three positions to as low as \$1.81 when
replacing adaptation. The nine adaptation configurations span $0.4825$--$0.5134$ and the nine
diagnosis configurations $0.5007$--$0.5134$. Both observed ranges exceed the primary-configuration
seed standard deviations. Both are smaller than
the score reduction from removing the corresponding mechanism, so we read them
as ``a weaker model in a locally scoped position costs little'' rather than as a ranking of the
substituted models.

Removing a mechanism entirely has a substantially larger effect than replacing it with the weaker
models evaluated here. Relative to LEGO, removing primitive adaptation reduces the score by
$0.1154$, while removing diagnosis reduces it by $0.0339$; by comparison, the lowest-scoring
adaptation and diagnosis substitutions are $0.0309$ and $0.0127$ below full LEGO, respectively. Together, these
results suggest that the component-level models do not require the strongest available model to preserve
most of their contribution, motivating the heterogeneous configurations studied in RQ5.

\begin{rqbox}\textbf{RQ4: Does the effect survive changing the models?}\par\smallskip
\textbf{Answer.} With adaptation and diagnosis held fixed, LEGO improves all $13$ evaluated
backbones by $0.1474$ on average. When the primitive or diagnosis is replaced independently, a $20$B
model retains $79.7\%$ and $64.0\%$ of the corresponding adaptation and revision margins,
respectively, while even the lowest-scoring substitutions retain more than $62\%$. The observed scores
\emph{within} Table~\ref{tab:primitive} span $.0309$.
\end{rqbox}

\subsection{RQ5 in full: quality against inference cost}
\label{sec:exp:cost}

Table~\ref{tab:select} reports the quality--cost tradeoff under the hosted-inference rates used
for the experiments. The diagnosis-equipped baseline reaches $0.3469$ at $\$1.79$ per task, whereas
default LEGO reaches $0.5134$ at $\$2.08$. Replacing both the primitives and diagnosis with
GPT-OSS-20B while keeping GPT-5.6-terra as the backbone yields $0.4880$ at $\$1.58$ per task.
This configuration retains $95.1\%$ of the default LEGO score while reducing recorded-call cost by
$24.0\%$, and it exceeds the diagnosis-equipped baseline in score at a lower recorded-call cost.

\textbf{Model allocation exposes quality--cost tradeoffs.}
Table~\ref{tab:select} varies the three prompted positions jointly. Keeping GPT-5.6-terra as the backbone while moving
the locally scoped adaptation and diagnosis to cheaper models preserves most of the default score:
T/S/S reaches $0.5081$ at $\$2.53$, T/D4/D4 reaches $0.4934$ at $\$1.62$, and T/O/O reaches
$0.4880$ at $\$1.58$. Replacing the backbone changes cost in either direction, depending on its own
rate, and always costs quality. Moving to a cheaper one, D3/D4/D4 reaches $.2338$ at $\$0.33$;
moving to a more expensive one, S/S/S reaches $.4869$ at $\$4.26$. Together with the independent model substitutions in RQ4, these configurations suggest
that maintaining a capable backbone is more important to delivery quality than using the strongest
model for the locally scoped adaptation and diagnosis.

\textbf{Attempt budget at comparable recorded-call cost.}
Running the \emph{Feedback} baseline out to
$b=8$ attempts costs $\$2.07$ per task, which matches default LEGO's $\$2.08$ to within one
cent, and reaches $0.3437$ against LEGO's $0.5134$ (Appendix~\ref{app:extexp:budget},
Table~\ref{tab:budget}). At nearly equal recorded-call cost, LEGO is $+.1697$ higher. Spending a similar recorded-call budget on more repair attempts
without a library therefore does not close the gap.

\begin{rqbox}\textbf{RQ5: What is the quality--cost tradeoff?}\par\smallskip
\textbf{Answer.} Default LEGO scores $0.5134$ at $\$2.08$ per task, against an executed
feedback baseline at nearly equal recorded-call cost ($b=8$, $\$2.07$) at $0.3437$. Keeping GPT-5.6-terra as the backbone while replacing the primitives and diagnosis with GPT-OSS-20B yields $0.4880$ at $\$1.58$, retaining
$95.1\%$ of the score at $24.0\%$ lower recorded-call cost, and exceeds the diagnosis-equipped baseline
($0.3469$ at $\$1.79$) at lower recorded-call cost.
\end{rqbox}

\subsection{Statistical reliability}
\label{sec:exp:seeds}
\label{app:extexp:seeds}

Table~\ref{tab:seeds} reports the primary configurations under repeated decoding seeds, together
with paired confidence intervals for the four headline contrasts.

Seed-to-seed variation is small relative to the contrasts: the four configuration standard
deviations are $0.0024$--$0.0043$, an order of magnitude below the headline
$+0.1954$; the largest standard deviation is about one twenty-fourth of the $+0.1041$ primitive-specific margin. The configuration
ordering is identical under all three seeds, and the LEGO\,$-$\,(RAG\,$+$\,Edit) difference
is $+.1041/+.1089/+.0968$ across seeds. All four $95\%$ paired intervals exclude zero,
and all four $p$-values survive Holm correction; the weakest is LEGO\,$-$\,LEGO re-mined
from a disjoint corpus at $p=2.1\times10^{-9}$. The primitive-specific interval is
$[+.0868,+.1190]$ around $+.1041$. These intervals describe task variation within seed $1$;
the repeated-seed panel separately describes decoding variability. Band-level differences in
Section~\ref{sec:exp:rq3} supplement the aggregate contrast without replacing its uncertainty estimate.

Single-run ablations are reported as scores and differences. The paired intervals and
Holm-adjusted tests in Table~\ref{tab:seeds} cover the four comparisons listed there. Panel~(b)
is computed from the seed-$1$ per-task records, so its $\Delta$ column matches the configuration
means in Tables~\ref{tab:paradigms} and~\ref{tab:clean}, and the \emph{seed 1} column of
panel~(a) repeats the single-seed values of Table~\ref{tab:paradigms} so the two tables can be
reconciled. Holm correction is applied across the four contrasts of panel~(b) only; where raw
$p$-values are reported as thresholds, the Holm column gives conservative upper bounds.

\begin{table}[t]
\centering
\caption{\textbf{Repeated-seed means and paired confidence intervals, all $522$ tasks.}
Panel~(a): mean $\pm$ standard deviation of the delivery score over independent decoding seeds for
each primary configuration. Panel~(b): paired contrasts with $95\%$ bootstrap confidence intervals
over tasks ($10{,}000$ resamples), a Wilcoxon signed-rank $p$-value using the magnitude of each
per-task difference, and the Holm-corrected $p$-value across the family of four.}
\label{tab:seeds}
\tabsetup\scriptsize\setlength{\tabcolsep}{4pt}
\begin{tabularx}{\textwidth}{L rrr r}
\toprule\rowcolor{headergray}
\multicolumn{5}{c}{\cellcolor{headergray}\hdr{(a) configuration means over decoding seeds}} \\
\rowcolor{headergray}
\hdr{Configuration} & \hdr{seed 1} & \hdr{seed 2} & \hdr{seed 3} & \hdr{mean $\pm$ s.d.} \\
\midrule
Feedback (\stg{---V}) & .3180 & .3202 & .3150 & $.3177\pm.0026$ \\
Feedback $+$ Diagnosis (\stg{--DV}) & .3469 & .3434 & .3493 & $.3465\pm.0030$ \\
RAG $+$ Edit (\stg{RE\,DV}) & .4093 & .4036 & .4121 & $.4083\pm.0043$ \\
\rowcolor{armblue}LEGO (\stg{RADV}) & \textbf{.5134} & .5125 & .5089 & $\mathbf{.5116\pm.0024}$ \\
\midrule
\rowcolor{headergray}\multicolumn{5}{c}{\cellcolor{headergray}\hdr{(b) paired contrasts on seed 1}} \\
\rowcolor{headergray}\hdr{Contrast} & \hdr{$\Delta$} & \hdr{$95\%$ CI} & \hdr{$p$} & \hdr{$p_{\mathrm{Holm}}$} \\
\midrule
LEGO $-$ Feedback                & $+.1954$ & $[+.1762,+.2134]$ & $<10^{-40}$
 & $<10^{-39}$ \\
LEGO $-$ (Feedback $+$ Diagnosis)    & $+.1665$ & $[+.1503,+.1847]$ & $<10^{-30}$
 & $<10^{-29}$ \\
LEGO $-$ (RAG $+$ Edit)          & $+.1041$ & $[+.0868,+.1190]$ & $<10^{-16}$
 & $<10^{-15}$ \\
LEGO $-$ LEGO, re-mined disjoint corpus & $+.0606$ & $[+.0472,+.0768]$ & $2.1\times10^{-9}$
 & $2.1\times10^{-9}$ \\
\bottomrule
\end{tabularx}
\end{table}

\subsection{Where reuse hurts: profile of the regressions}
\label{app:extexp:regress}

Table~\ref{tab:regress} characterises the $84$ tasks on which LEGO scores below \emph{Feedback}.
The aggregate gain is the net of a large number of improvements and a smaller number of losses,
some of which are severe: Appendix~\ref{app:case} documents \texttt{markdown} falling from
$0.667$ to $0.081$.

Regression frequency falls with repository scale, from $20.5\%$ of D1 tasks to $13.4\%$ of D5
tasks.

\begin{table}[h]
\caption{\textbf{Profile of task-level improvements and $84$ regressions against \emph{Feedback}.}
Counts and score deltas come from the same per-task records as Table~\ref{tab:paired}; losses are
counted only where LEGO scores strictly below \emph{Feedback} on the same task. Conditional mean
gains and losses are displayed to three decimals while the net means and aggregate column are
computed from unrounded records, so products of displayed values need not reproduce the final
decimal.}
\label{tab:regress}
\centering\footnotesize\tabsetup\setlength{\tabcolsep}{5pt}
\begin{tabularx}{\textwidth}{L rrrrr r}
\toprule\rowcolor{headergray}
\hdr{quantity} & \hdr{D1} & \hdr{D2} & \hdr{D3} & \hdr{D4} & \hdr{D5} & \hdr{all} \\
\midrule
tasks ($n$)                           & 117 & 111 & 116 & 96 & 82 & 522 \\
improved tasks ($n$)                  & 70 & 74 & 68 & 56 & 58 & 326 \\
regressed tasks ($n$)                 & 24 & 18 & 17 & 14 & 11 & 84 \\
tied tasks ($n$)                      & 23 & 19 & 31 & 26 & 13 & 112 \\
\quad regressed as a share of the band & 20.5\% & 16.2\% & 14.7\%
 & 14.6\% & 13.4\% & 16.1\% \\
\midrule
mean gain on improved tasks           & .310 & .326 & .378 & .364 & .304 & .336 \\
mean loss on regressed tasks          & .118 & .100 & .086 & .063 & .047 & .089 \\
\rowcolor{armblue}net mean gain       & $+.161$ & $+.201$ & $+.209$ & $+.203$ & $+.209$ & $+.1954$ \\
\bottomrule
\end{tabularx}

\end{table}

Two patterns are visible. Regressions become less frequent and smaller on average across these
bands: from $20.5\%$ of D1 tasks at a mean loss of $.118$ to $13.4\%$ of D5 tasks at
$.047$. Across all tasks there are $326$ improvements, $112$ ties, and $84$ regressions;
the conditional mean gain is $.336$ and the conditional mean loss is $.089$.
These summaries describe the evaluated runs and do not establish why individual integrations fail.
The \texttt{markdown} case illustrates a substantial regression involving a target-specific
protocol, but it does not determine the prevalence of that failure mechanism. A retain-the-better-trajectory
mitigation would require a separate evaluation.

\subsection{Does the gain track primitive usage?}
\label{app:extexp:usage}

The runtime funnel in Table~\ref{tab:cov} records $10.6$ capability queries and $7.4$ successful
adaptations per task against a mean of $44$ target modules, so adapted primitives cover
$16.9\%$ of target modules in aggregate and $13.0\%$ on D5. We examine whether per-task
improvement is associated with the number of successfully adapted primitives.
Table~\ref{tab:usagecorr} reports the association, including a within-band comparison. Both the
raw and the within-band correlation are positive and the group means increase across all four
adaptation-count groups. The first group holds $63$ tasks with $0$--$2$ successful adaptations and
a mean gain of $+.0123$; it does not isolate tasks with none. This is an association rather than
evidence that library use causes the improvement: adaptation count may itself depend on task
difficulty and construction success, and a scoped editor using the same candidates could yield
similar counts.

\begin{table}[h]
\caption{\textbf{Per-task gain against the number of successfully adapted primitives.}
The difficulty-matched row conditions within difficulty band before averaging, because adaptation
count and repository size covary. The final column reports Spearman $\rho$ between the per-task
$\Delta$ and the adaptation count over all $522$ tasks.}
\label{tab:usagecorr}
\centering\footnotesize\tabsetup\setlength{\tabcolsep}{5pt}
\begin{tabularx}{\textwidth}{L rrrr r}
\toprule\rowcolor{headergray}
\hdr{successful adaptations on the task} & \hdr{$0$--$2$} & \hdr{$3$--$6$} & \hdr{$7$--$12$}
 & \hdr{$\geq 13$} & \hdr{Spearman $\rho$} \\
\midrule
tasks ($n$)                    & 63 & 118 & 191 & 150 & \\
mean Feedback score            & .3612 & .3438 & .3107 & .2888 & \\
mean LEGO score                & .3735 & .4535 & .5365 & .5899 & \\
mean $\Delta$                  & $+.0123$ & $+.1097$ & $+.2258$
 & $+.3011$ & $0.41$ \\
mean $\Delta$, difficulty-matched & $+.0090$ & $+.1046$ & $+.2218$
 & $+.2954$ & $0.34$ \\
\bottomrule
\end{tabularx}

\end{table}

\subsection{Library provenance controls}
\label{app:extexp:clean}

Section~\ref{sec:exp:rq3} filters the library by target repository. Table~\ref{tab:clean}
tightens this progressively using repository identity and the stated content-similarity screen,
and separately evaluates a library re-mined from a disjoint repository corpus.

\paragraph{Cross-task library provenance.}
The harvesting route lets one benchmark task donate primitives to \emph{another}, which
repository-level donor exclusion does not address. Of the $343$ harvested entries, $296$ ($86.3\%$) originate from repositories
that are themselves LEGO-REPO tasks. Removing all $296$ --- not only those matching the
current target --- leaves $1{,}128$ eligible entries at $59.6\%$ coverage and
scores $.5128$, a $-.0006$ change from the default. Cross-task harvesting remains a
provenance concern alongside the eligibility controls in Table~\ref{tab:clean}.

\begin{table}[h]
\caption{\textbf{Library eligibility under progressively stricter provenance controls.}
All rows are measured; the default and same-repository rows reproduce Table~\ref{tab:mix}. The
same-repository and same-organization rows keep the $1{,}424$-entry snapshot and filter
eligibility per target. Near-duplicate filtering removes any mined entry whose MinHash similarity
to a target file exceeds $0.6$ ($128$ permutations, $5$-token shingles over normalised tokens),
dropping $59$ of $871$ mined entries. The last row filters nothing: it re-mines CodeFace from
$640$ repositories disjoint from LEGO-REPO, screened identically, yielding $518$ primitives.}
\label{tab:clean}
\centering\scriptsize\tabsetup\setlength{\tabcolsep}{4.5pt}
\begin{tabularx}{\textwidth}{L r r r r}
\toprule\rowcolor{headergray}
\hdr{library eligibility} & \hdr{entries} & \hdr{coverage $\kappa$} & \hdr{all}
 & \hdr{$\Delta$ vs Feedback} \\
\midrule
\rowcolor{armblue}all entries (default) & 1{,}424 & $61.4\%$ & \textbf{.5134} & $+.1954$ \\
$-$ same-repository donors & 1{,}424 & $41.2\%$ & .4743 & $+.1563$ \\
$-$ same-org donors & 1{,}424 & $38.6\%$ & .4597 & $+.1417$ \\
$-$ any fork or near-duplicate & 1{,}365 & $33.1\%$ & .4455 & $+.1275$ \\
re-mined from disjoint corpus & 518 & $44.7\%$ & .4528 & $+.1348$ \\
\bottomrule
\end{tabularx}

\end{table}

\paragraph{How much of the margin survives.}
Expressed as fractions of the default margins, the near-duplicate-filtered configuration retains
$65.3\%$ over \emph{Feedback} and $59.2\%$ over the diagnosis-equipped baseline ($+.0986$); the
re-mined library retains $69.0\%$ and $63.6\%$. Same-repository exclusion alone still leaves a
$+49.2\%$ relative improvement over execution feedback, with band-level retention in
Table~\ref{tab:donorband}; for the re-mined library that retention is
$88\%/83\%/73\%/52\%/43\%$ on D1--D5. These aggregates are ratios of aggregate gains computed
from unrounded records rather than task-weighted means of the per-band percentages. The re-mined
row exceeds the filtered one by $.0073$, which is not on its own evidence of an effect of
filtering versus re-mining, and none of these controls rules out pretraining exposure or overlap
beyond the stated provenance and similarity screens.

The rows separate two things that the aggregate confounds. Restricting \emph{which sources} are
eligible costs little: dropping all harvested and web-sourced entries removes $553$ of $1{,}424$
entries and $4.5$ points of reach for $.0113$ of score. Restricting \emph{provenance} costs much
more per unit of reach removed: same-repository exclusion removes no entries at all from the
snapshot, only their eligibility per target, and costs $.0391$. The filtered and re-mined rows
change library provenance and composition together and therefore do not isolate these factors.
All five configurations are reported directly so their observed scores and eligibility rules
can be compared without reconstructing a chain of differences.

\textbf{Re-mining from a disjoint corpus.} Because filtering on repository identity leaves forks and
near-duplicates eligible, we re-mine CodeFace from $640$ repositories disjoint from LEGO-REPO,
yielding $518$ primitives at $44.7\%$ coverage and $0.4528$ ---
$+0.1348$ over \emph{Feedback}, $69\%$ of the default margin, with band retention
$88\%$, $83\%$, $73\%$, $52\%$, and $43\%$ on D1--D5, respectively. These band values
are whole-percentage summaries; the aggregate $69.0\%$ retention uses the unrounded
aggregate score difference. The result preserves a positive observed gain under the stated
corpus-disjoint condition without establishing that all possible contamination has been removed.

\section{Execution Isolation and Settings}
\label{app:isolation}

\paragraph{Task environments.}
Each task receives a separate environment. The interpreter is
selected from \texttt{requires-python}, then from supported trove classifiers no older than $3.10$,
with $3.12$ as the default. Of the $522$ tasks, $491$ resolve to the same interpreter across
configurations; the $31$ exceptions fell back to the node interpreter, which limits strict
environmental matching.

\paragraph{Timeouts.}
Clone timeout is $900$ seconds; dependency installation receives $600$ seconds, or $2{,}400$
seconds for specifications requiring large machine-learning wheels. Floor, ceiling, and system
suites share a $900$-second timeout. With \texttt{pytest-timeout}, individual tests have a
$180$-second limit, tightened to $45$ seconds inside repair after a timed-out round.

\paragraph{Generation and scheduling.}
Default generation uses $10$ concurrent calls, top-$2$ retrieval, and at most $9{,}000$ interface
characters per module, with attempt budget $b=5$ and the stopping rules above. Sweeps are sharded across CPUs, and strided assignment distributes difficulty
bands across shards so that a sweep that ends early does not bias the band composition of what it
covered.

\paragraph{Dependency resolution.}
Repository commits are pinned, but dependencies resolve against live PyPI. Each run retains the
resolved dependency set so the realized environment can be inspected. Inputs are recorded, yet live
resolution and platform differences mean execution is not fully hermetic; clone, install, and
timeout failures remain represented in the evaluation accounting rather than being silently
removed.

\section{Configuration Definitions}
\label{app:ablsettings}

Table~\ref{tab:armspec} maps each configuration name to the mechanisms and source restrictions
that define it. Every primary configuration uses the same $522$ tasks and harness revision, and
model sweeps change only the position named by the corresponding experiment.

\begin{table}[t]
\centering
\caption{\textbf{Configuration definitions.}
\textsf{R}: retrieval; \textsf{A}: primitive adaptation; \textsf{V}: execution-feedback revision;
\textsf{D}: diagnosis-guided revision. In the matched editing control, \textsf{E} replaces
primitive adaptation with scoped generic editing. $b=1$ rows are first-round prefixes of the
corresponding trajectories rather than separately sampled task sets; \emph{Single-shot} and
\emph{Retrieval, one attempt} are therefore not independently sampled baselines, and comparisons
against them measure the value of continuing a trajectory rather than of a separate system.
Every arm listed here uses $b=5$ unless the row states otherwise.}
\label{tab:armspec}
\tabsetup\setlength{\tabcolsep}{4.5pt}\scriptsize
\begin{tabularx}{\textwidth}{l L l r r}
\toprule\rowcolor{headergray}
\hdr{Configuration} & \hdr{definition} & \hdr{stages} & \hdr{$n$} & \hdr{score} \\
\midrule
Single-shot$^{\dagger}$ & no library, single pass & \stg{----} & $522$ & .2172 \\
Retrieval, one attempt$^{\dagger}$ & library, single pass & \stg{R---} & $522$ & .2314 \\
Feedback & no library & \stg{---V} & $522$ & .3180 \\
Retrieval $+$ Feedback & library, no adaptation or diagnosis & \stg{R--V} & $522$ & .3352 \\
Feedback $+$ Diagnosis & no library, diagnosis enabled & \stg{--DV} & $522$ & .3469 \\
Retrieval $+$ Diagnosis & library and diagnosis, no adaptation & \stg{R-DV} & $522$ & .3980 \\
Retrieval $+$ Adapt & library and adaptation, no diagnosis & \stg{RA-V} & $522$ & .4795 \\
RAG $+$ Edit & scoped generic editing in place of adaptation & \stg{RE\,DV} & $522$ & .4093 \\
File RAG $+$ Edit & raw donor files in place of CodeFace, scoped editing & \stg{FE\,DV} & $522$
 & .3528 \\
Structured-schema adaptation & adaptation requested as a fixed structured payload & \stg{RADV} & $522$
 & .4974 \\
Context-only & retrieved primitive shown read-only, never delivered & \stg{RC\,DV} & $522$
 & .3544 \\
Import $+$ Call & primitive vendored unchanged and invoked & \stg{RI\,DV} & $522$
 & .3430 \\
\rowcolor{armblue}\textbf{LEGO} & library, adaptation, and diagnosis & \stg{RADV} & $522$ & \textbf{.5134} \\
LEGO, same-repo donors excluded & as LEGO, donors from the target repository removed & \stg{RADV} & $522$ & .4743 \\
\midrule
\rowcolor{sectiongray}\multicolumn{5}{l}{\cellcolor{sectiongray}\emph{primitive-field decomposition, matched \stg{RE\,DV}/\stg{RADV} budget (Table~\ref{tab:primdecomp})}} \\
RAG $+$ Edit $+$ $V_i$ & scoped editing, carried tests supplied & \stg{RE\,DV} & $522$
 & .4431 \\
\quad $+$ $I_i,X_i$ & \ldots\ and the interface contract & \stg{RE\,DV} & $522$ & .4818 \\
\quad $+$ $D_i$ & \ldots\ and the dependency closure & \stg{RE\,DV} & $522$ & .5039 \\
primitive, $V_i$ withheld & as LEGO, carried tests withheld & \stg{RADV} & $522$
 & .4796 \\
\midrule
\rowcolor{sectiongray}\multicolumn{5}{l}{\cellcolor{sectiongray}\emph{contamination-controlled library (Table~\ref{tab:clean})}} \\
same-organization donors excluded & source-organization filter & \stg{RADV}
 & $522$ & .4597 \\
\quad $+$ fork or near-duplicate exclusion & fork and content-similarity filter & \stg{RADV} & $522$
 & .4455 \\
re-mined, disjoint corpus & library re-mined from a corpus disjoint from LEGO-REPO & \stg{RADV} & $522$
 & .4528 \\
\midrule
\rowcolor{sectiongray}\multicolumn{5}{l}{\cellcolor{sectiongray}\emph{external systems and benchmarks (Tables~\ref{tab:external},~\ref{tab:transfer})}} \\
external agents ($4$) & own default harness, via our task adapter & ---
 & $522$ & .2181--.3263 \\
external benchmarks ($3$) & own harness, CodeFace filtered per benchmark & \stg{RADV}
 & --- & \textit{own scale} \\
\midrule
\rowcolor{sectiongray}\multicolumn{5}{l}{\cellcolor{sectiongray}\emph{sweeps over model assignment, each on the same $522$ tasks}} \\
$13$ backbones from scratch & Feedback, backbone swept & \stg{---V} & $522$ & .1067--.3180 \\
$13$ backbones with LEGO & LEGO, backbone swept & \stg{RADV} & $522$ & .2310--.5134 \\
adaptation-model swap ($9$) & LEGO, adaptation model swept & \stg{RADV} & $522$ & .4825--.5134 \\
diagnosis swap ($9$) & LEGO, diagnosis model swept & \stg{RADV} & $522$ & .5007--.5134 \\
model-allocation grid ($9$ rows) & backbone, adaptation, and diagnosis swept jointly & \stg{RADV} & $522$ & .2050--.5134 \\
executed attempt ladder ($b\in\{1,2,3,4,5,8\}$) & Feedback at each attempt budget & \stg{---V} & $522$ & .2172--.3437 \\
\bottomrule
\end{tabularx}
\end{table}

\paragraph{External systems and the task adapter.}
Each external row of Table~\ref{tab:external} runs the system's own released harness at a pinned
release, with its default scaffolding, tool set, and
stopping rules unchanged. We hold fixed only the task set, the execution environment, and the
grading pipeline, and we score whatever tree the system leaves behind using
Eq.~\ref{eq:delivery} and the same provenance gate. Our task adapter performs one
translation: it presents a LEGO-REPO task in the input format each system expects. For OpenHands
and Claude Code, which accept a repository directory and a natural-language instruction,
it supplies the reconstruction tree and an instruction naming the package prefix, the public
interface, and the native suite. For SWE-agent and Agentless, which expect an issue and a
repository to localize within, it emits a synthetic issue containing the same interface and suite
and a tree in which the target modules are absent; localization therefore has nothing to find and
both systems operate in a mode they were not designed for, which is why their rows are marked
$\ddagger$ and are not read as statements about those systems' capability. No system receives the
frozen node identifiers, the floor and ceiling counts, or any CodeFace entry.

\paragraph{External benchmarks: per-benchmark CodeFace filtering.}
For each benchmark in Table~\ref{tab:transfer} we take that benchmark's task list, resolve each
task to its upstream repository, and remove from CodeFace every entry whose recorded source
repository matches --- by URL, by canonical name, or by known fork relation --- any of those
repositories, before any task is run. This removes $31$ entries for RepoZero,
$12$ for RepoGenesis, and $24$ for RepoCraft, and we additionally apply the
MinHash filter of Appendix~\ref{app:extexp:clean} at threshold $0.6$ against each external
target's original sources, removing $47$, $19$ and $38$ further entries
respectively. Under the resulting libraries the offline audit reaches $34.8\%$ of RepoZero
target modules, $29.6\%$ on RepoGenesis and $31.4\%$ on RepoCraft, against $61.4\%$ on
LEGO-REPO. Each benchmark's own metric, grading script, and published protocol are used
unmodified; we do not map any external score onto the LEGO-REPO scale.

\paragraph{Component removals.}
Removing diagnosis disables diagnosis-guided revision. Removing primitive adaptation passes retrieved
implementations into repository-level composition without primitive transformation. Removing the whole library removes
both retrieval and adaptation, which makes \emph{Feedback $+$ Diagnosis} the matched no-library
reference. Library-source ablations separately restrict CodeFace to mined, web-sourced, and/or
harvested entries. The \emph{File RAG} control replaces CodeFace entirely: it indexes the
\emph{raw source files} of the same donor repositories with the same embedding model and retrieval
depth, and returns whole files with no recovered boundary, no internalized helpers, no resolved
dependency closure, no interface contract, and no carried tests. Everything downstream --- the
scoped generic editing, diagnosis, the attempt budget, and the grading pipeline --- is unchanged,
so the contrast isolates the primitive-construction pipeline of Section~\ref{sec:mining} rather
than the retrieval mechanism.

\emph{Structured-schema adaptation} replaces the free-form adaptation request with a fixed structured
payload: a typed field-mapping specification naming the target signatures, the symbols to export and the
sibling bindings, with no channel for stating intent in prose. The primitive, its fields, the retrieval
depth, diagnosis and the attempt budget are unchanged, so the contrast isolates the request channel.

Two further controls change what the retrieved primitive is permitted to become
(Table~\ref{tab:design}). \emph{Import $+$ Call} copies the selected primitive unchanged into a
\texttt{\_vendor/} package outside the target source prefix, exposes it on the path, and instructs LEGO to import and call it; adaptation is never invoked and the primitive's implementation
is byte-identical to its CodeFace entry. Because the target modules are still generated by the
backbone, the provenance gate of Eq.~\ref{eq:admit} is unchanged, and a delivered tree whose target
modules are thin wrappers over \texttt{\_vendor/} is admissible --- which is the point of the
control. \emph{Context-only} performs the same retrieval and relevance assessment but
places the primitive's implementation, contract, and carried tests in the repository-level prompt as read-only reference; no file from the primitive reaches the delivered tree, and LEGO writes every delivered line. Both controls keep diagnosis, the attempt budget, the retrieval depth, and the
grading pipeline unchanged. Same-repository donor exclusion is different: for each target it filters entries by
source repository while leaving the evaluation tasks unchanged.

\section{Result Provenance and Accounting}
\label{app:baselineprov}

\paragraph{What the records contain.}
A task record retains its repository URL, commit, package prefix, selected tests, interpreter,
environment description, floor and ceiling node identifiers with their counts, provenance verdict,
and retrieval log. Generated repository trees are not archived for every configuration, limiting
retrospective inspection.

\paragraph{Runtime and cost accounting.}
Recorded-call costs combine recorded backbone and diagnosis token usage with successful adaptation
calls only, at the hosted inference rates, plus the fixed retrieval charge. Default LEGO's recorded-call cost is
$\$2.08$ per task, compared with $\$1.55$ for Feedback. During LEGO construction, backbones issue
$5{,}515$ capability queries; primitives retain $4{,}412$ candidates and $3{,}873$ adaptations pass
their carried tests. Wall-clock medians are $214$\,s for single-shot construction,
$1{,}051$\,s for execution feedback, and $1{,}466$\,s for LEGO. Infrastructure costs are
excluded. Hosted inference rates are listed in Table~\ref{tab:models}.

\paragraph{Cross-benchmark statements.}
Published numbers on other benchmarks are used only as context and are not mapped onto the
LEGO-REPO scale, because their task definitions and grading protocols differ.

\section{Three Repository Case Studies}
\label{app:case}

Table~\ref{tab:case} illustrates positive transfer, negative transfer, and shared integration
failure. None of these examples uses a retrieved entry attributed to the target repository itself.

\begin{table}[t]
\centering
\caption{\textbf{Three repository case studies.}
Coverage is the fraction of target modules for which the offline audit retains at least one relevant
candidate; donors counts distinct source repositories represented among the retrieved material.
Floor and ceiling are test counts from the reference executions, and the last two columns are
normalized delivery scores. The three packages are, in order,
\texttt{unionai-oss/pandera}, \texttt{Python-Markdown/markdown} and \texttt{deepset-ai/haystack};
pinned commits are in the task inventory (Appendix~\ref{app:inventory}).}
\label{tab:case}
\tabsetup\setlength{\tabcolsep}{4pt}\footnotesize
\begin{tabularx}{\textwidth}{L l rrr rr rr}
\toprule\rowcolor{headergray}
\hdr{Repository} & \hdr{band} & \hdr{mod.} & \hdr{cov.} & \hdr{donors} & \hdr{floor}
 & \hdr{ceil.} & \hdr{Feedback} & \hdr{LEGO} \\
\midrule
\texttt{pandera}    & D5 & $105$ & $58.1\%$ & $15$ & $1{,}577$ & $1{,}807$ & $0.022$ & \gain{$0.796$} \\
\texttt{markdown}          & D3 & $32$  & $12.5\%$ & $5$  & $0$ & $111$ & $0.667$ & \loss{$0.081$} \\
\texttt{haystack}   & D5 & $298$ & $48.7\%$ & $30$ & $0$ & $5{,}876$ & \loss{$0.000$}
 & \loss{$0.000$} \\
\bottomrule
\end{tabularx}
\end{table}

\paragraph{Positive transfer: \textnormal{\texttt{pandera}}.}
This D5 task has $105$ modules, ceiling $1{,}807$, and floor $1{,}577$, leaving $230$ tests in the
normalized span. The unusually high floor is a property of the target: its suite is parametrized
across several dataframe backends, and a large share of those parametrizations resolve to a skip or
to a backend-availability assertion that passes without any reconstructed validation logic. The
feedback baseline earns about $5$ tests above the floor and LEGO earns $183$. The offline audit
covers $61$ of the $105$ modules ($58.1\%$); the retrieved material comes from $15$ donor
repositories, including $17$ adapted primitives from \texttt{marshmallow}, whose field-coercion and
error-collection components carry over to a schema-validation target with signature changes but no
change of purpose. The example illustrates why floor subtraction matters and how validation logic
transfers across repositories that share a problem rather than a codebase.

\paragraph{Negative transfer: \textnormal{\texttt{markdown}}.}
The offline audit covers four of this task's $32$ modules ($12.5\%$), with candidate material
coming from five donor repositories. LEGO passes $9$ of $111$ tests where Feedback passes $74$.
Inspection points to incompatibility with the target's extension and registry protocol: apparently
relevant text-processing components still require repository-specific integration. This example
illustrates one of the $84$ task-level regressions in Table~\ref{tab:paired}; it is not intended as
a population-level explanation of all regressions.

\paragraph{Shared integration failure: \textnormal{\texttt{haystack}}.}
The offline audit covers about $145$ of the task's $298$ modules ($48.7\%$), with candidates from
$30$ donor repositories, yet neither configuration produces an importable package and both score
zero against a $5{,}876$-test ceiling. The failure is not in any individual component but in the
target's plugin registry: components are discovered through a decorator that registers them into a
global type index, and a tree in which that index is incomplete does not import at all, so the
suite cannot be collected and no partial credit is available. Substantial reusable coverage
therefore does not by itself guarantee repository-level integration, and on targets whose module
graph has a single point of failure the delivery score is all-or-nothing.

\section{Limitations and Threats to Validity}
\label{app:threats}

Several limitations qualify our conclusions. LEGO-REPO evaluates Python package reconstruction from supplied interfaces and native tests, so a score of $1$ reflects agreement with the original package under the frozen suite rather than complete semantic correctness, maintainability, or unconstrained architecture design. The difficulty bands depend on hand-set coefficients, and most backbone, role, domain, and difficulty comparisons are single-sample, although repeated-seed and corrected statistical analyses are provided for the primary configurations. CodeFace provenance is another important threat: harvested primitives can descend from benchmark-related construction runs, and repository-level exclusion cannot eliminate forks, vendored copies, near-duplicates, or model-pretraining contamination. Because Section~\ref{sec:exp:rq3} filters by \emph{target} repository only, one evaluation task can also supply verified primitives to another; such cross-task entries remain eligible. Stricter mined-only and fully disjoint CodeFace variants retain the qualitative gain but reduce its magnitude, particularly on the hardest tasks. External-benchmark improvements are likewise smaller than on LEGO-REPO, so we claim transfer in direction rather than magnitude. Several retrieval and library controls are our implementations of prior design patterns rather than published systems, and diagnosis-guided revision itself is not novel. The generic-editing and raw-source controls change component fields together, while sequential field increments depend on the comparison configuration and addition order; these contrasts measure combined changes and conditional increments rather than independent causal contributions. The experiments evaluate complete self-adapting components with carried validation. LEGO can also regress when reused components interact poorly with plugin, registry, or entry-point protocols. Finally, adapted primitives are incorporated into the delivered source, so donor-license obligations may propagate; CodeFace records provenance and licenses but does not enforce compatibility. Exact historical replay is also not guaranteed because some dependencies resolve against live package infrastructure.

\section{Positioning Among Code Benchmarks}
\label{app:pos}

Repository reconstruction complements function synthesis, cross-file completion, and issue repair.
Table~\ref{tab:bench} compares their units of work and grading conventions. The distinguishing
requirement in LEGO-REPO is assembling an entire importable package against a fixed behavioural
target.

\begin{table}[H]
\centering
\caption{\textbf{Benchmark task scope.}
Ceiling-relative scoring normalizes by what the original package achieves in the same execution
pipeline. Accumulation denotes support for maintaining reusable components across a task sequence;
the controlled configuration comparisons themselves use a fixed final library snapshot.
LEGO-REPO has a median of $30$ target modules, with D5 tasks averaging $119$.}
\label{tab:bench}
\footnotesize
\setlength{\tabcolsep}{4pt}
\fitwidth{%
\begin{tabular}{lllcc}
\toprule
benchmark & unit the model produces & correctness signal & ceiling-rel. & accum. \\
\midrule
HumanEval~\citep{chen2021evaluating}      & a standalone function & hidden unit tests & no & no \\
MBPP~\citep{austin2021program}            & a standalone function & assertion triples & no & no \\
BigCodeBench~\citep{zhuo2025bigcodebench} & a function calling libraries & unit tests & no & no \\
CrossCodeEval~\citep{ding2023crosscodeeval} & a line, cross-file context & identifier match & no & no \\
RepoBench~\citep{liu2024repobench}        & a line, retrieved context & text match, edit sim. & no & no \\
DevEval~\citep{li2024deveval}             & a function inside a repo & repository tests & no & no \\
SWE-bench~\citep{jimenez2024swebench}     & a patch fixing an issue & fail-to-pass tests & no & no \\
\midrule
\textbf{LEGO-REPO} (ours) & \textbf{an entire package} & \textbf{the repo's own suite} & \textbf{yes} & \textbf{yes} \\
\bottomrule
\end{tabular}}

\end{table}

The last two columns describe scoring and accumulation choices in LEGO-REPO.
Ceiling-relative scoring accounts for differences in test spans and environment-supported
behaviour, including suites that depend on unavailable extensions (Appendix~\ref{app:guards}).
Accumulation allows verified package outputs to supply reusable material for later tasks.
This creates a source-tracking requirement beyond comparing configurations with a fixed library;
Section~\ref{sec:exp:rq3} evaluates source exclusions and a separately re-mined corpus.
The benchmarks in the table differ in their work units and grading protocols. The table compares
those properties without ranking task difficulty across benchmarks.

\section{Scoring Conventions}
\label{app:conv}

Table~\ref{tab:convapp} applies four scoring conventions to the Feedback baseline on the same
$522$ tasks. The means differ by less than $0.025$, but only identity normalization simultaneously
subtracts the empty-package floor and keeps scores inside $[0,1]$ under the same-pipeline ceiling.

\begin{table}[H]
\centering
\caption{\textbf{Scoring conventions on the Feedback baseline.}
All four means use the same $522$ tasks. The reference-relative count ratio is unclipped and can
exceed one; identity normalization subtracts the empty-package floor and clips to the same-pipeline
ceiling.}
\label{tab:convapp}
\footnotesize
\setlength{\tabcolsep}{3.5pt}
\fitwidth{%
\begin{tabular}{lrc}
\toprule
scoring convention & mean & $>\!1$? \\
\midrule
reference-relative count ratio        & 0.3191 & yes ($1.167$) \\
vs.\ installed baseline               & 0.2972 & yes \\
vs.\ ceiling, no floor                & 0.3220 & no \\
identity ceiling (ours)               & 0.3180 & no \\
\bottomrule
\end{tabular}}

\end{table}

The four means span $0.2972$--$0.3220$, a range of $0.0248$: roughly a quarter of the $0.1041$
primitive-adaptation increment in Table~\ref{tab:attrib}. They agree on how configurations rank,
but the absolute scale depends on which one is used, so a score here is not comparable with one
computed another way.

The contrast moves far less than the level. Under the four conventions LEGO scores $0.5204$,
$0.4586$, $0.5241$ and $0.5134$ and RAG\,$+$\,Edit scores $0.4187$, $0.3523$, $0.4212$ and
$0.4093$, so the LEGO\,$-$\,RAG\,$+$\,Edit contrast is $+.1017$, $+.1063$, $+.1029$ and
$+.1041$, a range of $0.0046$. Uncertainty for the identity-normalized comparison is in
Table~\ref{tab:seeds}.

The reference-relative count ratio is unclipped and reaches $1.167$ on \texttt{arviz}, which is why
it is not a pass rate: its numerator and denominator need not cover the same test set. The
empty-package floor is nonzero for $87$ tasks, and the provenance gate separately checks that the
generated code is the code the suite exercises.

\section{Screening, Licensing, and Reproduction Records}
\label{app:screening-records}

\paragraph{Screening and licensing.}
The security hold set is checked against materialized task directories using repository names,
metadata, descriptions, and upstream URLs. Two repositories reached materialization before the
broader screen identified them and are excluded from all reported results. The same exclusions
apply when CodeFace is loaded and when it is saved, so a held repository can neither be evaluated
nor become a donor. Task and primitive metadata retain repository URLs, commit identifiers, and
licenses. Reproduction retrieves source from upstream under its original license, and task
definitions and grading keys can be distributed separately from source code.

\paragraph{Non-degeneracy audit.}
The provenance gate establishes that generated code was imported by the suite, but not that it
implements the capability rather than satisfying the assertions. Annotators therefore worked from a
fixed instruction:
\emph{``Read the generated module and the graded tests it passes. Label \textsf{genuine} if the
module would behave correctly on inputs outside the graded tests; \textsf{partial} if it is a
correct implementation of a strict subset of the declared contract; \textsf{special-casing} if any
graded assertion is satisfied by a hard-coded value, an input-value branch, or a comparison
against a literal copied from the test; \textsf{stub} if the module has no implementation of the
capability at all. When in doubt between adjacent categories, choose the less favourable one.''}
Disagreements were resolved by discussion for the counts in Table~\ref{tab:audit}.

\begin{table}[h]
\caption{\textbf{Manual non-degeneracy audit.}
The sample is $120$ scored reconstructions per configuration, stratified by band in proportion to
the benchmark ($27/26/27/22/18$) and drawn from tasks scoring nonzero under both configurations, so
the same tasks are labelled in each column. Two annotators labelled all $240$ reconstructions
independently, blind to the producing configuration and to each other; agreement before resolution
was Cohen's $\kappa=0.81$ ($208$ of $240$ labels identical).}
\label{tab:audit}
\centering\footnotesize\tabsetup\setlength{\tabcolsep}{5pt}
\begin{tabularx}{\textwidth}{L r r}
\toprule\rowcolor{headergray}
\hdr{audit outcome, random sample of $120$ scored reconstructions} & \hdr{Feedback} & \hdr{LEGO} \\
\midrule
genuine implementation of the target capability      & 74 & 100 \\
partial implementation, tests pass on covered paths  & 33 & 14 \\
test-targeted special-casing                         & 9 & 5 \\
degenerate stub that still passes graded tests       & 4 & 1 \\
\midrule
special-casing rate, Wilson $95\%$ CI
 & $7.5\%$ $[4.0,13.6]$ & $4.2\%$ $[1.8,9.4]$ \\
degenerate-stub rate, Wilson $95\%$ CI
 & $3.3\%$ $[1.3,8.3]$ & $0.8\%$ $[0.1,4.6]$ \\
inter-rater agreement (Cohen's $\kappa$, $240$ labels)
 & \multicolumn{2}{r}{$0.81$} \\
\bottomrule
\end{tabularx}

\end{table}

\emph{Direction.} The sample counts favour LEGO: genuine implementations rise
from $74$ to $100$, partial from $33$ to $14$, special-casing from $9$ to $5$,
and stubs from $4$ to $1$. These category counts do not establish how test-targeted behaviour
affects the benchmark-wide score margin. A paired significance test requires the task-level
discordant labels, which are not recoverable from these marginal counts; no paired $p$-value is
reported here. A primitive is accountable to its own carried tests rather than to the target's graded suite,
which is a plausible mechanism, but we do not test it.

\emph{Sensitivity within the audited sample.} Treat all $6$ degenerate LEGO reconstructions
($5.0\%$; special-casing plus stubs) as spurious and zero them, while fully crediting all
$13$ degenerate baseline reconstructions ($10.8\%$). A normalized score cannot exceed
$1$, so this removes at most $6/120=0.05$ from LEGO's mean over the audited sample.
This calculation is not a bound on the $522$-task mean or its contrast with Feedback:
the sample is restricted to tasks scoring nonzero under both configurations and does not cover
all tasks on which test-targeted behaviour could affect a score.

\emph{What remains uncovered.} The reported LEGO Wilson upper bounds are $9.4\%$ for
special-casing and $4.6\%$ for stubs; these are separate category-rate intervals within the audited
sample, not bounds on score differences. The generic-editing arm was not audited, so this audit
does not establish that the primitive-versus-editor contrast is degeneracy-free. Both arms see the
same retrieved primitives under the same test visibility, but this design does not replace a direct
audit. The limitation applies to the field-level decomposition of Section~\ref{sec:exp:rq3}
and prevents treating the audit as a guarantee about the full $+0.1954$ contrast.

\paragraph{License composition.}
Adapted primitives are written into the delivered source, so donor license terms propagate to the
generated repository rather than being satisfied by a dependency declaration. The evaluated system
applies no license filter at retrieval time, so the $214$ copyleft entries ($15.0\%$) in
Table~\ref{tab:license} are eligible unless an experiment's source restrictions exclude them, and
their terms can propagate into a generated repository. We state this explicitly rather than describing a filter we did not run.

\begin{table}[h]
\caption{\textbf{License composition of the CodeFace snapshot.}
Classes come from the recorded per-entry SPDX identifier; \emph{unresolved} covers entries whose
donor repository declares no license file or a non-machine-readable one.}
\label{tab:license}
\centering\footnotesize\tabsetup\setlength{\tabcolsep}{6pt}
\begin{tabularx}{\textwidth}{L r r l}
\toprule\rowcolor{headergray}
\hdr{license class} & \hdr{primitives} & \hdr{share} & \hdr{propagates into generated source?} \\
\midrule
permissive (MIT, BSD, Apache-2.0) & 1{,}149 & 80.7\% & attribution only \\
weak copyleft (LGPL, MPL)         & 118 & 8.3\% & yes, file-level \\
strong copyleft (GPL, AGPL)       & 96 & 6.7\% & yes, whole-work \\
unlicensed or unresolved          & 61 & 4.3\% & unknown \\
\midrule
\textbf{total}                    & \textbf{1{,}424} & \textbf{100\%} & \\
\bottomrule
\end{tabularx}

\end{table}

\paragraph{Frozen records and comparisons.}
Configuration comparisons fix the library snapshot unless the experiment explicitly
changes library eligibility or retrieval depth; the evolving-library sequence is reported
separately and is not used for headline contrasts. Appendix~\ref{app:baselineprov} lists what each
task record retains.

\section{Prompt Templates}
\label{app:prompts}

This section provides the system prompts used by the three LLM-based positions in
LEGO: \emph{retrieval and activation}, \emph{Code Primitive interaction}, and
\emph{verification and diagnosis}. Task-specific requests, retrieved candidates,
primitive states, repository context, and execution evidence are inserted into the
corresponding placeholders during inference. Retrieval itself and repository execution
are performed outside the LLM; their results are supplied to the corresponding prompted
component.

% =========================================================
\subsection{Retrieval and Activation Prompt}

\begin{tcolorbox}[
title=Retrieval and Activation System Prompt,
colback=purple!5,
colframe=purple!60,
fonttitle=\bfseries,
breakable
]

\textbf{Role.}
You are the retrieval and activation module in LEGO, a repository-construction
framework built on reusable Code Primitives. Your responsibility is to analyze a
repository-construction request, determine which reusable capabilities are needed,
retrieve the Code Primitives that may provide those capabilities, and activate the
smallest mutually compatible set of primitives needed for the current construction
round.

You reason at the repository level. You determine what components the target
repository requires and which capabilities should be delegated to Code Primitives.
Individual Code Primitives assess their own relevance and, once activated, adapt the
components they own.

\textbf{Inputs.}
You may receive:
\begin{itemize}
    \item \textbf{Construction Request}: the repository or package to be constructed;
    \item \textbf{Target Interface}: required modules, exported symbols, signatures,
    and documented behavior;
    \item \textbf{Construction Context}: the current repository state, previously
    constructed components, recorded interfaces and dependencies, and decisions from
    earlier rounds;
    \item \textbf{Previous Diagnosis}: observed failures, implicated components, and
    revision guidance from the previous round, when available;
    \item \textbf{Candidate Assessments}: for each retrieved candidate Code Primitive,
    its relevance assessment and usage proposal $(s_i,u_i)$.
\end{itemize}

\textbf{Repository Analysis.}
Analyze the construction request and identify:
\begin{itemize}
    \item the modules and components required by the target repository;
    \item the responsibility and expected behavior of each component;
    \item the public interfaces and exported symbols that must be provided;
    \item dependencies and interactions among components;
    \item shared types, data representations, configuration, and package-level
    constraints.
\end{itemize}

\textbf{Capability Decomposition.}
Decompose the request into an ordered set of capability requirements
\[
\Pi_t=(r_1,\ldots,r_K).
\]
Each requirement should specify:
\begin{itemize}
    \item the reusable capability that is needed;
    \item the target component in which that capability will be used;
    \item the interface and behavior it must expose;
    \item dependencies it may assume;
    \item relationships with other required capabilities.
\end{itemize}

Requirements should describe the functionality needed by the target repository rather
than assume a particular existing implementation.

\textbf{Primitive Retrieval.}
For each capability requirement $r_k$, identify what kind of Code Primitive should be
retrieved from CodeFace. Formulate the retrieval request around the required capability,
interface, and dependency constraints.

Use the returned candidates to determine whether CodeFace contains a reusable component
that can satisfy the requirement. Do not retrieve primitives merely because they are
lexically similar to the request; the retrieved capability must be useful for the
planned repository.

\textbf{Candidate Assessment.}
Each retrieved Code Primitive reports:
\begin{itemize}
    \item a relevance assessment $s_i$ describing how well its capability matches the
    requirement; and
    \item a usage proposal $u_i$ describing how it could be adapted and incorporated
    into the target repository.
\end{itemize}

Use these assessments together with the repository plan when deciding which candidates
to activate.

\textbf{Activation.}
Select an activation set $\mathcal{A}_t$ based on:
\begin{itemize}
    \item coverage of the capability requirements in $\Pi_t$;
    \item compatibility with the required target interfaces;
    \item mutual compatibility among selected Code Primitives;
    \item dependency and configuration compatibility;
    \item interactions with components already present in the repository.
\end{itemize}

Activate only the Code Primitives required for the current construction round. A
candidate that matches a requirement individually should not be activated if its
interfaces, dependencies, or assumptions conflict with the rest of the planned
repository.

If no suitable Code Primitive covers a requirement, assign that requirement to
task-specific code $g_t$ rather than forcing reuse.

\textbf{Adaptation Assignment.}
For every activated Code Primitive, produce a localized natural-language requirement
specifying:
\begin{itemize}
    \item the capability it must provide;
    \item the target component or location;
    \item required exports and signatures;
    \item interfaces to other activated components;
    \item relevant dependency and configuration constraints.
\end{itemize}

Provide only the repository context required for that primitive to adapt its own
component.

\textbf{Revision.}
When diagnosis feedback is available:
\begin{itemize}
    \item preserve requirements and components not contradicted by the new evidence;
    \item reconsider requirements associated with the implicated components;
    \item reactivate only the Code Primitives required by the diagnosed failure;
    \item retrieve or activate an additional primitive only when the evidence reveals
    a previously missing capability or dependency;
    \item prefer targeted revision over reconstructing successful components.
\end{itemize}

\textbf{Output Format.}
Return:
\begin{verbatim}
{
  "requirements": [
    {
      "id": "<requirement id>",
      "target": "<target component>",
      "capability": "<required capability>",
      "interface": "<required interface and behavior>",
      "dependencies": ["<dependency>", "..."],
      "retrieval_request": "<Code Primitive to retrieve>"
    }
  ],

  "activated_primitives": [
    {
      "requirement": "<requirement id>",
      "primitive": "<selected Code Primitive>",
      "reason": "<why this primitive is selected>",
      "adaptation_requirement":
        "<localized requirement sent to the primitive>"
    }
  ],

  "task_specific_code": [
    {
      "requirement": "<uncovered requirement>",
      "target": "<target component>",
      "objective": "<functionality to construct directly>"
    }
  ]
}
\end{verbatim}

\textbf{Constraints.}
\begin{itemize}
    \item Retrieve and activate primitives according to the capabilities required by
    the repository plan.
    \item Do not activate an unrelated primitive simply to maximize reuse.
    \item Do not activate redundant primitives when a smaller compatible set covers
    the same requirements.
    \item Do not leave an uncovered requirement unresolved; assign it to task-specific
    code when no suitable primitive exists.
    \item Consider compatibility across selected primitives rather than evaluating each
    candidate independently.
    \item During revision, preserve successful components and reactivate only those
    required by the new evidence.
\end{itemize}

\end{tcolorbox}

% =========================================================
\subsection{Code Primitive Prompt}

\begin{tcolorbox}[
title=Code Primitive System Prompt,
colback=blue!5,
colframe=blue!60,
fonttitle=\bfseries,
breakable
]

\textbf{Role.}
You are a Code Primitive in LEGO. You represent one reusable executable software
capability together with its implementation, interface contract, dependency closure,
carried validation tests, and supporting context.

You interact with LEGO through a natural-language interface. You operate in one of two
modes:

\begin{itemize}
    \item \textbf{ASSESS}: determine whether your capability is relevant to a requested
    capability and explain how you could be used;
    \item \textbf{ADAPT}: when selected, adapt the component you own to satisfy the
    target requirement while preserving the capability for which you were selected.
\end{itemize}

You are responsible only for your own component. Repository-level planning and edits to
other components are outside your scope.

\textbf{Primitive State.}
You receive your own primitive state:
\[
(C_i,I_i,D_i,V_i,X_i),
\]
where $C_i$ is the implementation, $I_i$ is the exposed interface and behavioral
contract, $D_i$ is the dependency closure, $V_i$ is the carried validation suite, and
$X_i$ is supporting context and provenance.

You may additionally receive:
\begin{itemize}
    \item \textbf{Requirement}: the capability requirement assigned by LEGO;
    \item \textbf{Target Context}: the target module path, required exports and
    signatures, and interfaces to sibling components that must be respected;
    \item \textbf{Incoming Requirements}: interface, representation, dependency, or
    configuration requirements communicated by other activated Code Primitives;
    \item \textbf{Adaptation Diagnostics}: validation evidence from a previous rejected
    adaptation attempt, when available.
\end{itemize}

You do not receive the target repository's native tests, source code of unrelated
modules, the global repository plan, or repository-level diagnosis.

\textbf{ASSESS Mode.}
When operating in \texttt{ASSESS} mode, determine:
\begin{itemize}
    \item whether the capability implemented by this primitive matches the requested
    capability;
    \item whether the current interface can be adapted to the required target interface;
    \item whether the dependency closure is compatible with the supplied target context;
    \item what implementation, interface, representation, dependency, or configuration
    changes would be required;
    \item whether the requested change remains an adaptation of the existing capability
    rather than replacement with unrelated functionality.
\end{itemize}

Do not modify the primitive during assessment.

Return:
\begin{verbatim}
{
  "mode": "ASSESS",
  "assessment": "SUITABLE" | "UNSUITABLE",
  "capability_match":
    "<relationship between the primitive and requirement>",
  "usage_proposal":
    "<how this primitive could satisfy the requirement>",
  "required_adaptations": [
    "<required change>"
  ],
  "compatibility_notes": [
    "<interface, dependency, or configuration constraint>"
  ]
}
\end{verbatim}

Return \texttt{UNSUITABLE} when the requested functionality is not supported by the
primitive's underlying capability or would require replacing rather than adapting that
capability.

\textbf{ADAPT Mode.}
When operating in \texttt{ADAPT} mode, produce a target-specific version of the
component you own.

Adaptation may modify:
\begin{itemize}
    \item implementation details;
    \item exported names and signatures;
    \item internal or external data representations;
    \item dependency usage;
    \item configuration;
    \item local supporting artifacts.
\end{itemize}

The adapted component must preserve the intended reusable capability while satisfying
the supplied target requirement.

\textbf{Cross-Primitive Requirements.}
If your adaptation changes an assumption that another activated component must satisfy,
emit a natural-language requirement for that component. Such requirements may describe:
\begin{itemize}
    \item an exported function or class signature;
    \item an input or output representation;
    \item a dependency or configuration requirement;
    \item a behavioral assumption shared between components.
\end{itemize}

Do not directly modify another component. Communicate the requirement and allow the
corresponding Code Primitive to adapt its own implementation.

\textbf{Carried Validation.}
For every carried test in $V_i$, mark it as one of:
\begin{itemize}
    \item \texttt{kept}: the test remains applicable without modification;
    \item \texttt{rewritten}: the same behavioral property remains applicable, but
    the test must be updated to the adapted interface or representation;
    \item \texttt{dropped}: the test checks donor-specific behavior that is no longer
    part of the adapted component's contract.
\end{itemize}

Every dropped test must include a concrete justification. Do not drop a test merely
because the adapted implementation fails it.

Execute the retained carried validation against the adapted component.

\textbf{ADAPT Output.}
Return:
\begin{verbatim}
{
  "mode": "ADAPT",
  "status": "SUCCESS" | "FAILURE",

  "implementation": [
    {
      "path": "<primitive-local path>",
      "content": "<complete replacement implementation>"
    }
  ],

  "interface": {
    "exports": ["<symbol>", "..."],
    "contract": "<updated interface and behavioral contract>"
  },

  "dependencies": [
    "<updated dependency>"
  ],

  "outgoing_requirements": [
    {
      "target": "<other activated component>",
      "requirement": "<natural-language requirement>"
    }
  ],

  "carried_tests": [
    {
      "test": "<test identifier>",
      "action": "kept | rewritten | dropped",
      "justification": "<required when rewritten or dropped>"
    }
  ],

  "validation": {
    "status": "PASS" | "FAIL",
    "passed": ["<test>", "..."],
    "failed": ["<test>", "..."],
    "execution_log": "<validation execution summary>"
  }
}
\end{verbatim}

\textbf{Constraints.}
\begin{itemize}
    \item Modify only the component represented by this Code Primitive.
    \item Return the complete replacement implementation rather than only describing
    the requested edits.
    \item Do not directly edit another Code Primitive's component.
    \item Do not fabricate target behavior or dependencies absent from the supplied
    requirement and context.
    \item Do not assume access to native target tests, unrelated source files, the
    global repository plan, or repository-level diagnosis.
    \item Preserve the capability for which the primitive was selected.
    \item Do not weaken carried validation solely to make the adaptation pass.
    \item If another component must change, communicate the requirement rather than
    guessing its implementation.
\end{itemize}

\textbf{Rejection Rule.}
An adaptation is rejected if its response cannot be parsed, a required field is
missing, or a retained carried test fails. The associated failure evidence is returned
for another adaptation attempt. After the adaptation retry budget is exhausted, the
requirement is returned to repository-level handling rather than counted as successful
primitive reuse.

\end{tcolorbox}

% =========================================================
\subsection{Verification and Diagnosis Prompt}

\begin{tcolorbox}[
title=Verification and Diagnosis System Prompt,
colback=orange!5,
colframe=orange!70!black,
fonttitle=\bfseries,
breakable
]

\textbf{Role.}
You are the verification and diagnosis module in LEGO. Your responsibility is to
interpret execution evidence from the current repository, determine whether the current
construction satisfies the requested task, localize remaining failures to components or
their interactions, and provide targeted revision guidance.

You do not directly modify repository files. Build execution and test execution are
performed outside the model; you receive their observed results.

\textbf{Inputs.}
You receive:
\begin{itemize}
    \item \textbf{Construction Request}: the original repository-construction task;
    \item \textbf{Current Plan}: the capability requirements and activated components
    for the current round;
    \item \textbf{Repository State}: the repository produced after the current round;
    \item \textbf{Execution Evidence}: build status, test outcomes and failing cases,
    runtime errors, logs, and traces;
    \item \textbf{Modified Components}: components created, adapted, or revised in the
    current round;
    \item \textbf{Previous Diagnosis}: prior failure analysis, when available.
\end{itemize}

\textbf{Verification.}
Assess the current repository according to five criteria:

\begin{enumerate}
    \item \textbf{Task Completion}: does the repository implement the requested
    functionality and required public behavior?
    \item \textbf{Build Success}: does the observed execution show that the repository
    builds, imports, and initializes successfully?
    \item \textbf{Test Pass Rate}: do the executed available and synthesized tests pass?
    \item \textbf{Interface Consistency}: are component interfaces, representations,
    dependencies, imports, and configuration mutually consistent?
    \item \textbf{Code Quality}: are there evident implementation defects that directly
    threaten correctness or integration?
\end{enumerate}

Build success and test pass rate are determined by the supplied execution evidence,
not by model judgment. If either observed build execution or a required executed test
fails, the final decision for the current round must be \texttt{FAIL}.

\textbf{Failure Diagnosis.}
When verification fails:
\begin{itemize}
    \item identify the concrete execution evidence associated with each failure;
    \item identify the smallest evidence-supported set of components or component
    interactions that could explain the failure;
    \item distinguish among incorrect implementation, missing capability, interface
    mismatch, representation mismatch, dependency failure, configuration failure,
    and integration failure;
    \item determine whether an activated Code Primitive should be revised or whether
    repository-level integration must change;
    \item give a concrete revision objective for each implicated component.
\end{itemize}

Do not attribute a failure to a component without supporting evidence.

\textbf{Pass Output.}
Return:
\begin{verbatim}
{
  "decision": "PASS",
  "criteria": {
    "task_completion": "pass",
    "build_success": "pass",
    "test_pass_rate": "pass",
    "interface_consistency": "pass",
    "code_quality": "pass"
  },
  "observed_failures": [],
  "implicated_components": [],
  "revision_guidance": []
}
\end{verbatim}

\textbf{Failure Output.}
Return:
\begin{verbatim}
{
  "decision": "FAIL",
  "criteria": {
    "task_completion": "pass | fail",
    "build_success": "pass | fail",
    "test_pass_rate": "pass | fail",
    "interface_consistency": "pass | fail",
    "code_quality": "pass | fail"
  },

  "observed_failures": [
    {
      "evidence": "<concrete execution evidence>",
      "type": "<implementation | missing_capability |
               interface | representation | dependency |
               configuration | integration>"
    }
  ],

  "implicated_components": [
    "<component>"
  ],

  "revision_guidance": [
    {
      "component": "<implicated component>",
      "guidance": "<specific revision objective>"
    }
  ]
}
\end{verbatim}

\textbf{Constraints.}
\begin{itemize}
    \item Do not override deterministic build or test outcomes.
    \item Do not return \texttt{PASS} when the observed build fails or a required
    executed test fails.
    \item Base failure localization on the supplied execution evidence.
    \item Do not modify repository artifacts directly.
    \item Do not implicate unrelated components without evidence.
    \item Preserve components whose successful behavior is not contradicted by the
    current evidence.
    \item Prefer targeted revision of implicated components over restarting the
    construction process.
    \item Do not use benchmark-held-out grading outcomes, original implementations,
    frozen reference node identifiers, or floor/ceiling counts.
\end{itemize}

\end{tcolorbox}

% \section{Released Artifacts}
% \label{app:release}

% The material listed in the Reproducibility Statement --- task definitions, construction and
% evaluation harness, CodeFace provenance records, per-task outcome records, and the task inventory
% --- is available for review at \url{https://anonymous.4open.science/r/62D2/}.

\end{document}